\unless\ifcsname FFFvariant\endcsname
\fi
\documentclass[graybox, secnum]{svmult}

\usepackage{newtxtext,newtxmath} 
\usepackage{helvet}         
\usepackage{courier}        
\usepackage{type1cm}        
\usepackage{mathrsfs}
\usepackage{makeidx}         
\usepackage{graphicx}        
\usepackage{multicol}        
\usepackage[bottom]{footmisc}
\usepackage{hyperref}        
\usepackage{soul}            
\hypersetup{colorlinks=true,urlcolor=blue,citecolor=orange}
\usepackage[square,numbers]{natbib}
\makeindex             
\usepackage[textsize=tiny,backgroundcolor=yellow]{todonotes}
\usepackage{dsfont}%
\usepackage{booktabs}%
\usepackage{cancel}%
\usepackage{subcaption}%
\usepackage[inline]{enumitem}%
\usepackage{mathtools}
\usepackage{cleveref}
\usepackage[nolist]{acronym}
\usepackage{nicefrac}
\usepackage{xspace}
\usepackage{mleftright}
\mleftright
\makeatletter
\g@addto@macro\bfseries{\boldmath}
\makeatother
\newlist{goals}{enumerate*}{1}
\setlist[goals]{label=(\arabic*),ref=(\arabic*)}
\crefname{goalsi}{goal}{goals}
\Crefname{goalsi}{Goal}{Goals}
\newlist{challenges}{enumerate}{1}
\setlist[challenges]{label=\arabic*.,ref=\arabic*}
\crefname{challengesi}{challenge}{challenges}
\Crefname{challengesi}{Challenge}{Challenges}

\begin{document}
\title*{A Lightning Introduction to String Theory}
\author{Carlo Angelantonj and Ioannis Florakis}%
\institute{
Carlo Angelantonj\at Dipartimento di Fisica,
Universit\`a di Torino and INFN Sezione di Torino, Via Pietro Giuria 1, 10125 Torino, Italy, \email{carlo.angelantonj@unito.it}
\and
Ioannis Florakis \at Department of Physics,
University of Ioannina, 45110 Ioannina, Greece, \email{iflorakis@uoi.gr}
}
%
%
\maketitle 
\abstract{We give a lightning introduction to critical string theory, including the 26-dimensional bosonic string, the 10-dimensional superstrings and heterotic strings with and without spacetime supersymmetry. We also discuss open strings and D-branes, as well as the orientifold constructions, in ten dimensions. }

\section*{Keywords}
\textcolor{gray}{String Theory, Bosonic String, Superstrings, Heterotic Strings, D-branes, Orientifolds}
\clearpage
\tableofcontents
\markboth{Carlo Angelantonj and Ioannis Florakis}{Introduction to String Theory}

\acresetall

\section{Motivating String Theory}

String theory is currently regarded as a most promising framework in which quantum gravity is unified with gauge interactions and matter. Its origins can be traced back to the seminal work of Veneziano \cite{Veneziano:1968yb}, who proposed the simple expression
\begin{equation}
	A(s,t,u) = B(-\alpha(s),-\alpha(t))+B(-\alpha(s),-\alpha(u))+B(-\alpha(t),-\alpha(u))\,,
	\label{Ven1}
\end{equation}
for a relativistic $2\to 2$ amplitude  obeying the requirements of Regge trajectories and high-energy behaviour along with  crossing symmetry, in order to account for the large amount of experimental data on hadronic resonances produced throughout the '60s. 
$B(-\alpha(s),-\alpha(t))$ is the Euler beta function with $\alpha(z) = \alpha' z + \alpha_0$, clearly symmetric under the exchange of its arguments,
which guarantees the full $s\leftrightarrow t\leftrightarrow u$ crossing symmetry in \eqref{Ven1}, where $s,t,u$ are the familiar Mandelstam variables. Using standard properties of the Euler gamma function, it follows that $B(-\alpha(s),-\alpha(t))$ has an infinite number of equally spaced poles at $t=(n-\alpha_0)/\alpha' $, $n=0,1,2,\ldots$, with residues that are order-$n$ polynomials in $s$. This incorporates the linearly rising Regge trajectories 
\begin{equation}
	m^2 = (J-\alpha_0)/\alpha'  \,,
\end{equation}
observed in many hadronic resonances, where $J=n$ is the maximal spin of the particle exchanged in the $t$ channel, and a similar behaviour is clearly present in the $s$ and $u$ channels. Experimental fits required the Regge slope $\alpha'$ to be of order $\alpha' \sim 1\, \text{GeV}^{-2}$, while the Regge intercept $\alpha_0$ was left undetermined. Subsequent works dealing with unitarity and the absence of ghosts, led to no-ghost theorems fixing $\alpha_0 = 1$ and the dimension of spacetime to $D=26$.

In the high energy regime, the Stirling formula implies the asymptotic behaviour $\sim s^{\alpha(t)}$ for the beta function, valid in the Regge limit of large $s$ and fixed $t$. For sufficiently negative $t$, the high energy behaviour of the Veneziano amplitude is thus extremely soft, something which can only occur if an infinite number of particles are exchanged in the $t$ channel. Similarly,  the Veneziano amplitude at  high energy and fixed angle is exponentially suppressed $\sim f(\theta)^{-\alpha(s)}$, with $f$ a given function of the scattering angle $\theta$.

Soon after the work of the Veneziano, it was realised \cite{Nambu:1970xx, Nielsen:1970xx, Susskind:1970xm, Susskind:1970qz} that the linear Regge trajectory could be reproduced by the mechanical model of a vibrating quantum string. In this framework, the interactions are no longer localised in spacetime, which explains the softness of the high energy limit. 

This extreme UV softness of the dual models emerging from \eqref{Ven1}, together with their prediction of a massless spin-2 (hadronic) particle, and the surprising requirement for the dimensionality of spacetime, appeared to be incurable problems. The discovery of quantum chromodynamics (QCD) as the correct theory of strong interactions, eventually led to their demise.

Actually, the presence of a massless spin-2 particle, and the soft behaviour of the dual models turned out to be a blessing in disguise when, thanks to the seminal work of Yoneya \cite{Yoneya:1973ca, Yoneya:1974jg} and Scherk and Schwarz \cite{Scherk:1974ca, Scherk:1975xx}, it was proposed that the models of relativistic strings should rather be interpreted as a theory of quantum gravity, with $\alpha'$ now being related to the Planck scale. In this description, the apparently ``wrong" dimensionality of spacetime is no longer a problem, since spacetime becomes dynamical and can undergo a spontaneous compactification down to four dimensions.

Ever since, string theory underwent a series of momentous developments which contributed to sharpening our understanding of the subject and led to the many diverse steps which eventually shaped it into its present form. The string section of the handbook aims at giving an introduction to this exciting field, ranging from formal aspects to phenomenological implications in particle physics and cosmology. 

This chapter is organised as follows. In Section \ref{ST:Bosonic} we introduce the bosonic string, its quantisation and discuss its light spectrum in some detail. The partition function for closed bosonic strings is derived using the operatorial approach and its interpretation in terms of the torus amplitude is given. In Section \ref{ST:CFT} we focus on the salient properties of the conformal field theory living on the two-dimensional string world-sheet, including the Virasoro algebra and the $b,c$ ghost system.
In Section \ref{ST:PerturbationTheory}, we discuss the path integral and BRST quantisation, the vertex operators for physical fields are derived and the structure of string perturbation theory is presented. Conformal invariance on a general background and the associated low-energy effective action are briefly discussed. The Shapiro-Virasoro and Veneziano amplitudes are derived in some detail, along with a path integral evaluation of the torus vacuum amplitude. Fermionic strings and their light-cone quantisation are presented in Section \ref{ST:Fermionic}, with emphasis on modular invariance and the GSO projections, leading to the construction of the ten-dimensional type IIA, IIB, 0A, 0B theories. We briefly comment on the concept of orbifolds, within the context of constructing 0A, 0B from type IIA, IIB superstrings. Heterotic strings are introduced in Section \ref{ST:Heterotic}, along with  their ten-dimensional vacuum configurations with and without space-time supersymmetry. Section \ref{ST:Dbranes} contains a discussion of open strings and their boundary conditions, leading to the notion of D-branes. Their tensions and R-R charges are computed by studying the corresponding cylinder (transverse channel) amplitudes. Finally, the orientifold construction, involving orientifold planes and D-branes, is presented in Section \ref{ST:Orientifold}. The ten-dimensional Type I superstring and the Sugimoto vacuum, together with the various orientifolds of type 0B superstrings are also discussed.

The topics outlined in this review clearly do not exhaust this very rich field, and the style of the presentation reflects the authors' idiosyncrasies and biases. Given the lightning spirit of the exposition, the list of provided references is clearly incomplete and we apologise in advance for the inevitable omissions. We warmly encourage the interested reader to consult the many excellent books \cite{Green:1987sp, Green:1987mn, Lust:1989tj, Polchinski:1998rq, Polchinski:1998rr, Johnson:2003glb, Zwiebach:2004tj,  Ortin:2015hya, Becker:2006dvp, Kiritsis:2019npv, Gubser:2010zz,  Dine:2007zp, Ibanez:2012zz, West:2012vka, Blumenhagen:2013fgp, Schomerus:2017ycz, Szabo:2020ipd,  Mohaupt:2022uqx} and reviews \cite{Schwarz:1982jn, Green:1986wn, Peskin:1987rz, DHoker:1988pdl, Polchinski:1994mb, Polchinski:1996fm, Polchinski:1996na, Bachas:1996sc, Ooguri:1996ik, Vafa:1997pm, Dijkgraaf:1997ip, Kiritsis:1997gu, Sen:1998kr, Bachas:1998rg,  Schwarz:2000ew, Forste:2001ah, Angelantonj:2002ct,  Wadia:2008yn} available in the literature, many of which provide detailed references to the original papers. 

We also refer to the various other contributions to this handbook for an introduction to the tentacular developments and applications of string theory.


\section{The Bosonic String}
\label{ST:Bosonic}

Consider a point-particle of mass $m$ freely moving in $D$-dimensional Minkowski spacetime of signature $(-,+,\ldots,+)$. It traces a world-line $x^\mu(\tau)$ parametrised by the proper time $\tau$, whose 
invariant length $ds^2=-\eta_{\mu\nu} dx^\mu dx^\nu$ is defined by the Minkowski metric $\eta_{\mu\nu}$, where $\mu,\nu=0,\ldots,D-1$. The dynamics is controlled by the action
\begin{equation}
	S = -m \int  ds = -m \int d\tau \, \sqrt{-\eta_{\mu\nu}\, \dot{x}^\mu \dot{x}^\nu } \,,
	\label{PPNG}
\end{equation}
proportional to the invariant length of its world-line. For higher-dimensional objects, the natural generalisation is in terms of an action proportional to the area of the hypersurface traced by the object as it moves in spacetime.
In the case of a string of tension $T=1/2\pi\alpha'$, the action reads
\begin{equation}
	S = -T \int_{\Sigma} d\tau d\sigma \, \sqrt{-\det\left(\eta_{\mu\nu}\partial_a X^\mu \partial_b X^\nu \right)} \,,
	\label{STNG}
\end{equation}
as proposed by Nambu and Goto. Here, $\tau$ and $\sigma$ denote the proper time and the proper length on the worldsheet $\Sigma$, which is embedded into spacetime by the maps $X^\mu(\tau,\sigma)$.
Furthermore $\partial_a$, $a=0,1$ denotes the partial derivatives with respect to $\tau$ and $\sigma\in[0,1]$, respectively. One of the main differences with the point-particle case is the fact that
strings come in two different topologies: they may be either open or closed and, thus, require suitable periodicity or boundary conditions at $\sigma=0,1$. 

Notice that the actions \eqref{PPNG} and \eqref{STNG} are non-polynomial, which poses difficulties for quantising the theories. The situation may be remedied by introducing new non-dynamical auxiliary fields, which act as Lagrange multipliers and
remove the square roots. For the point particle, the new action reads
\begin{equation}
	S =  \frac{1}{2}\int d\tau \left( e^{-1} \eta_{\mu\nu}\dot{x}^\mu \dot{x}^\nu -e\,m^2 \right) \,.
\end{equation}
The equation of motion (e.o.m.) for the auxiliary field $e$ acts as the constraint $e^{-2}\,\eta_{\mu\nu}\dot{x}^\mu \dot{x}^\nu+m^2=0$, which one may solve for $e$ and plug back into the action to recover \eqref{PPNG}.
The constraint actually imposes the mass-shell condition $p^2+m^2=0$, which corresponds to the vanishing of the energy-momentum tensor of the one-dimensional theory living on the world-line of the point-particle.
This is so because the $e(\tau)$ is naturally identified as the einbein on the world-line, making reparametrisation invariance manifest.

Following a similar route in the case of the string, one introduces the world-sheet metric $g_{ab}$ of signature $(-,+)$, which plays the role of the Lagrange multipliers, and the action 
\begin{equation}
	S = -\frac{T}{2} \int_{\Sigma} d^2\sigma \, \sqrt{-g} \,g^{ab} \partial_a X^\mu \partial_b X^\nu \,\eta_{\mu\nu} \,,
	\label{STPol}
\end{equation}
as proposed by Brink, di Vecchia, Howe \cite{Brink:1976sc} and Deser, Zumino \cite{Deser:1976rb}, and known as the Polyakov action. Also in this case, the e.o.m. for $g_{ab}$ imposes the vanishing of the two-dimensional energy-momentum tensor
\begin{equation}
	T_{ab} = - \frac{2}{T} \frac{1}{\sqrt{-g}} \frac{\delta S}{\delta g^{ab}} = \partial_a X^\mu\partial_b X^\nu\, \eta_{\mu\nu} - \frac{1}{2} g_{ab} g^{cd}\partial_c X^\mu \partial_d X^\nu \eta_{\mu\nu} =0\,.
	\label{EMTensor}
\end{equation}
Again, solving the constraint and plugging it back into \eqref{STPol} reproduces \eqref{STNG}. Whether or not this equivalence persists at the quantum level is an open problem. In the following, we shall take the Polyakov action as the starting point for the quantisation of strings. 

Aside from $D$-dimensional Poincar\'e symmetry and two-dimensional diffeomorphisms $\sigma^a\to \sigma'^a(\sigma)$, the Polyakov action 
has the remarkable property of being invariant under Weyl rescaling of the two-dimensional metric, $g_{ab}\to e^{\omega(\sigma)}g_{ab}$, which is reflected in the vanishing of the trace of the energy-momentum tensor, $g^{ab}T_{ab}=0$.
This is a special property of two-dimensional world-sheets which plays a crucial role in the quantisation of the theory and singles out strings from higher-dimensional objects.

We can employ two-dimensional reparametrisation invariance and Weyl rescaling to locally gauge fix the world-sheet metric to $g_{ab}=\eta_{ab}$, usually called the conformal gauge.
With this choice, the e.o.m. for $X^\mu$ becomes simply the two-dimensional d'Alembert equation, $\partial_{+}\partial_{-}X=0$, whose general solution $X^\mu(\tau,\sigma)=X_L^\mu(\sigma^{+})+X_R^\mu(\sigma^{-})$ is the superposition of left-moving waves $X_L$ and right-moving waves $X_R$, with $\sigma^{\pm}=\tau\pm\sigma$. The tracelessness of the energy-momentum tensor implies that $T_{\pm\mp}=0$, while
\begin{equation}
	T_{\pm\pm} = \partial_{\pm} X^\mu \partial_{\pm} X^\nu \eta_{\mu\nu} \,.
\end{equation}
Its conservation then becomes $\partial_{\mp}T_{\pm\pm}=0$, so that each component is separately conserved.

In order to extract the e.o.m. from \eqref{STPol}, one needs to specify appropriate periodicity or boundary conditions. Closed strings clearly satisfy $X^\mu(\tau,\sigma+1)=X^\mu(\tau,\sigma)$ so that we can Fourier expand the solution as
\begin{equation}
	\begin{split}
		X^\mu_{L}(\sigma^+) &= \frac{1}{2}x_0^\mu + \pi\alpha' p^\mu\, \sigma^+ + i\sqrt{\frac{\alpha'}{2}} \sum_{n\neq 0} \frac{\alpha_n^\mu}{n} \, e^{-2\pi i n\sigma^+ } \,, \\
		X^\mu_{R}(\sigma^-) &= \frac{1}{2}x_0^\mu + \pi\alpha' p^\mu\, \sigma^- + i\sqrt{\frac{\alpha'}{2}} \sum_{n\neq 0} \frac{\tilde\alpha_n^\mu}{n} \, e^{-2\pi i n\sigma^- } \,.
	\end{split}
	\label{ModeExpBosonClosed}
\end{equation}
It is straightforward to see that $x_0^\mu$ corresponds to the centre of mass position of the string, which freely moves with momentum $p^\mu$. Furthermore, the reality of $X^\mu$ implies the relations $\alpha_{-n}^\mu = (\alpha_n^\mu)^*$, and similarly for the right-moving coefficients.

In the case of open strings, the boundary term arising from the Polyakov action is $\delta X^\mu \partial_\sigma X_\mu |_{\sigma=0}^{\sigma=1}=0$, and one has to distinguish among three inequivalent cases:
Neumann (N) boundary conditions $\partial_\sigma X^\mu=0$ at both endpoints, Dirichlet (D) boundary conditions $\partial_\tau X^\mu=0$ at both endpoints, or N boundary condition at one endpoint and D on the other.
In all cases, the left-moving waves are reflected into the right-moving ones so that $X_L^\mu$ and $X_R^\mu$ are no longer independent, and
\begin{equation}
	X^\mu(\sigma,\tau) = x_0^\mu + 2\pi\alpha' p^\mu\, \tau + i\sqrt{2\alpha'} \sum_{n\neq 0}\frac{\alpha_n^\mu}{n}\,e^{-i\pi n\tau}\cos(n\pi\sigma) \,,
	\label{XopenNN}
\end{equation}
for NN boundary conditions. In the DD case, the solution reads
\begin{equation}
	X^\mu(\sigma,\tau) = x_0^\mu + \delta^\mu\, \sigma + \sqrt{2\alpha'} \sum_{n\neq 0}\frac{\alpha_n^\mu}{n}\,e^{-i\pi n\tau}\sin(n\pi\sigma) \,,
	\label{XopenDD}
\end{equation}
which clearly shows that the centre of mass is no longer moving, while $x_0^\mu$ and $x_0^\mu+\delta^\mu$ are the fixed positions of the two endpoints $\sigma=0,1$.
In ND case there are no zero modes, while the frequencies are half-integral.

In canonical quantisation, one imposes the equal-time commutator relation $[X^\mu(\tau,\sigma), \Pi^\nu(\tau,\sigma')] = i\eta^{\mu\nu}\delta(\sigma-\sigma')$ between the coordinates $X^\mu$ and their conjugate momenta $\Pi^\nu=\partial_\tau X^\nu/2\pi\alpha'$. In terms of the Fourier coefficients, one finds
\begin{equation}
	[\alpha^\mu_m , \alpha^\nu_n] = m\,\delta_{m+n}\,\eta^{\mu\nu} \,, \quad  [\tilde\alpha^\mu_m , \tilde\alpha^\nu_n] = m\,\delta_{m+n}\,\eta^{\mu\nu} \,, \quad [\alpha^\mu_m , \tilde\alpha^\nu_n] = 0 \,,
	\label{AlphaCommutators}
\end{equation}
that, together with the Hermiticity conditions $\alpha^\mu_{-n}= (\alpha^\mu_n)^\dagger$, identifies them as creation and annihilation operators for an infinite number of harmonic oscillators.
The naive construction of the Fock space is plagued by negative-norm states, akin to the canonical quantisation of Maxwell theory. Indeed, $\| \alpha_{-n}^\mu |0\rangle\|^2 =n\,\eta^{\mu\mu}$ is negative in the temporal direction. 
As in the Gupta-Bleuler procedure, these ghost-like states are eliminated once the vanishing of the energy-momentum tensor is weakly imposed on the physical spectrum.
Actually, in string theory the construction of a ghost-free spectrum is more subtle since $T_{ab}$ has a quadratic dependence on the oscillators. A careful though tedious analysis reveals that it is only possible in $D=26$ dimensions.

An alternative way of quantising the theory which avoids negative-norm states is light-cone quantisation \cite{Goddard:1973qh}, that we shall now follow.  The conservation of $T_{ab}$ actually implies the existence of an infinite number of Noether charges, since
$\partial_{\mp}(f_\pm(\sigma^\pm)T_{\pm\pm})=0$ for arbitrary functions $f_{\pm}$. This suggests that the Polyakov action in the conformal gauge enjoys infinite residual symmetries. In fact, a combined action of Weyl rescaling $\delta g_{ab}=\omega\,g_{ab}$  and diffeomorphisms $\delta\sigma^\pm = \xi^\pm$ can leave the two-dimensional Minkowski metric invariant provided 
\begin{equation}
	\partial_{\mp} \xi^\pm = 0 \quad \text{and} \quad \omega= -\partial_{+}\xi^+ - \partial_{-}\xi^-  \,.
	\label{DiffWeyl}
\end{equation}
The finite transformations are then given by the arbitrary chiral reparametrisations $\sigma^\pm \to \sigma'^\pm(\sigma^\pm)$, which implies $\tau' = \sigma'^{+}(\sigma^+)+\sigma'^{-}(\sigma^-)$ also satisfies the d'Alembert equation.
The light-cone quantisation identifies the proper time $\tau$ with the $X^+ = (X^0+X^{D-1})/\sqrt{2}$ coordinate
\begin{equation}
	X^+ = x^+ +2\pi\alpha'p^+\tau \,, 
\end{equation}
thus eliminating the oscillators along this direction. This has the advantage of linearising the constraint \eqref{EMTensor}, which can be solved for $X^- =  (X^0-X^{D-1})/\sqrt{2}$,
\begin{equation}
	\partial_{\pm} X^- = \frac{1}{2\pi\alpha' p^+}\partial_{\pm} X^i \partial_{\pm} X^i \,,
\end{equation} 
with $i=1,\ldots,D-2$ labelling the transverse coordinates. In terms of oscillators,
\begin{equation}
	\alpha^-_n = \frac{1}{2\sqrt{2\alpha'}p^+}\sum_{m\in\mathbb{Z}} \alpha_{n-m}^i \alpha_m^i  \,,
	\label{AlphaMinusOpen}
\end{equation}
for open strings with NN boundary conditions, while
\begin{equation}
	\alpha^-_n = \frac{1}{\sqrt{2\alpha'}p^+}\sum_{m\in\mathbb{Z}} \alpha_{n-m}^i \alpha_m^i  \,,
	\label{AlphaMinusClosed}
\end{equation}
for closed strings, together with a similar equation for the right-movers. Here, we have introduced $\alpha_0^\mu=\sqrt{2\alpha'}\,p^\mu$ for open strings, while $\alpha_0^\mu=\tilde\alpha_0^\mu=\sqrt{\alpha'/2}\,p^\mu$.
 From eqs. \eqref{AlphaMinusOpen} and \eqref{AlphaMinusClosed}, it is clear that the physical oscillators are only those in the transverse directions and, hence, automatically build a Fock space of positive norm.
Among all modes, the $n=0$ one is special since it provides the mass-shell condition
\begin{equation}
	M_{\text{open}}^2 = 2p^+ p^- - p^i p^i = \frac{1}{2\alpha'}  \sum_{m\neq 0} \alpha^i_{-m}\alpha^i_m  \,,
\end{equation}
for open strings, and
\begin{equation}
		M_{\text{closed}}^2 =  \frac{1}{\alpha'}  \sum_{m\neq 0} \alpha^i_{-m}\alpha^i_m  =\frac{1}{\alpha'}  \sum_{m\neq 0} \tilde\alpha^i_{-m}\tilde\alpha^i_m  \,.
\end{equation}
We next employ the commutation relations \eqref{AlphaCommutators} to bring the sums into normal-ordered form,
\begin{equation}
	\sum_{m\neq 0} \alpha^i_{-m}\alpha^i_m = \sum_{m>0} \left( \alpha^i_{-m}\alpha^i_m + \alpha^i_{m}\alpha^i_{-m}\right) = 2N-\frac{D-2}{12}   \,,
\end{equation}
where $N=\sum_{m>0} \alpha^i_{-m}\alpha^i_m$ and we have used zeta function regularisation to evaluate the infinite contribution $\sum_{m>0} m$. Although the latter step appears to be a somewhat ad hoc procedure, it is nevertheless justified by introducing a suitable counter-term proportional to a cosmological constant in the world-sheet action. Putting everything together, the mass formulae for open strings reads
\begin{equation}
	M^2_{\text{open}} = \frac{1}{\alpha'} \left(N-\frac{D-2}{24}\right)  \,,
\end{equation}
while for closed strings
\begin{equation}
	M^2_{\text{closed}} = \frac{2}{\alpha'} \left(N+\tilde N-\frac{D-2}{12}\right)  \,,
\end{equation}
and must be supplemented by the level-matching condition $N=\tilde N$.

We are now ready to discuss the light spectrum, starting with open strings. The unique vacuum $|0\rangle$ is a space-time scalar with mass $M^2= -(D-2)/24\alpha'$. The first excitation is $\alpha_{-1}^i |0\rangle$ and transforms in the vectorial representation
of the little group $\text{SO}(D-2)$. This is compatible with Lorentz invariance if and only if this state is massless, which is the case only in $D=26$ dimensions. This implies that the vacuum corresponds to a tachyonic state.
The next levels $\alpha_{-2}^i |0\rangle$ and $\alpha_{-1}^i \alpha_{-1}^j |0\rangle$ are then massive with $M^2 = 1/\alpha'$. Although they are built out of the transverse oscillators, they can be re-organised into irreducible representations of the little group $SO(D-1)$, since a spin-2 representation of $\text{SO}(D-1)$ can be decomposed into a spin-2 tensor plus a vector and a scalar of $\text{SO}(D-2)$,
\begin{equation}
	\frac{(D-1)D}{2} -1 =  \frac{(D-2)(D-1)}{2} -1+ (D-2)+1 \,.
\end{equation}
 This is the Higgs mechanism applied to spin-2 fields. A similar pattern repeats for all higher levels. 
 
 Turning to closed strings one has to properly tensor together left movers and right movers, while respecting level-matching. The unique vacuum $|0,\tilde 0\rangle$ is again a space-time scalar with mass $M^2= -(D-2)/6\alpha'$. The first excited level is $\alpha_{-1}^i \tilde\alpha_{-1}^j |0,\tilde 0\rangle$ which decomposes into the symmetric traceless, the anti-symmetric and the singlet representations of $\text{SO}(D-2)$. As before, consistency with Lorentz symmetry requires that these states be massless, which occurs in $D=26$ dimensions. Therefore, these states correspond to a massless spin-2 field, to be identified with the space-time graviton $G_{\mu\nu}$, a massless rank-2 anti-symmetric tensor $B_{\mu\nu}$, known as the Kalb-Ramond field, and a massless scalar $\Phi$, known as the dilaton. The next level describes $104\,976$ massive degrees of freedom which, among others, include a massive spin-4 field.

A convenient way to count the massive degrees of freedom of open and closed strings is to compute the one-loop vacuum energy. In quantum field theory, it is not a quantity of particular relevance since it is simply a function of the particle masses. Different is the situation in string theory, since it describes infinite particles associated to the various harmonics. As we shall see, it imposes non-trivial constraints on the consistency of the theory, especially in the case of fermionic strings. For a real scalar field, the quantity we wish to compute is
\begin{equation}
	\mathscr{Z} = \int [\mathscr{D}\phi] \,e^{ \int d^D x\, \frac{1}{2}\phi (\Box-m^2)\phi } = \text{Det}^{-1/2}(-\Box+m^2) \,,
\end{equation}
or, rather, its logarithm. This can be most conveniently evaluated in the Schwinger representation
\begin{equation}
	\log \mathscr{Z} = -\frac{1}{2} \int_\epsilon^{\infty} \frac{dt}{t} \, \text{Tr}\, e^{-\pi t (-\Box+m^2)}  \,,
\end{equation}
where $t$ is the proper time for a point particle moving along a circle, and $\epsilon$ regulates the UV divergence. In the following, we shall be cavalier and set $\epsilon$ to zero. 
The contribution of the box operator is universal and reads
\begin{equation}
	\begin{split}
		\text{Tr} \, e^{\pi t \Box} =&  \int d^D p \,\langle p| e^{-\pi p^2 t} |p\rangle = \int d^D p \,e^{-\pi p^2 t}\int d^D x\, \langle p|x\rangle \langle x|p\rangle  \\
						    =& \frac{V_D}{(2\pi)^D}  \frac{1}{t^{D/2}}\,,
	\end{split}
\end{equation}
so that 
\begin{equation}
	\log \mathscr{Z} = -\frac{V_D}{2(2\pi)^D} \int_0^{\infty} \frac{dt}{t^{1+D/2}} \, \text{Tr}\, e^{-\pi t m^2}   \,.
\end{equation}
It is straightforward to extend this expression to string theory, by simply replacing the mass of the particles $m^2$ by the mass operator $M^2$.
For open strings, 
\begin{equation}
	\log \mathscr{Z} = -\frac{V_D}{2(2\pi)^D} \int_0^{\infty} \frac{dt}{t^{1+D/2}} \, \text{Tr}\, e^{-\pi t \left(N-\frac{D-2}{24}\right)/\alpha' }  \,.
	\label{PrePart}
\end{equation}
Recall that $N$ is the Hamiltonian of a system containing an infinite number of harmonic oscillators with frequencies $n=1,2,\ldots$. Therefore, setting $q=e^{-\pi t/\alpha'}$ and introducing the properly normalised ladder operators $a_n^i =\alpha_n^i/\sqrt{n}$ and $(a_n^i)^\dagger = \alpha_{-n}^i /\sqrt{n}$,
\begin{equation}
	\begin{split}
		\text{Tr} \, q^N =  \text{Tr} \, q^{ \sum_{i=1}^{D-2} \sum_{n>0} n (a_n^i)^\dagger a_n^i } = \left[ \prod_{n>0} \sum_{k\geq 0} q^{nk} \right]^{D-2} = \frac{1}{\prod_{n>0}(1-q^n)^{D-2} } \,,
	\end{split}
	\label{OscTrace}
\end{equation}
which indeed counts the number $P(n)$ of possible partitions of the total energy $n$ into the various oscillators. Inserting this result into \eqref{PrePart}, one gets
\begin{equation}
	\log \mathscr{Z} = -\frac{V_D}{2(2\pi)^D} \int_0^{\infty} \frac{dt}{t^{1+D/2}} \frac{1}{\eta(q)^{D-2}} \,,
	\label{OpenPart}
\end{equation}
written in terms of the Dedekind eta function
\begin{equation}
	\eta(q) = q^{1/24} \prod_{n>0} (1-q^n) \,.
\end{equation}
As promised, upon Taylor expanding the integrand around $q=0$, eq. \eqref{OpenPart} encodes the degrees of freedom of open-string states at all mass levels. Indeed, introducing $Z=1/\eta(q)^{24}$ in the case of $D=26$ dimensions, one finds
\begin{equation}
	Z(q) = \sum_{n} d(n) q^n = \frac{1}{q} + 24 + 324\,q+3200\,q^2 +\ldots \,,
\end{equation}
where $d(n)$ is the number of states at mass level $M^2 = n/\alpha'$. From this expansion we then recognise the real tachyon of the open string, the 24 d.o.f. of a massless vector, the 324 d.o.f.'s of a spin 2 field with mass $M^2=1/\alpha'$, and so on.

Moving to the closed string case, the mass operator receives contributions from both left-moving and right-moving oscillators, but physical states are those which obey the level-matching condition. Taking this into account, the closed string analogue of \eqref{PrePart} is
\begin{equation}
	\begin{split}
		\log \mathscr{Z} =& -\frac{V_D}{2(2\pi)^D} \int_0^{\infty} \frac{dt}{t^{1+D/2}} \, \text{Tr}\left[\delta_{N-\tilde N} \,e^{-2\pi t \left(N+\tilde N-\frac{D-2}{12}\right)/\alpha' }  \right] \\
					=&  -\frac{V_D}{2(2\pi)^D} \int_0^{\infty} \frac{dt}{t^{1+D/2}} \, \int_{-1/2}^{1/2}ds\, \text{Tr}\,q^{N-\frac{D-2}{24}} \, \bar q^{\tilde N-\frac{D-2}{24}}  \,,
	\end{split}
	\label{ClosedPartInit}
\end{equation}
where now $q=e^{2\pi i(s+it/\alpha')}$. A similar computation as in \eqref{OscTrace}, yields
\begin{equation}
	\log \mathscr{Z} = -\frac{V_D}{2(2\pi)^D} \int_{-1/2}^{1/2}ds \int_0^{\infty} \frac{dt}{t^{1+D/2}} \frac{1}{(\eta(q)\bar\eta(\bar q))^{D-2}} \,,
	\label{ClosedPart}
\end{equation}
where, by abuse of notation, we adopt the standard convention where the right-movers contribute with the anti-holomorphic function $\bar\eta(\bar q) \equiv (\eta(q))^*$. As before, Taylor expanding the integrand
\begin{equation}
	\begin{split}
	Z(q,\bar q) =& \sum_{n,m} d(n,m)\,q^n\,\bar q^m = \frac{1}{q\bar q} +24\left( \frac{1}{q} +\frac{1}{\bar q}\right) + 576 + 324 \left(\frac{q}{\bar q}+\frac{\bar q}{q}\right) \\
			&+ 7776(q+\bar q) +  3200\left(\frac{q^2}{\bar q}+\frac{\bar q^2}{q}\right) + 76800 (q^2+\bar q^2)+     104\,976\,q\bar q +\ldots \,,
	\end{split}
\end{equation}
and, upon imposing level-matching to select the physical excitations $M^2=4n/\alpha'$, we recognise the real tachyon, 576 massless states comprising the graviton, the dilaton and the Kalb-Ramond field, and so on.

Actually, eq. \eqref{ClosedPart} does not really take into account the extended nature of closed strings. The logic leading to \eqref{ClosedPart} was to follow the analogy with quantum field theory of point particles, where the one-loop vacuum-to-vacuum amplitude corresponds to a world-line with the topology of a circle. The same diagram in closed strings should, thus, correspond to a doughnut, \emph{i.e.} a worldsheet with the topology of a torus. As we shall see, the global properties of this worldsheet will impose non-trivial constraints on the consistency of a closed string vacuum. 

By analogy to the case of a circle, a flat torus can be constructed starting from two non-degenerate vectors $\boldsymbol{\omega}_1$, $\boldsymbol{\omega}_2$ in the complex plane, upon the identification $\Lambda:\ z\sim z+n\boldsymbol{\omega}_1+m\boldsymbol{\omega}_2$, for all integers $n,m$. Therefore, the torus $T^2=\mathbb{C}/\Lambda$ is an elementary cell, where opposite sides are identified. Contrary to the case of a circle, different choices of $\boldsymbol{\omega}_1$ and $\boldsymbol{\omega}_2$ do not necessarily define distinct torii. In fact, two choices of non-degenerate vectors $\boldsymbol{\omega}_i$ and $\tilde{\boldsymbol{\omega}}_i$ related by the linear transformation
\begin{equation}
	\begin{pmatrix}
		\tilde{\boldsymbol{\omega}}_1 \\
		\tilde{\boldsymbol{\omega}}_2 \\
	\end{pmatrix} = \begin{pmatrix}
					d & c \\ b & a \\ 
				\end{pmatrix} \begin{pmatrix}
		\boldsymbol{\omega}_1 \\
		\boldsymbol{\omega}_2 \\
	\end{pmatrix} \,,
	\label{OmegaTransf}
\end{equation}
define the same torus provided that $a,b,c,d\in\mathbb Z$ and $ad-bc=1$, which preserves the volume of the torus. Using the rotational and scaling symmetries, one may always bring one side of the parallelogram to lie on the real axis and normalise its length to one, so that the torus is identified by a complex number $\tau = \tau_1+i\tau_2= \boldsymbol{\omega}_2/\boldsymbol{\omega}_1$, known as the complex structure. The maps \eqref{OmegaTransf} act on $\tau$ as the fractional linear transformations
\begin{equation}
	\tilde\tau = \frac{a\tau+b}{c\tau+d}\,.
	\label{ModTau}
\end{equation}
The set of matrices in \eqref{OmegaTransf} are elements of the modular group $\text{SL}(2;\mathbb{Z})$, generated by
\begin{equation}
	T= \begin{pmatrix}
		1 & 1 \\
		0 & 1 \\
		\end{pmatrix}  \,, \qquad S=\begin{pmatrix}
							0 & -1\\
							1 & 0 \\
						\end{pmatrix} \,,
\end{equation}
which act as translations, $T:\ \tau\to\tau+1$, and inversions, $S:\ \tau\to-1/\tau$, on the complex structure.

Returning to the partition function of closed strings \eqref{ClosedPart}, it is natural to identify $\tau=s+it/\alpha'$. Indeed, using the definition of $q$, the trace in \eqref{ClosedPartInit} can be conveniently rewritten as
\begin{equation}
	\text{Tr} \left[ e^{-2\pi \tau_2(N+\tilde{N} -2)} \,e^{2\pi i\tau_1(N-\tilde{N})} \right] \,,
\end{equation}
where $N+\tilde{N}-2$ is recognised as the two-dimensional Hamiltonian $H$ associated to time-evolution, while $N-\tilde{N}$ is the two-dimensional momentum operator $P$ generating translations along the spatial world-sheet direction. 
This has an interesting physical interpretation as a closed string state propagating for proper time $\tau_2$ to form a cylinder, whose two ends are to be then glued together by the trace, after a relative rotation by an angle $2\pi\tau_1$.
The resulting object is clearly a torus with metric $ds^2 = \tau_2^{-1}|d\sigma^1+\tau d\sigma^0|^2$, which therefore justifies the aforementioned identification.

Since $\tau$ and $\tilde\tau$ related by \eqref{ModTau} correspond to different parameterisations of the same worldsheet, consistency requires that the integrand of \eqref{ClosedPart} be invariant under the action of the modular group.
For the simple example of the closed bosonic string discussed so far, this is guaranteed by the modular properties of the Dedekind eta function
\begin{equation}
	\eta(\tau+1)=e^{i\pi/12}\, \eta(\tau) \,, \qquad \eta(-1/\tau) = \sqrt{-i\tau}\, \eta(\tau)\,,
	\label{EtaTransform}
\end{equation}
as first noticed by Shapiro \cite{Shapiro:1972ph}.
In more general situations, however, we shall see that the requirement of modular invariance is non-trivial and actually provides the rationale for the construction of consistent string vacua.
The integration over the $s$ and $t$ variables in \eqref{ClosedPart} is now interpreted as a collective contribution of all world-sheet geometries with the topology of the torus. However, to avoid over-counting we should restrict the integration only over those complex structures that correspond to gauge-inequivalent torii. This effectively reduces the integration domain to 
\begin{equation}
	\mathscr{F} = \mathbb{H}/\text{SL}(2;\mathbb{Z}) = \{ |\tau|\geq 1,\ -\tfrac{1}{2}\leq \tau_1< \tfrac{1}{2},\ \tau_2>0 \} \,,
	\label{FundDomain}
\end{equation}
known as the fundamental domain of $\text{SL}(2;\mathbb{Z})$, where $\mathbb{H}$ is the Poincar\'e upper-half plane, endowed with the hyperbolic metric.

A similar interpretation can be given for open strings, although in this case there is a single Schwinger parameter and modular invariance is no longer present. We defer this discussion to Section \ref{ST:Orientifold}, since a deeper interpretation of the amplitude can be best achieved only after world-sheet fermions have been introduced.


\section{Two-Dimensional Conformal Field Theory}
\label{ST:CFT}

The residual symmetry \eqref{DiffWeyl} that allowed us to quantise string theory in the light-cone actually plays a prominent role in the world-sheet description of the theory.
Already the fact that the energy momentum tensor is traceless is an indication that the two-dimensional theory is invariant under conformal transformations. In general, the conformal group is the subgroup of general coordinate transformations, $\delta x^\mu= \xi^\mu(x)$, which leave the metric invariant up to an overall local rescaling
\begin{equation}
	\delta g_{\mu\nu}(x) = \omega(x) \, g_{\mu\nu}(x) \,.
\end{equation}
In $D$ dimensions, and specialising to the Minkowski metric this implies the relation
\begin{equation}
	\partial_\mu \xi_\nu +\partial_\nu \xi_\mu = \frac{2}{D}\eta_{\mu\nu}\nabla\cdot\xi \,.
	\label{ConformalD}
\end{equation}
In $D>2$ the general solution involves $\frac{1}{2}(D+1)(D+2)$ independent parameters, comprising the Poincar\'e transformations $\xi^\mu={\Lambda^\mu}_\nu\,x^\nu + a^\mu$, together with scale transformations $\xi^\mu =\lambda\, x^\mu$, and special conformal transformations $\xi^\mu = b^\mu\, x^2 -2x^\mu b\cdot x$. In $D=2$, instead, eq. \eqref{ConformalD} reduces to the Cauchy-Riemann equations admitting an infinite number of solutions. These are nothing but the transformations $\sigma^\pm \to 
\sigma'^\pm(\sigma^\pm)$ that we discussed in Section \ref{ST:Bosonic}.

In two-dimensional conformal field theories (CFTs) \cite{Friedan:1983xq, Friedan:1984rv, Belavin:1984vu, Friedan:1985ge}  (see \cite{Ginsparg:1988ui, Itzykson:1989sx, Itzykson:1989sy, DiFrancesco:1997nk, Gaberdiel:1999mc, Blumenhagen:2009zz} for an introduction), it is useful to work in Euclidean space and introduce the complex coordinates $z=e^{\tau+i\sigma}$, $\bar z=e^{\tau-i\sigma}$, which map the cylinder spanned by a closed string to the whole complex plane. In terms of these variables, the proper time $\tau$ becomes the radius of circles centred around the origin and, therefore, the generator of scale transformations on $z$ is identified with the Hamiltonian of the system. This is referred to as \emph{radial quantisation}. The conservation of the energy momentum tensor
\begin{equation}
	\bar\partial T_{zz} + \partial T_{z\bar z} = 0 \,,
\end{equation}
together with the scale invariance property $T_{z\bar z}=0$, imply that $T_{zz}\equiv T(z)$ is (classically) an holomorphic function of $z$ so that, for any holomorphic function $\xi(z)$,
\begin{equation}
	T_\xi = \oint \frac{dz}{2\pi i} \, \xi(z)\,T(z)\,,
\end{equation}
generates the infinitesimal conformal transformations $\delta z= \xi(z)$, once the contour is taken to encircle the origin. Similar arguments can be made for the transformations $\delta \bar z=\bar\xi (\bar z)$ generated by the anti-holomorphic energy-momentum tensor $\bar T(\bar z) \equiv T_{\bar z\bar z}$.

On general grounds, an infinitesimal conformal transformation on an operator $\mathscr{O}(z,\bar z)$ is given by the equal-time commutator $[T_\xi + \bar T_{\bar\xi} , \mathscr{O}(z,\bar z)]$, which in QFT is to be suitably averaged in the path integral sense. However, path integrals return the time-ordered correlator and, therefore, in order to compute a commutator one needs to slightly deform the time of the Noether charges. In radial quantisation, this amounts to 
\begin{equation}
	[T_\xi, \mathscr{O}(z,\bar z)] = \oint \frac{dw}{2\pi i} \, \xi(w)\, T(w)\,\mathscr{O}(z,\bar z) \,,
\end{equation}
where the integration contour encircles $z$, and similarly for its anti-holomorphic counterpart. This integral is determined by the singular behaviour of the operator product expansion (OPE) of $T(w)\,\mathscr{O}(z,\bar z)$ as $w \to z$. The latter depends on the properties of $\mathscr{O}$. For a tensorial operator $\mathscr{O}_{z\ldots \bar z\ldots}(z,\bar z)$ of rank $(h,\bar h)$, the transformation under $z\to z'$ and $\bar z\to \bar z'$ reads
\begin{equation}
	\mathscr{O}_{z\ldots \bar z\ldots}(z,\bar z) \to \left(\frac{dz'}{dz}\right)^h\,\left(\frac{d\bar z'}{d\bar z}\right)^{\bar h} \, \mathscr{O}_{z\ldots \bar z\ldots}(z',\bar z') \,,
\end{equation}
which implies the OPE
\begin{equation}
	\begin{split}
		T(w)\,\mathscr{O}(z,\bar z) &= \frac{h}{(w-z)^2}\,\mathscr{O}(z,\bar z) + \frac{1}{w-z} \, \partial \mathscr{O}(z,\bar z) + \text{regular terms} \,, \\
		\bar T(\bar w)\,\mathscr{O}(z,\bar z) &= \frac{\bar h}{(\bar w- \bar z)^2}\,\mathscr{O}(z,\bar z) + \frac{1}{\bar w- \bar z} \, \bar\partial \mathscr{O}(z,\bar z) + \text{regular terms} \,.
	\end{split}
	\label{OPEprimary}
\end{equation}
Fields that satisfy \eqref{OPEprimary} are called \emph{primary fields} of conformal weight $(h,\bar h)$, but do not exhaust the fields present in a CFT. Secondary or descendant fields, which are derivatives of the primaries, have higher-order singularities in their OPEs with the energy-momentum tensor.

We are now ready to apply these techniques to the Polyakov action and, for simplicity, we shall first consider a single free boson $X(z,\bar z)$ described by the Lagrangian
\begin{equation}
	S = \frac{1}{2\pi\alpha'} \int d^2 z \, \partial X\,\bar\partial X \,,
\end{equation}
with energy-momentum tensor
\begin{equation}
	T(z) = - \frac{1}{\alpha'} \partial X\,\partial X(z) \ , \qquad \bar T(\bar z) = - \frac{1}{\alpha'} \bar\partial X\,\bar\partial X(\bar z) \,,
\end{equation}
and we suppress the explicit display of the normal ordering symbol.
It is straightforward to work out the 2-point function
\begin{equation}
	\langle X(z,\bar z)\,X(w,\bar w)\rangle = - \frac{\alpha'}{2} \log|z-w|^2\,,
\end{equation}
which is ill-defined in the IR and, thus, $X$ itself is not a conformal field. Its holomorphic derivative, however, is a primary field of conformal weight $(1,0)$, as shown by the OPE
\begin{equation}
	\begin{split}
		T(z)\,\partial X(w) &= -\frac{2}{\alpha'}\,\partial X(z)\,\langle \partial X(z)\,\partial X(w)\rangle \\
					&= \frac{1}{(z-w)^2}\, \partial X(w) + \frac{1}{z-w}\, \partial^2 X(w) + \text{regular terms} \,.
	\end{split}
\end{equation}
Similarly, $\bar\partial X(\bar z)$ is primary with weight $(0,1)$, while higher-order derivatives of $X$ are not primary, as may be easily verified. There is, however, another important class of primary fields that may be constructed out of a free scalar, obtained by the normal ordered exponentials of the form $e^{ipX(z,\bar z)}$, where $p$ is a real parameter. Its OPE with the energy-momentum tensor 
\begin{equation}
	T(z)\,e^{ipX(w,\bar w)} = \frac{\alpha' p^2/4}{(z-w)^2}\,e^{ipX(w,\bar w)} + \frac{1}{z-w}\,\partial \left(e^{ip X(w,\bar w)}\right) + \text{regular terms}
\end{equation}
reveals that it has conformal weight $(\alpha' p^2/4, \alpha' p^2/4)$. Notice that $e^{ipX}$ classically has zero scaling dimension and it is only at the quantum level that it acquires a non-trivial conformal weight.

In the space of holomorphic functions, the monomials $z^{n+1}$ constitute a basis, each corresponding to a different conformal transformation. The algebra of the corresponding generators 
\begin{equation}
	L_n = \oint \frac{dz}{2\pi i}\, z^{n+1} T(z)\,,
\end{equation}
requires the 2-point function of the energy-momentum tensor with itself. On dimensional grounds, since the naive conformal dimension of $T(z)$ is $(2,0)$, one would expect 
\begin{equation}
	T(z)T(w) = \frac{c/2}{(z-w)^4} + \frac{2}{(z-w)^2}\,T(w) + \frac{1}{z-w}\, \partial T(w) + \text{regular terms}\,,
\end{equation}
for some constant $c$. This form of the $TT$ OPE can be shown to be true for a general CFT, and in the case of a single free scalar $c=1$. A direct computation gives the Virasoro algebra
\begin{equation}
	[L_n,L_m] = (n-m) L_{n+m} + \frac{c}{12}\,n(n^2-1)\,\delta_{n+m} \,.
	\label{Virasoro}
\end{equation}
The constant $c$ is the \emph{central charge} and, as we shall see, reflects a quantum violation of Weyl symmetry.
From eq. \eqref{Virasoro}, it is also clear that only $L_{0},L_{\pm 1}$ close into a finite-dimensional sub-algebra and, together with their right-moving counterparts $\bar L_0, \bar L_{\pm 1}$ they form the $\text{SL}(2;\mathbb{C})$ subgroup of the conformal group. 

We may build the Hilbert space of the CFT by defining suitable in- and out-states. Because of the map $z=e^{\tau+i\sigma}$ taking the cylinder to the sphere, the past $\tau\to -\infty$ corresponds to $z\to 0$ while the future $\tau\to \infty$ corresponds to $z\to\infty$. Given a primary field $\phi(z,\bar z)$ of conformal weight $(h,\bar h)$, the in-state is defined as
\begin{equation}
	| \phi\rangle = \lim_{z,\bar z\to 0} \phi(z,\bar z) |0\rangle \,,
\end{equation}
where $|0\rangle$ is the $\text{SL}(2;\mathbb{C})$ invariant vacuum annihilated by the Virasoro operators $L_n$, with $n\geq -1$.
Clearly, the $T\phi$ OPE implies
\begin{equation}
	L_0 |\phi\rangle = h |\phi\rangle \,, \qquad L_n|\phi\rangle = 0 \quad \text{for } n \geq 1 \,.
	\label{L0Phi}
\end{equation}
The action of raising operators $L_n$, $n<0$ on $|\phi\rangle$ builds the descendant states associated to the primary field and define the Verma module of $\phi$.
As a simple example, take $\phi(z) =\partial X$, which admits the Laurent expansion
\begin{equation}
	i\partial X(z) = \sqrt{\alpha'/2}\sum_{n\in\mathbb Z} \frac{\alpha_n}{z^{n+1}} \,.
	\label{LaurentX}
\end{equation}
The corresponding in-state is, hence,
\begin{equation}
	\lim_{z\to 0} \partial X(z) |0\rangle \propto \alpha_{-1} |0\rangle \,,
\end{equation}
and is clearly identified with the first excited state (the vector) of the open string.
The field $\phi(z,\bar z)$ is known as the string vertex operator and plays a crucial role in string amplitudes.
The definition of the out-states requires a careful analytic continuation to the Minkowski space cylinder (for a review, see \cite{Friedan:1985ge, Ginsparg:1988ui, Gaberdiel:1999mc, Blumenhagen:2009zz}).

The scalar field we have described so far does not exhaust the list of interesting CFTs with a free-field realisation. Another notable example important for string theory is the so-called $(h,1-h)$ system \cite{Friedan:1985ge}. This is made up of anti-commuting holomorphic primary fields $b,c$ with conformal weight $h$ and $1-h$, respectively. As we shall see, this system is rich enough to contain the reparametrisation ghosts as well as world-sheet fermions. The system is defined by the action
\begin{equation}
	S = \frac{1}{2\pi} \int d^2 z\, b\bar\partial c \,,
\end{equation}
from which one may extract the non-trivial correlator
\begin{equation}
	\langle b(z)\, c(w)\rangle = \frac{1}{z-w}  \,,
\end{equation}
and the energy-momentum tensor
\begin{equation}
	T_{bc}(z) = -h\,b\partial c +(1-h)\,\partial b\, c \,.
\end{equation}
From these expressions it is possible to extract the central charge of the $bc$ system by simply evaluating the most singular term in $T(z)T(w)$. The result is
\begin{equation}
	c = 12\,h(1-h)-2 \,.
\end{equation}
An analogous construction can be made for the analogous $(\bar h,1-\bar h)$ system involving the anti-holomorphic fields $\bar b, \bar c$, with right-moving central charge $\bar c=12\,\bar h(1-\bar h)-2$.

Another interesting family of $(h,1-h)$ systems is built out of commuting holomorphic primary fields $\beta$ and $\gamma$. The change of statistics simply implies an extra minus sign in the correlator and in the central charge. 


\section{String Perturbation Theory}
\label{ST:PerturbationTheory}

We return now to the study of the Polyakov action \eqref{STPol}, seen as a two-dimensional QFT of $D$ free scalars. The naive path integral quantisation
\begin{equation}
	\mathscr{Z} = \int [\mathscr{D}X]\,[\mathscr{D}g]\, e^{-S[X,g]} \,,
\end{equation}
is ill-defined since the action is clearly invariant under the local diffeomorphisms and Weyl rescalings. Therefore, the correct way \cite{Polyakov:1981rd} to compute the path integral is via the Faddeev-Popov procedure, which we now review in the simple case where the Riemann surface is taken to be the sphere (see \cite{DHoker:1988pdl} for a detailed exposition). 
Under reparametrisations, the ``off-diagonal'' components of the metric transform as
\begin{equation}
	\delta g_{zz} = 2\nabla_z \xi_z \,, \qquad \delta g_{\bar z\bar z} = 2\nabla_{\bar z} \bar\xi_{\bar z} \,,
\end{equation}
so that the appropriate conformal gauge fixing condition is $\delta g_{zz}=\delta g_{\bar z\bar z} =0$. As usual, this is achieved by inserting 
\begin{equation}
	1 = \int [\mathscr{D}\gamma] \, \delta(g^{\gamma}_{zz}) \,\delta(g^{\gamma}_{\bar z\bar z})\, \det\left(\frac{\delta g^{\gamma}_{zz}}{\delta\gamma}\right) \det\left(\frac{\delta g^{\gamma}_{\bar z\bar z}}{\delta\gamma}\right) \,,
\end{equation}
into the path integral, where $g^\gamma$ denotes the new metric into which $g$ is transformed by a reparametrisation $\gamma$ and the integral over the group manifold.
Since both the action and measures are invariant under the diffeomorphisms $\gamma$, the integration over the group manifold factorises and yields an irrelevant (infinite) volume factor. Furthermore, upon converting the determinants into Berezin integrals, one obtains
\begin{equation}
	\mathscr{Z} = \int  [\mathscr{D}\omega]\, \int [\mathscr{D}X]\,[\mathscr{D}b]\,[\mathscr{D}c]\, e^{-S_X[X] - S_{\text{ghost}}[b,c] } \,,
	\label{FPpathInt}
\end{equation}
with
\begin{equation}
	S_X[X] = \frac{1}{2\pi\alpha'} \int d^2 z\, \partial X^\mu \,\bar\partial X^\nu\,\eta_{\mu\nu} \,, \qquad S_{\text{ghost}}[b,c] = \frac{1}{2\pi} \int d^2 z\, (b\bar\partial c + \bar b\partial \bar c) \,.
\end{equation}
Here, $b$ and $c$ are the anti-commuting ghost fields associated to diffeomorphisms and carry conformal weight $2$ and $-1$, respectively. They are a particular realisation  of the $(h,1-h)$ system discussed previously, with $h=2$ and $c_{\text{ghost}}=-26$.
The function $\omega(z,\bar z)$ parametrises the ``diagonal'' components of the metric $g$, which is conformal to the two-dimensional Minkowski metric, $g_{z\bar z}=e^{\omega}\, \eta_{z\bar z}$.
Classically, the integrand in \eqref{FPpathInt} does not depend on $\omega$ and, again, its integral would give an irrelevant multiplicative (infinite) constant. However, this is not true quantum-mechanically, since in arbitrary dimension $D$, there is a Weyl anomaly proportional to $c_{\text{tot}} = c_{X}+c_{\text{ghost}}$. Although \eqref{FPpathInt} may define a sensible theory even when $c_{\text{tot}}$ is different than zero, critical string theory requires the vanishing of the Weyl anomaly and selects the dimension of spacetime $D=26$. Indeed, in $D=26$, the Virasoro algebra associated to the total energy-momentum tensor does not have a central extension, and conformal symmetry is exact at the quantum level.

As in the case of Yang-Mills theory, $S_X+S_{\text{ghost}}$ has a residual BRST symmetry generated by the Hermitian, nilpotent charge $Q_{\text{BRST}}= Q+\bar{Q}$, with
\begin{equation}
	Q = \oint \frac{dz}{2\pi i}\,\left[ c(z) \left( T_X(z)+\frac{1}{2} T_{\text{ghost}}(z) \right) +\frac{3}{2}\partial^2 c(z)\right]\,,
\end{equation}
and $\bar Q$ similarly given in terms of anti-holomorphic fields \cite{Friedan:1985ge}. Note that the products in the integrand are assumed to be normal ordered, while the nilpotency of $Q$ and $\bar Q$ is a consequence of the vanishing of the total central charge.
Clearly, physical (gauge invariant) states $|\psi\rangle$ must be closed under the action of the BRST operator, $Q_{\text{BRST}}|\psi\rangle = 0$. A trivial way to satisfy this requirement is to consider exact states, $|\chi\rangle = Q_{\text{BRST}} |\eta\rangle$, but these have zero norm and, therefore, decouple from physical processes, 
\begin{equation}
	( \langle \psi_f | + \langle \chi_f | )\, S\, (  |\psi_i \rangle + | \chi_i \rangle ) =  \langle \psi_f |\, S\,  |\psi_i \rangle \,.
\end{equation}
Therefore, if the $S$-matrix defines a unitary theory on the full Hilbert space, it is also unitary once restricted on the BRST cohomology.
On general grounds, one expects that physical states do not contain any ghost excitation and, thus, should be proportional to the ghost vacuum. Notice that the zero modes $c_0$ and $b_0$ of the ghost fields commute with the Hamiltonian and satisfy the Clifford algebra $\{ c_0, b_0\} = 1$, which admits a two-dimensional representation $|\pm\rangle$ satisfying $b_0|+\rangle=|-\rangle$ and $c_0|-\rangle=|+\rangle$. It turns out that the correct definition of physical states involves $|-\rangle$, so that $|\psi\rangle = \psi(0)\,|0\rangle\otimes|-\rangle$ and 
\begin{equation}
	Q |\psi\rangle = \left(c_0(L_0^X - 1)+\sum_{n>0} c_{-n}L_n^{X} \right) |\psi\rangle = 0\,,
\end{equation}
yields the physical conditions \eqref{L0Phi} with conformal weight $h=1$. 

A generic physical state in string theory carries spacetime momentum $p_\mu$, which is injected by the operator $e^{ip\cdot X}$. Indeed, from \eqref{LaurentX} it is clear that the momentum operator $\alpha_0^\mu$ acts on the state as
\begin{equation}
	\alpha_0^\mu\,e^{ip\cdot X}|0\rangle=  i\sqrt{2/\alpha'} \oint \frac{dz}{2\pi i}\, \partial X^\mu(z)\,e^{ip\cdot X(0)} |0\rangle = \sqrt{\alpha'/2}\, p^\mu \,|0\rangle \,.
\end{equation} 
The BRST condition then implies that the state created by the vertex operator $e^{ip\cdot X}$ is a tachyon with $p^2=4/\alpha'$, which coincides with the vacuum discussed previously in the context of light-cone quantisation of closed strings.
The first excited states correspond to the vertex operator $V(\zeta,p)= \zeta_{\mu\nu}\, \partial X^\mu \bar\partial X^\nu\,e^{ip\cdot X}$, which has conformal dimension $(1,1)$ provided $p^2 = 0$. The BRST condition further implies transversality, $\zeta_{\mu\nu}p^\mu=\zeta_{\mu\nu}p^\nu=0$, so that it describes the massless graviton, the dilaton and the Kalb-Ramond field. The construction of physical massive states proceeds in a similar fashion.

Until now, we have considered the propagation of strings on a simple Minkowski spacetime and it is natural to ask what happens when the background geometry is non-trivial \cite{Alvarez-Gaume:1980zra, Alvarez-Gaume:1981exa, Alvarez-Gaume:1981exv, Fradkin:1981dd, Fradkin:1982ge, Lovelace:1983yv, Callan:1985ia, Fradkin:1985qd}. This implies that the Minkowski metric $\eta_{\mu\nu}$ in the Polyakov action should be replaced by a general pseudo-Riemannian metric $G_{\mu\nu}(X)$. Actually in string theory one has the freedom to deform the background by introducing non-trivial configurations for the Kalb-Ramond field $B_{\mu\nu}(X)$ and the dilaton $\Phi(X)$. This gives rise to a non-trivial sigma model described by
\begin{equation}
	\begin{split}
		S =& -\frac{1}{4\pi\alpha'} \int_{\Sigma} d^2\sigma \, \left(\sqrt{-g} \,g^{ab} \partial_a X^\mu \partial_b X^\nu \,G_{\mu\nu}(X) +\epsilon_{ab}\partial_a X^\mu \partial_b X^\nu \,B_{\mu\nu}(X) \right. \\
				& \qquad \qquad \left. -\alpha' \sqrt{-g}\,R^{(2)}\,\Phi(X)\right) \,,
	\end{split}
	\label{BackgroundAction}
\end{equation} 
where the last term is the coupling of the dilaton to the two-dimensional Ricci scalar. An important point is that not all backgrounds admit a consistent propagation of strings, due to the occurrence of the Weyl anomaly at the quantum level. It is possible to show that consistent backgrounds are those which yield vanishing beta functions for the three above action terms. A long and tedious computation yields the conformal invariance conditions
\begin{equation}
	\begin{split}
		0 &= \beta_{\mu\nu}^G = R_{\mu\nu}+\frac{1}{4}H_\mu{}^{\lambda\rho} H_{\nu\lambda\rho} - 2\nabla_\mu \nabla_\nu\Phi + \mathcal{O}(\alpha')\,, \\
		0 &= \beta_{\mu\nu}^B = \nabla_\lambda H^\lambda{}_{\mu\nu} -2\nabla_\lambda \Phi \, H^\lambda{}_{\mu\nu} + \mathcal{O}(\alpha')\,, \\
		0 &= \beta^\Phi = 4\nabla_\lambda \Phi \nabla^\lambda \Phi -4 \nabla_\lambda \nabla^\lambda \Phi +R + \frac{1}{12}\,H_{\lambda\rho\sigma}\,H^{\lambda\rho\sigma}+ \mathcal{O}(\alpha')\,,
	\end{split}
	\label{betaEquations}
\end{equation}
where $\nabla$ is the standard covariant derivative acting on tensors and $H_{\mu\nu\rho}=\partial_{\mu}B_{\nu\rho}+\partial_{\nu}B_{\rho\mu}+\partial_{\rho}B_{\mu\nu}$ is the field strength for the Kalb-Ramond field.
Therefore, only backgrounds which satisfy these conditions are conformal and give rise to consistent string theories. It is not a coincidence that these equations correspond to the Euler-Lagrange equation emanating from the effective action
\begin{equation}
	S_{\rm eff} = -\frac{1}{2\kappa^2} \int d^{26}x \, \sqrt{-G} \,e^{-2\Phi} \left( R - 4\nabla_{\mu}\Phi\,\nabla^\mu \Phi + \frac{1}{12}H_{\mu\nu\lambda} H^{\mu\nu\lambda} + \mathcal{O}(\alpha') \right) \,,
	\label{Seffective}
\end{equation}
where, on dimensional grounds, $\kappa^2$ is proportional to $(\alpha')^{12}$.
The $\mathcal{O}(\alpha')$ terms contain higher derivative couplings and describe string corrections to Einstein gravity. The overall factor $e^{-2\Phi}$  is a characteristic of string theory and originates from the way the dilaton couples to the two-dimensional Ricci tensor in \eqref{BackgroundAction}. Although technically this factor could be removed by a suitable Weyl rescaling of the spacetime metric so that one recovers the canonical Einstein-Hilbert term, it is actually useful in the sense that it identifies the quantum corrections to the effective action. Indeed, expanding \eqref{BackgroundAction} around a constant dilaton background, which trivially solves the beta function equations \eqref{betaEquations},  the path integral weights each world-sheet topology by the factor $e^{-\langle\Phi\rangle \chi(\Sigma)}$, with $\chi(\Sigma)$ being the Euler characteristic of the world-sheet $\Sigma$. This way, we can identify $e^{\langle\Phi\rangle}$ with the string coupling constant $g_s$ and the path integral implies a sum over all topologies of increasing Euler number. For closed oriented strings, $\chi(\Sigma)=2-2g$ with the genus $g$ counting the number of handles of $\Sigma$. In open (and possibly unoriented) strings, the Riemann surfaces involve boundaries (and possibly cross-caps) so that the Euler number becomes $\chi(\Sigma)=2-2g-b-c$ where $b$ counts the number of boundaries and $c$ the number of cross-caps, not to be confused with the $b,c$ ghosts. This way,  the topological expansion parallels the loop expansion of QFT, where $\chi$ is the stringy analogue of the loop order. As a result, the effective action $S_{\rm eff}$ is a double expansion in both $\alpha'$ which contains the corrections due to the extended nature of strings, and in $g_s$ which incorporates quantum effects. In this sense, the terms weighted by $e^{-2\Phi}$ in \eqref{Seffective} are the tree-level contributions associated to the world-sheet topology of the sphere.

At fixed genus $g$, a generic scattering amplitude of interest will involve $N$ insertions of vertex operators associated to the external legs, whose positions typically have to be integrated over the corresponding Riemann surface. It turns out that for $g=0,1$ there is a residual symmetry which is not fixed by our choice of conformal gauge. This is known as the Conformal Killing Group (CKG) and is $\text{SL}(2;\mathbb{C})$ in the case of the sphere, and translations in the case of the torus. For higher genera the CKG is trivial. This residual symmetry is reflected in the presence of zero modes for the $c$-ghost, and there are $C_0=3$ of them on the sphere, only $C_1=1$ on the torus, while $C_{g>1}=0$. Also the $b$-ghost can have non-trivial zero modes which  count the number of conformally invariant complex moduli that describe a surface $\Sigma$ of genus $g$. The Riemann-Roch theorem relates the number $B_g$ of the $b$-ghost zero modes to $C_g$ by 
\begin{equation}
	B_g-C_g = 3g-3 \,,
\end{equation}
so that there are no $b$ zero modes on the sphere, which indeed has no moduli, one $b$ zero mode on the torus which is characterised by its complex structure $\tau$, and $3g-3$  $b$-zero modes on a genus-$g$ surface.
Given that $b,c$ are Grassmann variables, the path integral is non-trivial only provided a suitable number of $c$ and $b$ fields are inserted. For instance, on the sphere, the $\text{SL}(2;\mathbb{C})$ symmetry allows one to fix the positions of three vertex operators, conventionally chosen as $0,1,\infty$, while dressing each of them with a $c$-ghost. Diffeomorphism invariance then requires that the positions of the remaining $N-3$ vertex operators be integrated over the sphere. 

As an example, we can study the simplest non-trivial scattering amplitude on the sphere involving four external tachyons of incoming momenta $p_i$. According to the previous discussion, we fix the positions of three of them $z_{1,2,3}$ accompanying their vertex operators by $c$-ghosts, while the fourth one does not carry any ghost and its position $z_4$ is integrated over the sphere, 
\begin{equation}
	\begin{split}
		\mathscr{A}_c(p_i) \sim g_s^2 \, & \Big\langle c(z_1)\bar c(\bar z_1)e^{ip_1\cdot X(z_1,\bar z_1)} c(z_2)\bar c(\bar z_2)e^{ip_2\cdot X(z_2,\bar z_2)} c(z_3)\bar c(\bar z_3)e^{ip_3\cdot X(z_3,\bar z_3)}\, \\
					& \quad \times \int d^2 z_4 \, e^{ip_4\cdot X(z_4,\bar z_4)} \Big\rangle \,.
	\end{split}
\end{equation}
The $g_s$ dependence of the amplitude can be justified as follows. This is a tree-level diagram involving the topology of a sphere, and is therefore weighted by $g_s^{-2}$. For each external leg, there is a cylinder (the propagator) which connects the scattered states to the sphere, therefore introducing a boundary which by the previous argument is weighted by $g_s$. Upon mapping the sphere to the complex plane, each external leg shrinks to a puncture, where the vertex operator is inserted. In the case at hand we have four external states, so that the overall factor is $g_s^{-2}g_s^4 = g_s^2$.

The correlators involving the holomorphic and anti-holomorphic ghosts factorise and can be computed independently. Requiring that the conformal symmetry generated by $c(z)\partial$ be regular at infinity implies that only the three generators $c_{-1}z^2 \partial$, $c_{0}z \partial $ and $c_{1}\partial$ corresponding to the $\text{SL}(2;\mathbb{C})$ subgroup contribute. Taking into account the fact that the operators $c_n$ anti-commute, one finds
\begin{equation}
	\langle c(z_1)c(z_2)c(z_3)\rangle = z_{12}z_{13}z_{23} \,,
\end{equation}
and similarly for the anti-holomorphic ghosts. Here and in the following, we adopt the standard notation $z_{ij}=z_i-z_j$. Evaluating the $X$ correlators also produces the standard momentum conserving delta function so that
\begin{equation}
	\mathscr{A}_c(p_i) \sim g_s^2 \, \delta^{(26)}(p_1+p_2+p_3+p_4)\, |z_{12}z_{13} z_{23}|^2 \, \int d^2 z_4\, \prod_{i<j=1}^{4} |z_{ij}|^{\alpha' p_i\cdot p_j} \,.
\end{equation}
It is conventional to set $z_{1,2,3}$ to $0,1,\infty$, respectively, and introduce the Mandelstam variables $s= -(p_1+p_2)^2$, $t=-(p_1-p_3)^2$, and $u=-(p_1-p_4)^2$. Using the mass-shell condition $p_i^2 = -4/\alpha'$ one arrives at the celebrated Shapiro-Virasoro amplitude \cite{Virasoro:1969me, Shapiro:1969km}
\begin{equation}
	\mathscr{A}_c(p_i) \sim g_s^2 \, \delta^{(26)}(p_1+p_2+p_3+p_4)\, \frac{ \Gamma(-1-\frac{\alpha'\, s}{4}) \, \Gamma(-1-\frac{\alpha'\, t}{4}) \, \Gamma(-1-\frac{\alpha'\, u}{4}) }{ \Gamma(2+\frac{\alpha'\, s}{4}) \, \Gamma(2+\frac{\alpha'\, t}{4}) \, \Gamma(2+\frac{\alpha'\, u}{4}) } \,.
\end{equation}
In the field theory limit, this single amplitude describes the three processes associated to the $s$, $t$ and $u$ channels, so that crossing symmetry is built in. Using the properties of the $\Gamma$ functions, one may show that $\mathscr{A}_c(p_i)$ enjoys the Regge behaviour 
\begin{equation}
	\mathscr{A}_c(p_i) \propto s^{2+\alpha' \, t/2} \frac{\Gamma(-1-\frac{\alpha'\, t}{4})}{\Gamma(2+\frac{\alpha'\, t}{4})} \,,
\end{equation} 
for large $s$ and fixed $t$, and is exponentially suppressed 
\begin{equation}
	\mathscr{A}_c(p_i) \propto e^{-\frac{\alpha'}{2} ( s\log s+t\log t+u\log u) } \,,
\end{equation}
for $s,t,u \to \infty$. 

For completeness, we can briefly discuss the scattering of four open string tachyons. The endpoints of open strings trace boundaries upon their propagation, so that the relevant Riemann surface is now a disk with vertex operators attached to its boundary.
Conformal transformations map this surface to the upper complex plane with vertex operators inserted on the real axis. The corresponding amplitude reads
\begin{equation}
	\mathscr{A}_o(p_i) \sim g_s\, \Big\langle c(x_1) e^{ip_1\cdot X(x_1)}\,c(x_2) e^{ip_2\cdot X(x_2)}\,c(x_3) e^{ip_3\cdot X(x_3)}\, \int dx_4 \,e^{ip_4\cdot X(x_4)} \Big\rangle \,.
\end{equation}
The calculation of the correlators follows a similar procedure and, by fixing $x_{1,2,3}$ to $0,1,\infty$, respectively, and integrating $x_4$ over $[0,1]$ one obtains
\begin{equation}
	\mathscr{A}_o(p_i) \sim g_s \,\delta^{(26)}(p_1+p_2+p_3+p_4)\, B(-1-\alpha' s,-1-\alpha' t) \,,
\end{equation}  
where $B(p,q)$ is the Euler beta function. Clearly this is only one out of six possible choices of ordering the positions $x_i$. Summing over all possibilities, one obtains the celebrated Veneziano amplitude \cite{Veneziano:1968yb}
\begin{equation}
	\begin{split}
		\mathscr{A}_o(p_i) \sim & g_s \,\delta^{(26)}(p_1+p_2+p_3+p_4) \, \left[ B(-1-\alpha' s,-1-\alpha' t)   \right. \\
				& \qquad \left. + B(-1-\alpha' s,-1-\alpha' u) + B(-1-\alpha' t,-1-\alpha' u) \right] \,,
	\end{split}
\end{equation}  
which is fully crossing symmetric.

Higher genus amplitudes can be computed following a similar pattern, but involve the CFT correlators on genus-$g$ Riemann surfaces, as well as the integration over the corresponding moduli.
In general, this can quickly become involved and, in the following, we shall focus on the genus one vacuum energy \cite{Polchinski:1985zf}.

In this case, the path integral over the worldsheet metric reduces to a finite dimensional integral of the complex structure $\tau$ of the torus, parametrising gauge inequivalent metrics.
The CKG of the torus is Abelian and contains two translations. Its volume is finite and given by $\int d^2 z = \tau_2$ and, thus, 
\begin{equation}
	\int \frac{[\mathscr{D}g]}{\text{vol(CKG)}}  \to \int_{\mathscr{F}} \frac{d^2\tau}{\tau_2} \,,
\end{equation}
with $\mathscr{F}$ being the fundamental domain \eqref{FundDomain}. Notice that for scattering amplitudes, one instead fixes the position of one vertex operator and, therefore, the $\tau_2$ factor associated to $\text{vol(CKG)}$ is now absent.
One is left to perform the Gaussian integrals over the scalar fields $X^\mu$ and the reparametrisation ghosts. 

Let us start with the contribution of a single scalar $X$ with periodic boundary conditions 
along both cycles of the worldsheet torus with metric
\begin{equation}
	g_{ab} = \frac{1}{\tau_2} \begin{pmatrix} 1 & \tau_1 \\ \tau_1 & |\tau|^2 \end{pmatrix} \,.
\end{equation}
We can now expand $X$ onto the orthonormal basis of eigenmodes $\phi_{m,n}(\sigma^1,\sigma^2)$ of the Laplace operator $\Box$ on the torus, 
\begin{equation}
	X = \sum_{m,n} c_{m,n} \phi_{m,n} \,,
\end{equation}
where $\Box \,\phi_{m,n} = -\lambda_{m,n}\phi_{m,n}$ with 
\begin{equation}
	\lambda_{m,n} = \frac{\pi^2}{\tau_2^2}|m-\tau n|^2 \,.
\end{equation}
Notice that the zero mode $\phi_{0,0}$ is actually constant, due to the periodicity conditions, and it is fixed by the normalisation condition $\| \phi_{0,0} \| = 1$ to be $\phi_{0,0}=\tau_{2}^{-1/2}$.
The path integral over $X$ now turns into an integral over the Fourier modes $c_{m,n}$. In particular, the integral over the zero mode is
\begin{equation}
	\int dc_{0,0} = \int \frac{dx}{\phi_{0,0}} = L \sqrt{\tau_2} \,,
\end{equation}
where $L$ is the infinite linear volume spanned by the centre of mass $x$ of the string. The integral over the remaining Fourier modes produces $({\det}' \Box)^{-1/2}=\prod'_{m,n} \lambda_{m,n}^{-1/2}$, where the primes indicate that the zero eigenvalue $\lambda_{0,0}$ is excluded. This infinite product is clearly divergent and must be carefully defined using zeta function regularisation. The main ingredients needed for this calculation are 
\begin{equation}
	\prod_{n\neq 0} a = a^{-2\zeta(0)} \,,\quad \prod_{n>0} n^\alpha = e^{-\alpha \zeta'(0)} \,, \quad \prod_{n>0}\left(1-\frac{a^2}{n^2}\right) = \frac{\sin \pi a}{\pi a}\,.
\end{equation}
A straightforward computation then yields
\begin{equation}
	{\det}' \Box = (2\tau_2 \,\eta\,\bar\eta)^2 \,.
\end{equation}
Putting everything together, one finds $L/(\sqrt{\tau_2}\,\eta\bar\eta)$, up to an overall constant.

The contribution of the reparametrisation ghosts to the vacuum energy proceeds in a similar fashion, although one has to properly treat the zero modes. From the Riemann-Roch theorem we know that there is one zero mode for each ghost and, therefore, one should compute 
\begin{equation}
	\int [\mathscr{D}b]\,[\mathscr{D}c]\,[\mathscr{D}\bar b]\,[\mathscr{D}\bar c] \,bc\bar b\bar c\,e^{-S_{\text{ghost}}} \,.
\end{equation}
The ghost insertions simply soak up the zero modes from the integration measure, so that their contribution to the path integral no longer vanishes, and gives $\phi_{0,0}^4 = \tau_2^{-2}$. The non-zero modes instead contribute with 
\begin{equation}
	{\det}' \nabla_z \, {\det}' \nabla_{\bar z} \sim {\det}' \Box = (2\tau_2\,\eta\,\bar\eta)^2 \,.
\end{equation}
Assembling the contributions of the 26 bosonic coordinates $X^\mu$, together with those of the ghosts, one recovers the result
\begin{equation}
	\log{\mathscr{Z}} = V_{26} \int_{\mathscr{F}} \frac{d^2\tau}{\tau_2^{14}}\, \frac{1}{\eta^{24}\,\bar\eta^{24}} \,,
\end{equation}
which precisely matches eq. \eqref{ClosedPart} for $D=26$ upon restriction of the integral to the fundamental domain.


\section{Fermionic Strings}
\label{ST:Fermionic}

In spite of successfully providing a quantum theory of gravity, the simple bosonic string model discussed so far clearly has a number of drawbacks. Its spectrum necessarily contains a tachyon, while spacetime fermions are absent.
In order to overcome these problems, one is lead to introduce additional degrees of freedom on the string worldsheet \cite{Polyakov:1981re}. The natural choice is two dimensional Majorana fermions $\psi^\mu$, so that the Polyakov action, in a suitable generalisation of the conformal gauge, reads
\begin{equation}
	S = -\frac{1}{4\pi\alpha'} \int d^2\sigma \left( \partial_a X^\mu \partial^a X_\mu - i \bar\psi^\mu \rho^a \partial_a \psi_\mu \right) \,,
	\label{SuperPol}
\end{equation}
where $\rho^0=\sigma^2$ and $\rho^1=i \sigma^1$ being the two-dimensional Dirac matrices satisfying the Clifford algebra $\{ \rho^a, \rho^b\} = -2\eta^{ab}$ and the index $\mu=0,\ldots,D-1$ runs over $D$-dimensional spacetime. In addition to global Poincar\'e symmetry and conformal invariance on the worldsheet, this action is also invariant under the supersymmetry transformation \cite{Gervais:1971ji}
\begin{equation}
	\delta X^\mu = \bar\epsilon \psi^\mu \,, \qquad \delta\psi^\mu = -i \rho^a \epsilon\,\partial_a X^\mu \,,
\end{equation}
with real parameter $\epsilon$, transforming as a Majorana spinor\footnote{Actually, this transformation only realises a faithful representation of the supersymmetry algebra on-shell, where indeed the bosonic and fermionic degrees of freedom match. An off-shell realisation would require the introduction of the auxiliary real scalar field $B^\mu$, entering the action via the bilinear $-B^\mu B_\mu$. The transformation of the fermion then involves the additional term $B^\mu \epsilon$, while $\delta B^\mu=-i\bar\epsilon\rho^a\partial_a\psi^\mu$.}. The Dirac equation of motion implies that the two components of $\psi^\mu$ propagate independently as left or right movers, in accordance with the fact that, in two dimensions, irreducible spinorial representations of the Lorentz group are Majorana-Weyl fermions. As a consequence, the supersymmetry transformations independently rotate the left and right moving fields
\begin{equation}
	\delta X^\mu_{\text{L,R}}(\sigma^\pm) = \mp i\epsilon_{\pm}\psi^\mu_{\pm} \,,\qquad \delta\psi^\mu_{\pm}(\sigma^\pm) = \pm 2\epsilon_{\pm}\partial_{\pm}X^\mu \,,
\end{equation}
thus, respecting the factorisation of two-dimensional conformal symmetry. The energy-momentum tensor $T$ and the supercurrent $G$ read
\begin{equation}
	T_{\pm\pm} = \partial_{\pm}X^\mu \partial_{\pm}X_\mu + \frac{i}{2} \psi^\mu_{\pm}\partial_{\pm}\psi_{\pm\, \mu} \,,\qquad G_{\pm} = \psi^\mu_{\pm}\partial_{\pm}X_\mu \,.
\end{equation}
The vanishing of energy-energy momentum tensor, together with the structure of the supersymmetry algebra $GG\sim T$, also requires the vanishing of the supercurrent. This constraint can be also seen to arise from the consistent coupling of the worldsheet fermions to two-dimensional supergravity. 

As usual, the equations of motion have to be supplemented by appropriate boundary conditions. For the bosonic coordinates they were already discussed in Section \ref{ST:Bosonic}, together with the corresponding mode expansions. 
In the case of closed strings, the vanishing of the boundary terms for the fermions simply implies periodicity or anti-periodicity in the $\sigma$ variable. In the case of open strings, the boundary conditions imply 
 $\psi^\mu_{+} \delta\psi_{+\,\mu} =\psi^\mu_{-} \delta\psi_{-\,\mu}$ at each endpoint; this leads to the two possibilities $\psi^\mu_{+}=\pm \psi^\mu_{-}$ at $\sigma=1$, upon conventionally fixing $\psi^\mu_{+}=\psi^\mu_{-}$ at $\sigma=0$.

In the closed string case, the mode expansions, thus, read
\begin{equation}
	\psi^\mu_{+} = \sqrt{2\pi\alpha'}\sum_{n\in\mathbb Z} d^\mu_{n} \,e^{-2\pi i n\sigma^{+}} \,,\qquad  \psi^\mu_{+} = \sqrt{2\pi\alpha'}\sum_{r\in\mathbb Z+\frac{1}{2}} b^\mu_{r} \,e^{-2\pi i r\sigma^{+}}  \,,
	\label{ModeExpClosedPsi}
\end{equation}
for periodic and anti-periodic fermions, respectively. Similar expressions hold for the right movers $\psi^\mu_{-}$, with $d\to \tilde d$, $b\to \tilde b$ and $\sigma^{+} \to \sigma^{-}$.
In the open string case, we have
\begin{equation}
	\psi^\mu_\pm = \sqrt{\pi\alpha'} \sum_{n\in \mathbb Z} d^\mu_{n} \, e^{-i\pi n\sigma^{\pm}} \,, \qquad \psi^\mu_\pm = \sqrt{\pi\alpha'} \sum_{r\in \mathbb Z+\frac{1}{2}} b^\mu_{r} \, e^{-i\pi r\sigma^{\pm}}
\end{equation}
for the two cases $\psi^\mu_{+} = \pm \psi^\mu_{-}$ at $\sigma=1$, respectively. For both closed and open strings, integer modes define the Ramond (R) sector \cite{Ramond:1971kx, Ramond:1971gb}, while the half-integer ones define the Neveu-Schwarz (NS) sector \cite{Neveu:1971rx, Neveu:1971iv, Neveu:1971iw}.

The canonical quantisation of the fermions amounts to imposing the equal-time anti-commutation relations $\{\psi^\mu_{\pm}(\sigma), \psi^\nu_{\pm}(\sigma')\} = \pi \eta^{\mu\nu}\delta(\sigma-\sigma')$.
Similarly to the bosonic string, the Fock space contains negative norm states, generated by the Fourier modes of $\psi^0$ in both NS and R sectors. These are non-propagating degrees of freedom, which are removed by imposing that physical states lie in the cohomology of $G_\pm$, and it can be shown that this happens only in $D=10$ spacetime dimensions.

An alternative way to quantise the action \eqref{SuperPol} is to resort to light-cone quantisation. As in the bosonic case, there is an infinite number of conserved Noether charges associated to $T_{\pm\pm}$ and $G_{\pm}$. Indeed, any current of the form $f(\sigma^\pm)T_{\pm\pm}$ and $g(\sigma^\pm)G_\pm$ is trivially conserved for any $f$ and $g$. This implies the existence of an infinite number of generators that extend the bosonic conformal symmetry to the infinite-dimensional superconformal group.
Making use of these symmetries, we can set $X^{+} = x^{+}+2\pi\alpha' p^{+}\tau$ and $\psi^{+} = 0$, so that the energy-momentum tensor and supercurrent constraints can be solved for
\begin{equation}
		\partial_{\pm} X^{-} = \frac{1}{2\pi\alpha' p^{+}} \left( \partial_{\pm} X^i \partial_{\pm} X^i + \frac{i}{2} \psi_{\pm}^i \partial_{\pm} \psi_{\pm}^i \right) \,, \qquad   \psi^{-}_{\pm} = \frac{1}{\pi\alpha' p^{+}} \psi^i_{\pm}\partial_{\pm} X^i \,.
\end{equation}
As a result, only the transverse (super)coordinates $X^i, \psi^i$ are physical and generate a Fock space with positive norm.
Indeed, plugging in the mode expansions, we find
\begin{equation}
	\alpha^{-}_{m} = \frac{1}{\sqrt{2\alpha'}p^{+}} \left[ \sum_{n \in \mathbb{Z} } \alpha^i_{n}\alpha^i_{m-n} + \sum_{r \in \mathbb{Z}+\frac{1}{2}} \left( \frac{m}{2}-r\right) \,b^i_r b^i_{m-r}  \right]
	\label{SuperstringLC}
\end{equation}
and
\begin{equation}
	b^{-}_{r} = \sqrt{\frac{2}{\alpha'}} \frac{1}{p^{+}} \sum_{s \in \mathbb{Z}+\frac{1}{2}} \alpha^i_{r-s} b^i_{s}  \,,
\end{equation}
for the NS sector of the left-movers of closed strings. Similar expressions can be derived for the right-movers, as well as for the $R$ sectors. In the open string case one simply multiplies the r.h.s. of both equations by a factor of $1/2$.

As in the bosonic case, among the infinite relations \eqref{SuperstringLC}, the $m=0$ one plays a special role, since it provides the mass-shell condition. Recalling that $\alpha_0^\mu=\sqrt{2\alpha'}\,p^\mu$ for open strings, while $\alpha_0^\mu=\tilde\alpha_0^\mu=\sqrt{\alpha'/2}\,p^\mu$ for closed strings, one finds
\begin{equation}
	M^2_{\text{open}} = \frac{1}{\alpha'} \left(N_X + N_\psi - \frac{D-2}{16} \right)  \,,
\end{equation}
in the NS sector, and
\begin{equation}
	M^2_{\text{open}} = \frac{1}{\alpha'} \left(N_X + N_\psi \right)  \,,
\end{equation}
in the R sector.  In the closed string case, the left movers contribute with
\begin{equation}
	M^2_{\text{closed}, L} = \frac{4}{\alpha'} \left(N_X+ N_\psi - \frac{D-2}{16} \right)  \,,
\end{equation}
in the NS sector, and
\begin{equation}
	M^2_{\text{closed}, L} = \frac{4}{\alpha'} \left(N_X+ N_\psi \right)  \,,
\end{equation}
in the R sector, with similar expressions for the right moving mass. Closed string states are then built by combining the left and right-moving oscillators while respecting the level matching condition $M^2_{\text{closed}, L}=M^2_{\text{closed}, R}$.
In these expressions, we used the definition of $N_X$ given in Section \ref{ST:Bosonic}, while
\begin{equation}
	N_\psi = \begin{cases}
					\sum_{r=1/2}^{\infty} r b^i_{-r} b^i_{r} & \text{in the NS sector} \,, \\
					\sum_{n=1}^{\infty} n d^i_{-n} d^i_{n} & \text{in the R sector} \,.
			\end{cases}
\end{equation}
Furthermore, the zero point energy inside the brackets is computed via zeta function regularisation and each periodic boson contributes with $-1/24$, each R fermion contributes with $+1/24$, while each NS fermion contributes with $-1/48$.
The overall vanishing of the zero point energy in the R sector reflects the fact that the boundary conditions of bosons and fermions respect supersymmetry. In the NS sector, instead, the anti-periodicity of the fermions ``spontaneously break" worldsheet supersymmetry and represent the simplest realisation of the Scherk-Schwarz mechanism \cite{Scherk:1979zr}.

We now have all the ingredients to discuss the light spectrum, starting from the open strings. In the NS sector, the vacuum $|0\rangle$ is a spacetime scalar with mass $M^2= -(D-2)/16\alpha'$ and is, thus, tachyonic (for $D>2$). The first excited level is $b^i_{-1/2}|0\rangle$ transforming in the vectorial representation of the little group $\text{SO}(D-2)$. This is clearly compatible with Lorentz invariance if and only if this state is massless, which is the case only in $D=10$ dimensions.
Indeed, it can be shown that the spacetime Lorentz algebra is properly realised in this critical dimension. Massive states can be similarly constructed by the (repeated) action of $\alpha_{-n}^i$ and $b^i_{-r}$ and correspond to higher spin fields. In the R sector, instead, the vacuum is massless, but is no longer a singlet. This can be easily seen from the fact that $d^i_0$ commutes with the Hamiltonian and satisfies the Clifford algebra $\{d_0^i, d_0^j\} = 2\delta^{ij}$. As a result, the vacuum transforms in the spinorial representations $\textbf{8}_s$ and $\textbf{8}_c$ of the spacetime little group $\text{SO}(8)$, corresponding to the Majorana-Weyl fermions of opposite chirality $s$ and $c$. This implies that the superstring in the R sector describes spacetime fermions.

The closed string spectrum can be built by combining left and right movers and one has to distinguish between four possibilities. In the NS-NS sector is bosonic and its lightest state is the vacuum $|0\rangle_L\otimes |0\rangle_R$, which clearly satisfies level matching and describes a tachyonic scalar of mass $M^2= -(D-2)/4\alpha'$. The next level-matched states originate from $b^i_{-1/2}\tilde b^j_{-1/2}$ acting on the vacuum. This decomposes into the symmetric traceless, the anti-symmetric and the singlet representations of $\text{SO}(D-2)$, and consistency with Lorentz symmetry requires that these states be massless, which again fixes $D=10$. These states are identified as the graviton $G_{\mu\nu}$, the rank-2 anti-symmetric tensor $B_{\mu\nu}$, and the dilaton $\Phi$. Also the R-R sector is bosonic, and its tower of states starts already at the massless level. To identify the spacetime representations of the latter states, we have to consider the tensor product decomposition of $(\textbf{8}_s \oplus \textbf{8}_c)\otimes (\textbf{8}_s \oplus \textbf{8}_c)$. As a result, $\textbf{8}_s\otimes \textbf{8}_s = \textbf{1} \oplus \textbf{28} \oplus \textbf{35}_{+}$ yields a scalar, a 2-form and a 4-form whose field strength is self-dual in $D=10$ dimensions, 
$\textbf{8}_c\otimes \textbf{8}_c = \textbf{1} \oplus \textbf{28} \oplus \textbf{35}_{-}$ yields another scalar, another 2-form and a 4-form whose field strength is now anti self-dual, while
$\textbf{8}_s\otimes \textbf{8}_c = \textbf{8}_v \oplus \textbf{56}$ yields a vector and a 3-form field, and similarly for $\textbf{8}_c\otimes \textbf{8}_s$. The NS-R sector is fermionic and its level-matched spectrum starts at the massless level. It is obtained by the tensor product decompositions $\textbf{8}_v\otimes \textbf{8}_s=\textbf{56}_s + \textbf{8}_c$ and  $\textbf{8}_v\otimes \textbf{8}_c = \textbf{56}_c + \textbf{8}_s$. Therefore, it comprises two spin $3/2$ fields of $s$ and $c$ chirality, known as the gravitini, and two spin $1/2$ fields of $c$ and $s$ chirality, known as the dilatini. The R-NS sector gives an additional copy of the NS-R spectrum.

Actually, the closed string spectrum discussed so far yields an inconsistent string vacuum. One way to see this is the fact that  keeping all the above states, together with their massive excitations, is incompatible with modular invariance. As we shall see, only certain sub-sectors will give a consistent string theory \cite{Gliozzi:1976qd}. For open strings, the argument is slightly more involved and we will return to this in the next sections. In order to construct consistent vacua, it is useful to package the states of the NS and R sectors according to their worldsheet fermion parity $(-1)^{F_{\text{ws}}}$. The latter is defined such that states built out of an even (resp. odd) number of $b_r$ and/or $d_n$ oscillators have $(-1)^{F_{\text{ws}}} = +1$  (resp. $-1$). The parity of R vacua is positive for $|\textbf{8}_s\rangle$ and negative for $|\textbf{8}_c\rangle$ since, conventionally, $|\text{8}_s\rangle$  (resp. $|\textbf{8}_c\rangle$) is built out of an even (resp. odd) number of $d_0$ oscillators. This packaging yields the four traces
\begin{equation}
	\begin{split}
		O_8(q) &\equiv \text{Tr}\,{}_{\text{NS}}\left[ \, \frac{1+(-1)^{F_{\text{ws}}} }{2} \,q^{N_X+N_\psi - 1/2} \, \right]  \,,\\
		V_8(q) &\equiv \text{Tr}\,{}_{\text{NS}}\left[ \, \frac{1-(-1)^{F_{\text{ws}}} }{2} \,q^{N_X+N_\psi - 1/2} \, \right]  \,,\\
		S_8(q) &\equiv \text{Tr}\,{}_{\text{R}}\left[ \, \frac{1+(-1)^{F_{\text{ws}}} }{2} \,q^{N_X+N_\psi } \, \right]  \,,\\
		C_8(q) &\equiv \text{Tr}\,{}_{\text{R}}\left[ \, \frac{1-(-1)^{F_{\text{ws}}} }{2} \,q^{N_X+N_\psi } \, \right] \,,
	\end{split}
\end{equation}
associated, respectively, to the singlet/adjoint, vectorial, and the two spinorial conjugacy classes of $\text{SO}(8)$. In general, these traces can actually be identified with the characters \cite{Bianchi:1990yu} of the level-one current algebra of $\text{SO}(2n)$, given by
\begin{equation}
	\begin{split}
		O_{2n}(q) &= \frac{ \vartheta_3(0|\tau)^n + \vartheta_4(0|\tau)^n}{2\eta^n(\tau) } \,,\\
		V_{2n}(q) &=\frac{ \vartheta_3(0|\tau)^n - \vartheta_4(0|\tau)^n}{2\eta^n(\tau) }\,,\\
		S_{2n}(q) &= \frac{ \vartheta_2(0|\tau)^n +i^{-n} \vartheta_1(0|\tau)^n}{2\eta^n(\tau) }\,,\\
		C_{2n}(q) &= \frac{ \vartheta_2(0|\tau)^n -i^{-n}\vartheta_1(0|\tau)^n}{2\eta^n(\tau) }\,.
	\end{split}
\end{equation}
For a review on affine current algebras, see \cite{Goddard:1986bp}.
 Here, we have introduced the four Jacobi theta functions
\begin{equation}
	\begin{split}
		\vartheta_1(z|\tau) &= 2\sin \pi z \, q^{1/8} \prod_{n>0} (1-q^n)(1-q^n\,e^{2\pi i z})(1-q^n\,e^{-2\pi i z}) \,, \\
		\vartheta_2(z|\tau) &= 2\cos \pi z \, q^{1/8} \prod_{n>0} (1-q^n)(1+q^n\,e^{2\pi i z})(1+q^n\,e^{-2\pi i z}) \,, \\
		\vartheta_3(z|\tau) &=  \prod_{n>0} (1-q^n)(1+q^{n-1/2}\,e^{2\pi i z})(1+q^{n-1/2}\,e^{-2\pi i z}) \,, \\
		\vartheta_4(z|\tau) &=  \prod_{n>0} (1-q^n)(1-q^{n-1/2}\,e^{2\pi i z})(1-q^{n-1/2}\,e^{-2\pi i z}) \,, \\
	\end{split}
\end{equation}
evaluated at $z=0$. Notice that $\vartheta_1(0|\tau)$ vanishes identically, reflecting the fact that at each mass level we have an equal number of fermionic states of opposite chirality. This degeneracy can be lifted by turning on suitable background magnetic fields, so that one may then really distinguish between the $S_8$ and $C_8$ traces. The generators $T$ and $S$ of the modular group $\text{SL}(2;\mathbb{Z})$ have the well-defined action
\begin{equation}
	\begin{split}
		&T: \quad \vartheta_1 \to e^{i\pi/4} \,\vartheta_1 \,, \quad  \vartheta_2 \to e^{i\pi/4} \,\vartheta_2 \,, \quad \vartheta_{3,4} \to \vartheta_{4,3} \,, \\
		&S:\quad \vartheta_1 \to \sqrt{-i\tau} \,\vartheta_1 \,,\quad \vartheta_{2,4} \to \sqrt{-i\tau}\, \vartheta_{4,2} \,, \quad \vartheta_3 \to \sqrt{-i\tau}\, \vartheta_3 \,,
	\end{split}
\end{equation}
on the Jacobi theta functions with $z=0$ which, together with the transformation properties \eqref{EtaTransform} of the Dedekind eta function, implies the matrix representations 
\begin{equation}
	T = e^{-i\pi n/12}\,\text{diag}(1,-1,e^{i\pi n/4},e^{i\pi n/4}) \,,\qquad S = \frac{1}{2}\,\begin{pmatrix} 1 & 1 & 1 & 1 \\ 1 & 1 & -1 & -1 \\ 1 & -1 & i^{-n} & -i^{-n} \\  1 & -1 & -i^{-n} & i^{-n}  \end{pmatrix} \,.
\end{equation}
on the space of the $\text{SO}(2n)$ characters $\chi=\{O_{2n},V_{2n},S_{2n},C_{2n}\}$.

Consistent closed string vacua thus correspond to the sesquilinear combinations 
\begin{equation}
	Z = \frac{1}{\eta^8\,\bar\eta^8}\, \mathscr{T} \equiv \frac{1}{\eta^8\,\bar\eta^8} \sum_{i,j}\bar\chi_i \,\mathscr{N}_{ij}\, \chi_j  \,,
\end{equation}
of left moving and right moving characters,  with the matrix $\mathscr{N}_{ij}$ enforcing the Gliozzi-Scherk-Olive (GSO) projection \cite{Gliozzi:1976qd}. Its entries can be either $\pm 1$ or $0$, and are required to satisfy spin-statistics and  the (genus one) modular invariance constraints
\begin{equation}
	T^\dagger \mathscr{N} T = \mathscr{N} \,,\qquad  S^\dagger \mathscr{N} S = \mathscr{N} \,.
\end{equation}
It can be shown that these conditions automatically guarantee higher genus modular invariance.  There are only four inequivalent solutions, 
\begin{equation}
	\begin{split}
		\mathscr{T}_{\text{IIA}} =& (V_8-S_8)(\bar{V}_8-\bar{C}_8)    \,, \\
		\mathscr{T}_{\text{IIB}} =& |V_8-S_8|^2   \,, \\
		\mathscr{T}_{\text{0A}} =& |O_8|^2 + |V_8|^2 + S_8 \bar{C}_8 +C_8 \bar{S}_8   \,, \\
		\mathscr{T}_{\text{0B}} =& |O_8|^2 + |V_8|^2 + |S_8|^2 +|C_8|^2    \,, 
	\end{split}
	\label{10dSuperstrings}
\end{equation}
corresponding to the type IIA, type IIB, type 0A and type 0B superstring theories in $D=10$ \cite{Gliozzi:1976qd, Dixon:1986iz, Seiberg:1986by}.

The type IIA and type IIB superstrings enjoy $\mathscr{N}=(1,1)$ and $\mathscr{N}=(2,0)$ spacetime supersymmetry, respectively, and their corresponding partition functions vanish identically, as implied by the \emph{equatio identica satis abstrusa} identity of Jacobi
\begin{equation}
	\vartheta_3^4(0|\tau) -\vartheta_4^4(0|\tau)-\vartheta_2^4(0|\tau) = 0\,.
\end{equation}
 Their massless excitations, thus, correspond to the field content of the associated supergravity multiplets. In type IIA they comprise $G_{\mu\nu}$, $B_{\mu\nu}$ and $\Phi$ from the NS-NS sector, a 1-form $C_1$ and a 3-form $C_3$ from the R-R sector, together with a pair of opposite-chirality gravitini and dilatini. In type IIB they comprise $G_{\mu\nu}$, $B_{\mu\nu}$ and $\Phi$ from the NS-NS sector, a 0-form $C_0$,  a 2-form $C_2$ and a 4-form (with self-dual field strength) $C_4^{(+)}$ from the R-R sector, together with two gravitini of the same ($S$) chirality and two dilatini of opposite ($C$) chirality. This chiral spectrum is actually anomaly-free \cite{Alvarez-Gaume:1983ihn}, as guaranteed by modular invariance \cite{Schellekens:1986yi,Schellekens:1986xh,Lerche:1987sg}.  
 The type 0A and type 0B superstrings are non-supersymmetric and only contain bosonic excitations. The NS-NS sector is common to both theories and comprises a tachyon, $G_{\mu\nu}$, $B_{\mu\nu}$ and $\Phi$. In the R-R sectors, type 0A comprises  two 1-forms and two 3-forms from the R-R sectors, while type 0B comprises two 0-forms, two 2-forms, and one unconstrained 4-form.

Although the partition functions in \eqref{10dSuperstrings} define four independent string theories in ten dimensions there is, nevertheless, an interesting way to relate them. As an example, let us start from type IIB and observe that it is invariant under the action of the spacetime fermion parity $(-1)^F$. This discrete symmetry can then be gauged, by restricting the Hilbert space $\mathscr{H}_{\text{IIB}}$ to only those states invariant under this transformation. This amounts to computing 
\begin{equation}
	\begin{split}
		\mathscr{T}' &= \text{Tr}_{\mathscr{H}_{\text{IIB}}} \,\left[\, \frac{1+(-1)^{F}}{2} \,q^{L_0}\,\bar{q}^{\bar L_0} \,\right]   \\
				&= \frac{1}{2}\,|V_8-S_8|^2 + \frac{1}{2}\, |V_8+S_8|^2 \,.
	\end{split}
\end{equation}
In accordance with the gauging of the spacetime fermion parity, the resulting spectrum only contains bosonic states from the NS-NS and R-R sectors of the type IIB superstring theory. However, $\mathscr{T}'$ is not modular invariant and, therefore, cannot describe a consistent string vacuum. A way out is to restore modular invariance with the addition of new ``twisted" states, which are not part of the original Hilbert space. In the case at hand, the only solution is
\begin{equation}
	\begin{split}
		\mathscr{T}' &\to  \frac{1}{2}\,|V_8-S_8|^2 + \frac{1}{2}\, |V_8+S_8|^2 + \frac{1}{2} |O_8-C_8|^2 + \frac{1}{2} |O_8+C_8|^2 \,,
	\end{split}
\end{equation}
which is nothing but the partition function of the type 0B theory. This construction \cite{Dixon:1986iz,Seiberg:1986by} is simplest instance of an orbifold, whereby discrete symmetries can be gauged at the cost of introducing new ``twisted" states which restore modular invariance.
Similar constructions involving the full spacetime fermion parity or its right moving analogue $(-1)^{F_{\text{R}}}$, relate the four theories in \eqref{10dSuperstrings}. As we shall see in the next section, this procedure is an efficient way to construct new consistent closed string vacua.


\section{Heterotic Strings}
\label{ST:Heterotic}

In the previous section, we saw that an interesting way to generalise the bosonic string construction was to introduce additional fermionic degrees of freedom described in terms of Majorana-Weyl spinors $\psi^\mu_{\pm}$ carrying the same Lorentz index as the bosonic coordinates $X^\mu$. This lead to a natural factorisation of the corresponding super-CFT into a left and right moving sector, where $\psi^\mu_+$ and $\psi^\mu_{-}$ are independently rotated into $\partial_{\pm}X^\mu$ by supersymmetry transformations. This factorisation lies at the heart of string constructions and actually allows for an asymmetric generalisation. One may replace $\psi^\mu_{-}$ by $N$ free fermions $\lambda^A_{-}$, which are now invariant under the spacetime Lorentz group but transform in the fundamental representation of $\text{SO}(N)$ \cite{Gross:1984dd, Gross:1985fr, Gross:1985rr}. The resulting worldsheet action reads
\begin{equation}
	S = -\frac{1}{4\pi\alpha'} \int d^2\sigma \left( \partial_a X^\mu \partial^a X_\mu - 2i \psi^\mu_{+} \partial_- \psi_{+\, \mu} -2i \lambda^A_{-} \partial_{+} \lambda^A_{-} \right) \,,
\end{equation}
and the quantisation proceeds as in the fermionic string with an important difference: the right-moving sector is no longer supersymmetric and, therefore, there is no associated conserved supercharge. One is, thus, left with the constraints
\begin{equation}
	\begin{split}
		T_{++} &=   \partial_{+}X^\mu \partial_{+}X_\mu + \frac{i}{2} \psi^\mu_{+}\partial_{+}\psi_{+\, \mu}  = 0\,, \\
		T_{--} &=    \partial_{-}X^\mu \partial_{-}X_\mu + \frac{i}{2} \lambda^A_{-}\partial_{-}\lambda^A_{-}  = 0\,, \\
		G_{+} &=  \psi^\mu_{+}\partial_{+}X_\mu = 0 \,.
	\end{split}	
\end{equation}
As usual, in light-cone quantisation, the infinite conserved charges associated to $T_{++}$ and $T_{--}$, allow us to eliminate the oscillators in $X^{+} = x^{+} + 2\pi\alpha' p^{+} \tau$. The chiral nature of worldsheet supersymmetry, however, now implies that the  infinite conserved charges associated to $G_{+}$ only allow us to set $\psi^{+}_{+}=0$, while all right-moving $\lambda^A_{-}$ oscillators must be retained. Taking this into account, the above constraints may be solved for $X^{-}$ and $\psi^{-}_{+}$, 
\begin{equation}
	\begin{split}
		\partial_{+} X^{-} &= \frac{1}{2\pi\alpha' p^{+}} \left( \partial_{+} X^i \partial_{+} X^i + \frac{i}{2} \psi_{+}^i \partial_{+} \psi_{+}^i \right) \,,\\
		\partial_{-} X^{-} &= \frac{1}{2\pi\alpha' p^{+}} \left( \partial_{-} X^i \partial_{-} X^i + \frac{i}{2} \lambda_{-}^A \partial_{-} \lambda_{-}^A \right) \,,\\
		\psi^{-}_{+} &= \frac{1}{\pi\alpha' p^{+}} \psi^i_{+}\partial_{+} X^i \,.
	\end{split}
	\label{HetConstraints}
\end{equation}
These relations are enough to remove the negative-norm states associated to $X^0$ and to the left-moving $\psi^0_{+}$, and include the mass-shell conditions for the heterotic string. 

Using the mode expansions for the  bosons \eqref{ModeExpBosonClosed} and the left-moving fermions \eqref{ModeExpClosedPsi}, together with the mode expansions
\begin{equation}
	\lambda^A_{-} = \sqrt{2\pi\alpha'}\sum_{n\in\mathbb Z} \tilde\lambda^{A}_{n} \,e^{-2\pi i n\sigma^{-}} \,,\qquad  \lambda^A_{-} = \sqrt{2\pi\alpha'}\sum_{r\in\mathbb Z+\frac{1}{2}} \tilde{\lambda}^{A}_{r} \,e^{-2\pi i r\sigma^{-}}  \,,
\end{equation}
for the periodic and anti-periodic right-moving fermions $\lambda^A$, respectively, and assuming that all $\lambda^{A}$'s carry the same periodicity conditions, one finds
\begin{equation}
	M^2_{\text{L}} = \begin{cases}
					\frac{4}{\alpha'}\left( N_X + N_\psi - \frac{D-2}{16} \right) & \text{in the NS sector}\,, \\
					\frac{4}{\alpha'}\left( N_X + N_\psi  \right) & \text{in the R sector}\,, \\
				\end{cases}
\end{equation}
for the left-moving mass, and
\begin{equation}
	M^2_{\text{R}} = \begin{cases}
					\frac{4}{\alpha'}\left( N_X + N_\lambda - \frac{2D+N-4}{48} \right) & \text{in the anti-periodic $\lambda$ sector}\,, \\
					\frac{4}{\alpha'}\left( N_X + N_\lambda -\frac{D-N-2}{24} \right) & \text{in the periodic $\lambda$ sector}\,, \\
				\end{cases}
\end{equation}
for the right-moving mass. Clearly, physical states require the level-matching condition $M^2_{\text{L}} = M^2_{\text{R}}$. Notice that also in the right-moving periodic $\lambda$ sector, the zero modes $\tilde\lambda_0^{A}$ commute with the Hamiltonian and satisfy the $\text{SO}(N)$ Clifford algebra $\{ \tilde\lambda_{0}^{A} , \tilde\lambda_{0}^{B} \} = 2\delta^{AB}$. Therefore, the periodic $\lambda$ vacuum transforms as a spinor of $\text{SO}(N)$. Compatibility of the light spectrum with the spacetime Lorentz symmetry, requires also in this case $D=10$ dimensions, while fixing $N=32$. As in the type IIA/IIB superstrings, in the left-moving sector, the GSO projection removes the NS tachyonic vacuum and the Ramond $\textbf{8}_c$ vacuum together with their excitations. In the right-moving sector, the GSO projection removes instead the states belonging to the conjugacy classes of the fundamental and one of the spinorial representations of $\text{Spin}(32)$. The modular invariant partition function, thus, reads
\begin{equation}
	\mathscr{T}_{\text{SO}(32)} = (V_8-S_8)(\bar O_{32}+ \bar S_{32}) \,.
	\label{PartFunSO32}
\end{equation}
The massless states comprise the graviton, the $B$-field and the dilaton from $b^i_{-1/2}\tilde\alpha_{-1}^j |0\rangle_{L}\otimes |0\rangle_{R}$, the Rarita-Schwinger field of chirality $s$ and a Majorana-Weyl spinor of chirality $c$ from $\tilde \alpha^j_{-1} |\textbf{8}_s\rangle_{L}\otimes |0\rangle_{R}$, together with 496 gauge bosons and $s$ Majorana-Weyl fermions from $b^i_{-1/2}\tilde\lambda^A_{-1/2}\tilde\lambda^B_{-1/2} |0\rangle_{L}\otimes |0\rangle_{R}$ and $\tilde\lambda^A_{-1/2}\tilde\lambda^B_{-1/2}|\textbf{8}_s\rangle_{L}\otimes |0\rangle_{R}$, respectively. This spectrum enjoys $\mathscr{N}=(1,0)$ supersymmetry in ten dimensions, and the aforementioned states form the gravity multiplet and the gauge multiplet of $\text{SO}(32)$.
This spectrum is free of irreducible gravitational and gauge anomalies \cite{Alvarez-Gaume:1983ihn}, while the reducible anomalies are cancelled by the Green-Schwarz mechanism \cite{Green:1984sg} (see \cite{Bilal:2008qx,Alvarez-Gaume:2022aak} for a review of anomaly cancellation in string theory).

In the above, we imposed the same periodicity conditions for all 32 $\lambda$'s, which gave rise to the SO(32) heterotic string with $\mathscr{N}=(1,0)$ spacetime supersymmetry. This is not the only allowed choice, and one could instead split the right-moving fermions into sets obeying different periodicity conditions. If one insists on preserving spacetime supersymmetry in $D=10$, the GSO projection implies that the partition function factorises into the holomorphic contribution $V_8-S_8$  times an anti-holomorphic one associated to the gauge degrees of freedom. As a result, the latter must be modular invariant by itself and, it turns out, that there are exactly two choices involving 32 fermions: the $\bar{O}_{32}+\bar{S}_{32}$ combination discussed above, and $(\bar{O}_{16}+\bar{S}_{16})(\bar{O}_{16}+\bar{S}_{16})$ which gives rise to the celebrated $\text{E}_8\times \text{E}_8$ heterotic string with partition function 
\begin{equation}
	\mathscr{T}_{\text{E}_8\times\text{E}_8} = (V_8-S_8)(\bar{O}_{16}+\bar{S}_{16})(\bar{O}_{16}+\bar{S}_{16}) \,.
	\label{PartFunE8E8}
\end{equation}
Aside from the $\mathscr{N}=(1,0)$ gravitational multiplet, the light spectrum comprises a vector multiplet transforming under the 248-dimensional adjoint representation of $\text{E}_8\times\text{E}_8$. Indeed, an $\text{E}_8$ gauge group is isomorphic to $\text{Spin}(16)/\mathbb{Z}_2$, where one retains the conjugacy classes associated to the adjoint and one spinorial representation of $\text{Spin}(16)$. Notice that we have exactly $248+248=496$ copies of a vector multiplet, as in the case of the $\text{SO}(32)$ heterotic string, which is essential for gravitational anomaly cancellation. 

It is a property of two-dimensional CFTs that a pair of real free fermions $\lambda^{1,2}$ can be bosonised \cite{Coleman:1974bu, Mandelstam:1975hb} into a chiral compact scalar $\Phi$ at radius $\sqrt{\alpha'/2}$, known as the fermionic point.  Indeed, the dictionary 
\begin{equation}
	\lambda^{\pm} \equiv \frac{\lambda^2 \pm i \lambda^1}{\sqrt{2}} \to  e^{\pm i \sqrt{2/\alpha'} \Phi} \ , \qquad \lambda^{+} \lambda^{-} \to i \bar\partial \Phi \,,
\end{equation}
consistently reproduces the OPE relations. As a result, one may give an alternative description of the heterotic strings whereby the right-moving sector involves 16 compact scalars $\Phi^a$. The chiral factorisation induced by the requirement of spacetime supersymmetry, together with the modular invariance of the holomorphic and anti-holomorphic sectors, implies that the compact $\Phi^a$'s are associated to a 16-dimensional even, self-dual chiral lattice. The only two such cases are the root lattices of $\text{Spin}(32)/\mathbb{Z}_2$ and $\text{E}_8\times\text{E}_8$, corresponding to the partition functions \eqref{PartFunSO32} and \eqref{PartFunE8E8}, respectively \cite{Gross:1984dd, Gross:1985fr, Gross:1985rr}.

Although these cases are the only possible choices of heterotic theories in ten dimensions enjoying spacetime supersymmetry, they do not exhaust the space of all consistent, modular invariant, heterotic vacua. The most notable example is the non-supersymmetric $\text{SO}(16)\times\text{SO}(16)$ theory of \cite{Alvarez-Gaume:1986ghj, Dixon:1986iz,Seiberg:1986by}. In this case, the modular invariant partition function 
\begin{equation}
	\begin{split}
		\mathscr{T}_{\text{SO}(16)\times\text{SO}(16)} &= V_8 (\bar{O}_{16}\bar{O}_{16} + \bar{S}_{16}\bar{S}_{16}) - S_8 (\bar{O}_{16}\bar{S}_{16} + \bar{S}_{16}\bar{O}_{16}) \\
					&+ O_8 (\bar{V}_{16}\bar{C}_{16} + \bar{C}_{16}\bar{V}_{16}) - C_8 (\bar{V}_{16}\bar{V}_{16} + \bar{C}_{16}\bar{C}_{16}) 
	\end{split}
	\label{PartFunSO16SO16}
\end{equation}
no longer factorises, and the light spectrum comprises the universal graviton, $B$-field and dilaton, gauge bosons in the adjoint representation of $\text{SO}(16)\times\text{SO}(16)$, left-handed fermions in the $(\textbf{128},\textbf{1}) + (\textbf{1},\textbf{128})$  and right-handed fermions in the $(\textbf{16},\textbf{16})$ representations. The would-be tachyonic state originating from $O_8$ is actually massive, because of level-matching, which makes this non-supersymmetric closed string theory unique in $D=10$.

As in the superstring case, the $\text{SO}(16)\times\text{SO}(16)$ theory can be constructed from the supersymmetric $\text{E}_8\times\text{E}_8$ one by employing the $\mathbb{Z}_2$ orbifold, generated by $(-1)^{F+F_1+F_2}$. Here $F$ is the usual spacetime fermion number, while $F_{1,2}$ are the analogous ``fermion numbers" for each $\text{E}_8$ factor. Also in this case, modular invariance requires the presence of a twisted sector respecting the orbifold symmetry, which is precisely encoded in the second line of \eqref{PartFunSO16SO16}.

The only other choices are non-supersymmetric and have gauge groups $\text{SO}(32)$, $\text{SO}(16)\times\text{E}_8$, $(\text{SU}(2)\times\text{E}_7)^2$, $\text{SO}(8)\times\text{SO}(24)$, $\text{U}(16)$  \cite{Dixon:1986iz,Seiberg:1986by} and $\text{E}_8$  \cite{Kawai:1986vd}.  However, all of them involve tachyonic states, which render them classically unstable. 

Among these theories, the $\text{E}_8$ one is special, in that it has reduced rank and involves a current algebra of level 2. It can be constructed as a permutation orbifold, where the exchange of the two $\text{E}_8$'s is accompanied by $(-1)^F$. The partition function reads
\begin{equation}
	\begin{split}
		\mathscr{T}_{\text{E}_8} &= \frac{1}{2}\left[ (V_8-S_8) \bar\chi_8(\bar q)\bar\chi_8(\bar q) + (V_8+S_8)\bar\chi_8(\bar q^2) \right. \\
			& \left. + (O_8-C_8)\bar\chi_8(\sqrt{\bar{q}}) + (O_8+C_8) \bar{\hat{\chi}}_8(-\sqrt{\bar{q}}) \right] \,,
	\end{split}
\end{equation}
where, for convenience, we denote by $\bar\chi_8 = \bar{O}_{16}+\bar{S}_{16}$ the chiral $\text{E}_8$ character and we have also introduced the hatted characters 
\begin{equation}
	\hat{\chi}(-\sqrt{q}) = e^{-i\pi (h-c/24)}\, \chi(-\sqrt{q}) \,,
	\label{HattedCharacters}
\end{equation}
of conformal weight $h$ and associated central charge $c$. The light spectrum comprises the universal graviton, $B$-field and dilaton, a vector and a Majorana fermion in the adjoint representation of $\text{E}_8$, together with a singlet tachyon. This vacuum is a prototype example of a larger class of lower-dimensional constructions with reduced rank involving (freely-acting) permutation orbifolds, known as CHL strings \cite{Chaudhuri:1995fk}. 


\section{Open Strings and D-branes}
\label{ST:Dbranes}

Until now, we have mainly focused on the construction of closed string vacua. As we shall see, open strings also lead to interesting, although less straightforward, constructions. Closed strings naturally require periodicity conditions for the $X^\mu$ coordinates, which imply that their centre of mass is free to move in the whole ten-dimensional spacetime. Open superstrings, instead, require the specification of Neumann or Dirichlet boundary conditions at the two endpoints, and this affects their propagation.
The main difference between NN and DD strings lies in the zero modes of their $X^\mu$ coordinates, whereby only the former admit a centre of mass momentum, while the latter are stuck, as can be seen from eqs. \eqref{XopenNN} and \eqref{XopenDD}. Therefore, open strings with $p+1$ NN and $9-p$ DD boundary conditions specify a $(p+1)$-dimensional hypersurface along which the open strings are free to move, and naturally break $\text{SO}(1,9)$ down to $\text{SO}(1,p)\times \text{SO}(9-p)$. 
In this way, the Lorentz index $\mu$ splits into $\mu\to (a,i)$, where $a=0,\ldots,p$ spans the directions along the D$p$ brane, while $i=p+1,\ldots,9$ labels the transverse coordinates.
This hypersurface is known as a D$p$ brane \cite{Dai:1989ua,Leigh:1989jq,Horava:1989ga} (for reviews, see \cite{Polchinski:1996na, Polchinski:1996fm, Bachas:1996sc, Bachas:1998rg, Angelantonj:2002ct})  and fully specifies the set of boundary conditions of open strings. Following similar steps as in Section \ref{ST:Fermionic}, light-cone quantisation yields the mass formula
\begin{equation}
		M^2_{\text{open}} = \frac{1}{\alpha'} \left(N_X + N_\psi + \Delta\right)  \,,
\end{equation}
for open strings whose endpoints live on the same D$p$ brane, with $\Delta$ being the zero point energy which vanishes in the R sector, whereas it equals $-1/2$ in the NS sector. 
In the NS sector, the light spectrum contains the tachyonic vacuum $|0\rangle$, together with a massless vector in $(p+1)$ dimensions from $b_{-1/2}^a |0\rangle$ and $9-p$ scalars from $b_{-1/2}^i |0\rangle$.
The latter are actually associated to the position of the D-brane along the transverse directions. In the R sector, the vacuum is massless and, as usual, describes the $\textbf{8}_s$ and $\textbf{8}_c$ spinors of the original Lorentz group $\text{SO}(1,9)$, which are then to be properly decomposed. All these fields are free to propagate only along the world-volume of the D$p$ brane.

\begin{figure}
	\begin{center}
		\includegraphics[width=0.8\textwidth]{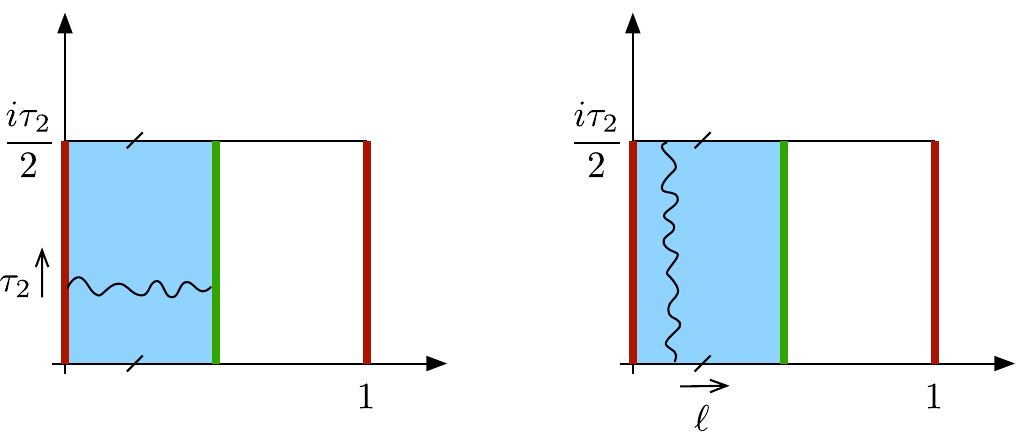}
	\end{center}
	\caption{The left figure shows the one-loop vacuum diagram of an open string which propagates for a (vertical) proper time $\tau_2$; its end-points trace the two boundaries of an annulus. The right figure illustrates the tree-level propagation of a closed string bouncing between the two boundaries; the proper time $\ell$ now flows horizontally. Both figures also display the double covering torus with modulus $i\tau_2/2$.}
	\label{FigAnnulus}
\end{figure}

Also in this case, one has to properly truncate the Hilbert space in order to construct a consistent open string theory. The way to proceed, however, is drastically different from the closed string case, since the one-loop vacuum diagram associated to an open string does not enjoy modular invariance. Indeed, it is given by a Riemann surface with the topology of an annulus, where the two boundaries are traced by the string endpoints, as shown in Fig. \ref{FigAnnulus}. Such a Riemann surface has vanishing Euler characteristic and may be built from a double-covering torus via the anti-holomorphic involution $z\to 2-\bar z$. The compatibility of this involution with the torus identification $z \sim z+m\tau+2n$, where $m,n \in \mathbb{Z}$, implies that the modulus of the double-covering torus is purely imaginary, a requirement clearly incompatible with modular invariance. Actually, the fact that the complex structure must be purely imaginary is not surprising since, in the case of open strings, one does not need to impose level-matching and, therefore, the Schwinger proper time $\tau_2$ is sufficient to parametrise the vacuum diagram.

Another important difference is that, although the torus always describes the one-loop propagation of closed strings, independently of the choice of elementary cell, in open strings an inversion of $\tau_2$ implies that time now flows horizontally as shown in Fig. \ref{FigAnnulus} and calls for a completely different interpretation in terms of closed strings freely propagating between the two boundaries of a cylinder \cite{Polchinski:1987tu, Sagnotti:1987tw, Bianchi:1988fr, Horava:1989vt}. The main lesson to be drawn from this observation is that open strings alone do not define a unitary theory, since their endpoints may join to form a closed string and/or the loop diagrams admit dual descriptions in terms of open/closed propagation. Given this fact, the open string spectrum encoded in the annulus partition function must be compatible with the closed string spectrum propagating along the dual cylinder. It is this constraint that essentially replaces modular invariance and selects the correct GSO projection of the open string Hilbert space. Therefore, if we wish to consistently describe open strings together with type II (closed) superstrings, we must employ the same supersymmetric GSO projection \cite{Gliozzi:1976qd}, so that, aside from an irrelevant multiplicative volume factor, the annulus partition function reads
\begin{equation}
	\mathscr{A} = \int_0^\infty \frac{d\tau_2}{\tau_2}\, \frac{1}{\tau_2^{(p+1)/2}\, \eta^8(\frac{i\tau_2}{2}) } \,(V_8 - S_8) (\tfrac{i\tau_2}{2}) \,,
	\label{AnnulusPFdirect}
\end{equation}
where Dedekind and Jacobi functions naturally depend on the modulus $i\tau_2/2$ of the double-covering torus.
Here, we have assumed that open strings have NN boundary conditions along $p+1$ coordinates, so that the massless spectrum corresponds to the dimensional reduction of a ten-dimensional vector and left-handed spinor on the $(p+1)$-dimensional world-volume of the D$p$ brane. This is the field content of a vector supermultiplet in a theory with 16 supercharges. 

The representation \eqref{AnnulusPFdirect} of the open string vacuum amplitude is normally referred to as the \emph{loop} or \emph{direct channel amplitude}. The transformation $\tau_2 \to \ell = 2/\tau_2$ defines what is called the \emph{tree-level} or \emph{transverse channel amplitude}
\begin{equation}
	\tilde{\mathscr{A}} = 2^{-(p+1)/2}\,\int_0^\infty \frac{d\ell}{\ell^{(p-9)/2}}\, \frac{1}{\eta^8(i\ell) } \,(V_8 - S_8) (i\ell) \,,
	\label{AnnulusPFtransverse}
\end{equation}
which describes the propagation of NS-NS and R-R states for proper time $\ell$ between the two boundaries, \emph{i.e.} the D$p$ branes.
In the limit of an infinitely long cylinder, $\ell \to \infty$, only massless excitations survive and the resulting diagram involves 1-point functions coupling the massless fields to the boundaries, and an on-shell propagator evaluated at zero momentum.
This shows that D$p$-branes are physical objects carrying tension as well as charge for the R-R potentials \cite{Polchinski:1995mt}, which identifies them as the BPS solitons of type II supergravity \cite{Hull:1994ys, Townsend:1995kk, Witten:1995ex} (see also \cite{Duff:1994an} for a review), preserving 16 of the original 32 supercharges. Since a $(p+1)$-form potential naturally couples to a $p$-dimensional (static) source, D$p$ branes exist in type IIA (IIB) superstring theory for $p$ even (odd).

Notice that, if the space transverse to the D-brane is compact, the theory of closed and open oriented  strings discussed so far cannot yield a consistent vacuum. This is because, on a compact space, Gauss' law requires a neutral configuration of charges, so that Faraday lines emitted from positively charged sources are absorbed by negatively charged ones. A consistent vacuum configuration may still be built, but it requires the inclusion of unoriented strings, and we defer the relevant discussion to the next section.

\begin{figure}
	\begin{center}
		\includegraphics[width=0.8\textwidth]{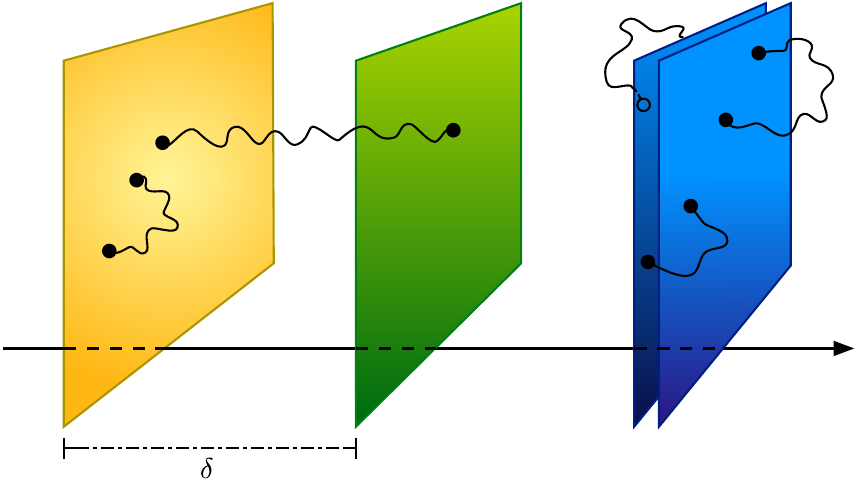}
	\end{center}
	\caption{D-branes and their open strings. The open strings stretching between the yellow and green branes include a massive vector that becomes massless when the relative distance $\delta$ goes to zero. The stack of the blue branes yields a non-Abelian gauge group.}
	\label{FigDbranes}
\end{figure}

Before we conclude this section, it is instructive to consider the situation depicted in Fig. \ref{FigDbranes}, where two parallel D-branes are separated by a distance $\delta$. Here we can identify two types of open strings: those which start and end on the same D-brane and those which stretch between the two different D-branes. The former case is similar to what has been discussed so far, and the light spectrum contains a pair massless Abelian vectors $A_{\mu}^{1,2}$, each living on the world-volume of a D-brane. In the latter case, the mass operator is shifted by the distance $\delta$,
\begin{equation}
		M^2_{\text{open}} = \frac{1}{\alpha'} \left(N_X + N_\psi + \Delta\right) + \frac{\delta^2}{(2\pi\alpha')^2} \,,
\end{equation}
so that the stretched strings produce a pair of massive vectors $A_\mu^{\pm}$, charged with respect to $A_{\mu}^{1,2}$. In the limit where $\delta\to 0$ and the D-branes form a stack, the $A_{\mu}^{\pm}$ become massless  and it can be shown that the original gauge symmetry $\text{U}(1)^2$ is enhanced to $\text{U}(2)$ \cite{Bianchi:1991eu,Witten:1995im}. This exercise can be repeated in the case where $N$ D-branes are involved. When they form a single stack, the gauge symmetry is maximal, while the various separations induce the breaking $\text{U}(N)\to \text{U}(N_1)\times \text{U}(N_2)\times \ldots \times \text{U}(N_m)$, with $N_1+N_2+\ldots+N_m=N$. This way of describing non-Abelian gauge symmetries in open strings is equivalent to the introduction of Chan-Paton factors $\lambda_{ij}^a$ dressing string states. It has been shown \cite{Schwarz:1982md, Marcus:1982fr} that the proper factorisation of scattering amplitudes restricts the consistent gauge group factors that can appear in open string constructions to $\text{U}(N)$, $\text{SO}(N)$ and $\text{USp}(2N)$, with the last two requiring unoriented strings. Note that one may alternatively recover these gauge groups from the dynamics of additional degrees of freedom living at the two endpoints of an open string \cite{Marcus:1986cm}.


\section{Orientifolds}
\label{ST:Orientifold}

In Sections \ref{ST:Fermionic} and \ref{ST:Heterotic}, we saw that a useful way to construct new string vacua is to gauge discrete symmetries. This method was employed in building the type 0 theories from type II superstrings, or the non-supersymmetric heterotic $\text{SO}(16)\times \text{SO}(16)$ using the spacetime fermion number, as well as the level-two $\text{E}_8$ theory where the discrete symmetry involves the permutation of the two $\text{E}_8$ factors of the supersymmetric heterotic string. Even the GSO projection itself can be seen as an orbifold construction, where the full spectrum is modded out by the world-sheet fermion number. Following this paradigm, one may wonder if there are additional discrete symmetries whose gauging could lead to new constructions. 

To this end, recall that the type IIB theory employs the same GSO projection for left and right movers and thus, although being chiral in spacetime, is non-chiral on the world-sheet. We may then use world-sheet parity $\Omega: \sigma\to-\sigma$ to restrict its spectrum. This operation, known as the \emph{orientifold construction} \cite{Polchinski:1987tu, Sagnotti:1987tw, Pradisi:1988xd, Horava:1989ga, Horava:1989vt, Bianchi:1990yu, Bianchi:1990tb, Bianchi:1991eu} (for a detailed review with applications to lower-dimensional vacua see \cite{Angelantonj:2002ct} and \cite{Dudas:2000bn}), exchanges the left and right waves on a closed string via the natural action of mapping left-moving oscillators into their right-moving counterparts, \emph{i.e.} $\alpha_{n}^\mu \leftrightarrow \tilde\alpha_{n}^\mu$ and similarly for the fermion modes\footnote{Actually, the simplest instance of a closed string invariant under world-sheet parity is the bosonic string itself. However, the resulting theory is not rich enough to reveal the salient features of orientifold constructions and will not be discussed here. We refer the interested reader to the original literature \cite{Douglas:1986eu, Weinberg:1987ie, Marcus:1986cm, Bianchi:1988fr} and to \cite{Angelantonj:2002ct} for a review.}. Under this operation, the graviton and the dilaton, corresponding to the symmetric part of $b_{-1/2}^\mu\, \tilde b_{-1/2}^\nu |0\rangle_{\text{L}}\otimes |0\rangle_{\text{R}}$, are clearly invariant, while the $B$-field, corresponding to the anti-symmetric part, is odd and thus it is projected away. A similar projection is also at work in the R-R sector, while only one combination of the fermions in the NS-R and R-NS sectors survives. As usual, a natural way to encode the spectrum of the theory is to compute the partition function, which now involves a projector onto $\Omega$-invariant states, 
\begin{equation}
	Z = \text{Tr}\,\left[ \, \frac{1+\Omega}{2} \, P_{\text{GSO}}\, q^{L_0} \, \bar{q}^{\bar L_0} \, \right] \equiv \frac{T+K}{2}\,,
\end{equation}
where $P_{\text{GSO}}$ enforces the GSO projection of type IIB. The first term is nothing but the familiar torus partition function,
\begin{equation}
	T =  \frac{1}{\eta^8(\tau)\bar\eta^8(\bar\tau)}\, |V_8(\tau)-S_8(\tau)|^2\,.
\end{equation}
 whereas the second one involves the insertion of the $\Omega$ operator in the trace and reads
\begin{equation}
	K =  \frac{1}{\eta^8(2i\tau_2)}\, (V_8-S_8)(2i\tau_2)\,.
	\label{KleinBottle}
\end{equation}
Notice that $K$ only receives contributions from the NS-NS and R-R sectors. Taken together with $T$, it symmetrises the contribution of the NS-NS sector, thus eliminating the $B$-field from the massless spectrum, and it anti-symmetrises the contribution of the R-R sector, thus eliminating the 0-form and the self-dual 4-form. $\Omega$ exchanges the NS-R and R-NS sectors so that only their diagonal combination survives the projection, which is reflected in the $1/2$ factor multiplying $T$.
All in all, the massless spectrum enjoys $\mathscr{N}=(1,0)$ supersymmetry and comprises the fields in the supergravity multiplet: the graviton, the dilaton, the R-R 2-form potential, the left-handed gravitino and the right handed dilatino. This chiral spectrum is, however, anomalous and does not define a consistent theory in ten dimensions. An alternative way to see this inconsistency is to understand the topology of the Riemann surface associated to $K$.

\begin{figure}
	\begin{center}
		\includegraphics[width=0.8\textwidth]{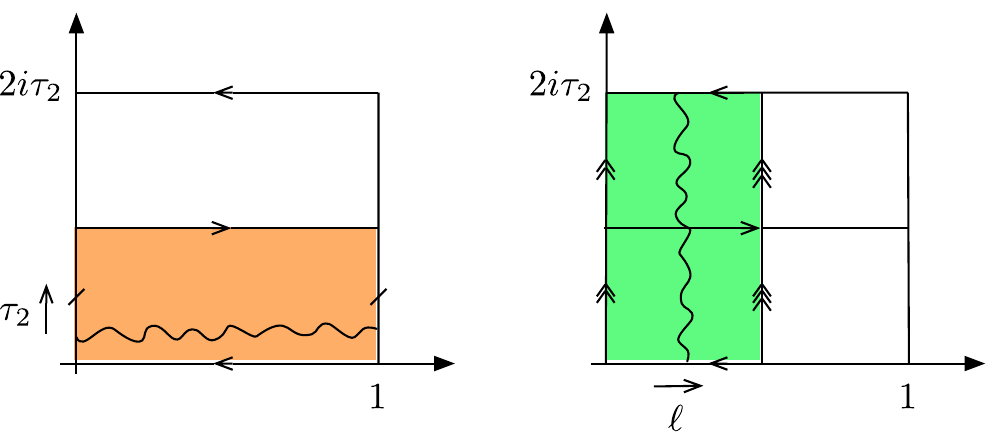}
	\end{center}
	\caption{The left figure shows the one-loop vacuum diagram of a closed string which propagates for a (vertical) proper time $\tau_2$ and flips its orientation. The right figure illustrates the tree-level (horizontal) propagation of a closed string bouncing between the two cross-caps. Both figures also display the double covering torus with modulus $2i\tau_2$.}
	\label{FigKlein}
\end{figure}

\begin{figure}
	\begin{center}
		\includegraphics[width=0.8\textwidth]{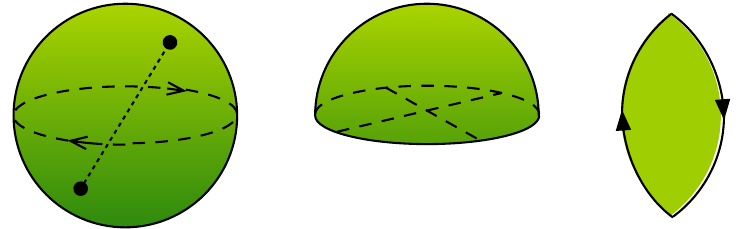}
	\end{center}
	\caption{The cross-cap is obtained by identifying antipodal points on the sphere. The sides of the fundamental domain are identified following the direction of the arrows, so that the surface has no boundaries and is unoriented. }
	\label{FigCrosscap}
\end{figure}

Indeed, the amplitude \eqref{KleinBottle} describes the loop propagation of closed strings with the exchange of left and right movers and, therefore, corresponds to a non-orientable closed Riemann surface of vanishing Euler characteristic, known as the Klein bottle. This can be built from the double-covering torus via the anti-holomorphic involution $z\to 1-\bar{z}+i\tau_2$, which is compatible with the equivalence relations $z\sim z+n\tau+m$ only if the torus is rectangular with $\tau= 2i\tau_2$.
As a result, modular invariance is lost and the Klein bottle admits two alternative descriptions, as depicted in Fig. \ref{FigKlein}: if the proper time $\tau_2$ is taken to flow vertically, it describes the one-loop propagation of closed strings with an exchange of their left and right waves and the integrated amplitude reads
\begin{equation}
	\mathscr{K} = \int_0^\infty \frac{d\tau_2}{\tau_2^6}\, \frac{1}{\eta^8(2i\tau_2)}\, (V_8-S_8)(2i\tau_2)\,.
	\label{OrientKlein}
\end{equation}
 If the proper time $\ell$ is instead taken to flow horizontally, which is obtained by the $S$-modular transformation $\ell = 1/2\tau_2$, the integrated amplitude
\begin{equation}
	\tilde{\mathscr{K}} = 2^5\, \int_0^\infty d\ell\, \frac{1}{\eta^8(i\ell)}\, (V_8-S_8)(i\ell) 
\end{equation} 
describes the tree-level propagation of the NS-NS and R-R states between two cross-caps\footnote{A cross-cap is the simplest unoriented Riemann surface with Euler characteristic $\chi=1$ and can be built from the double-covering sphere by identifying antipodal points, as shown in Fig. \ref{FigCrosscap}. The resulting surface has no boundaries and is unoriented.} \cite{Polchinski:1987tu, Sagnotti:1987tw, Bianchi:1988fr, Horava:1989vt}, also known as \emph{orientifold planes} (O-planes, for short), see Fig. \ref{FigKlein}. In the limit of an infinitely long tube, $\ell \to \infty$, only massless excitations survive and the resulting diagram involves 1-point functions coupling the massless fields to the cross-caps, and an on-shell propagator evaluated at zero momentum. This shows that also orientifold planes carry tension as well as charge for the R-R potentials. Since the $\Omega$ involution preserves the ten-dimensional Lorentz symmetry, the O-planes invade the whole space-time and, therefore, can only couple to the R-R 10-form potential $C_{10}$. 
 This field is (trivially) non-dynamical and its equation of motion implies that the net charge of its sources must vanish, which is not the case if only orientifold planes are present. This problem can be solved by the introduction of D9 branes, which are also charged with respect to $C_{10}$ and may yield a neutral configuration.

Since the closed strings considered in this section are unoriented, so should be the open strings living on the D9 branes. For unoriented open strings, the two end-points are equivalent and, hence, the worldsheet parity exchanges them and acts as $\Omega: \sigma\to 1-\sigma$. This action translates to
\begin{equation}
	\alpha_n \to (-1)^n \, \alpha_n \,,  \qquad d_n \to (-1)^n \, d_n \,,\qquad b_r \to (-1)^{r-1/2}\, b_r \,.
	\label{OmegaOsc}
\end{equation}
As a result, the spectrum of unoriented open strings is encoded into
\begin{equation}
	\text{Tr}\,\left[ \, \frac{1+\Omega}{2} \, P_{\text{GSO}} \, q^{L_0} \, \right] \,.
\end{equation}
The trace involving the identity corresponds to the annulus amplitude computed in Section \ref{ST:Dbranes} and reads
\begin{equation}
	\mathscr{A} = N^2 \,\int_0^\infty \frac{d\tau_2}{\tau_2^6} \, \frac{1}{\eta^8(\frac{i\tau_2}{2})} \, (V_8 - S_8)(\tfrac{i\tau_2}{2}) \,,
	\label{OrientAnnulus}
\end{equation}
where we have introduced a stack of $N$ D9 branes. The multiplicity factor $N^2$ is due to the fact that open strings can start and end on any of the D-branes in the stack and reflects the possibility of introducing Chan-Paton charges at the string endpoints \cite{Paton:1969je, Marcus:1982fr, Marcus:1986cm}.
The trace involving the world-sheet parity $\Omega$ reads
\begin{equation}
	\mathscr{M} = \epsilon\, N \,\int_0^\infty \frac{d\tau_2}{\tau_2^6} \, \frac{1}{\hat\eta^8(\frac{1}{2}+\frac{i\tau_2}{2})} \, (\hat V_8 - \hat S_8)(\tfrac{1}{2}+\tfrac{i\tau_2}{2}) \,.
\end{equation}
Here, the $1/2$ shift in the argument reflects the action \eqref{OmegaOsc} of $\Omega$ on the oscillators, the hatted characters defined as in \eqref{HattedCharacters} remove the fictitious overall phase, while the sign $\epsilon$ is ascribed to the
parity ambiguity of the vacuum and will be determined shortly. In this case, the multiplicity scales like $N$ since the open strings must start and end on the same D-brane so that the system respects the world-sheet parity that exchanges the two endpoints. 

\begin{figure}
	\begin{center}
		\includegraphics[width=0.8\textwidth]{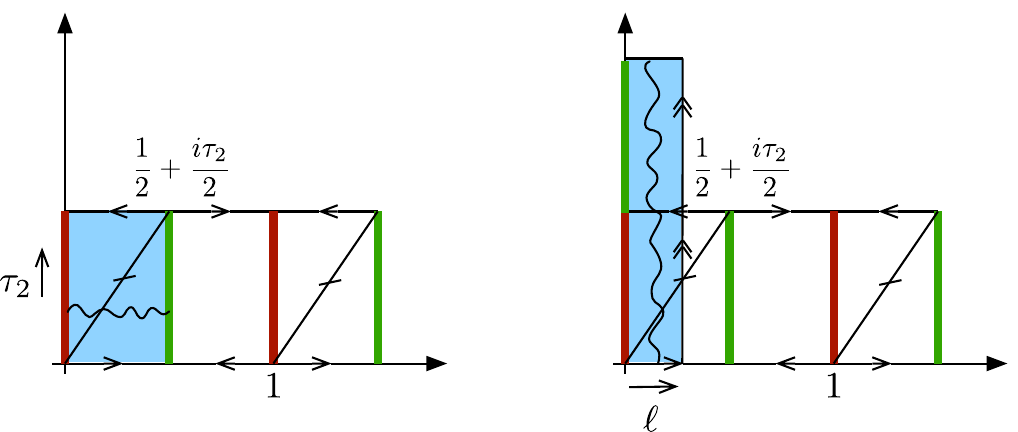}
	\end{center}
	\caption{The left figure shows the one-loop vacuum diagram of an open string which propagates for a (vertical) proper time $\tau_2$ and flips its orientation; the end-points trace the single boundary of the M\"obius strip composed by the red and green vertical lines. The right figure shows the tree-level (horizontal) propagation of a closed string bouncing between a boundary and a cross-cap. Both figures also display the double covering torus with modulus $\frac{1}{2}+\frac{i\tau_2}{2}$.}
	\label{FigMoebius}
\end{figure}

Also this amplitude has a geometrical interpretation in terms of a Riemann surface spanned by an open string which exchanges its endpoints. This unoriented Riemann surface has a boundary and it is known as the M\"obius strip, see Fig. \ref{FigMoebius}. As in the previous cases, it can be built via an anti-holomorphic involution of a double-covering torus whose modulus now has a fixed real part $\tau=\frac{1}{2}+\frac{i\tau_2}{2}$. Also here, there is no notion of modular invariance and, in fact, also the M\"obius strip admits two alternative descriptions depending on whether the proper time is taken to flow vertically or horizontally. In the former case, it describes the one-loop propagation of open strings which exchange their end-points while, in the latter, closed strings freely propagate between a boundary and a cross-cap \cite{Polchinski:1987tu, Sagnotti:1987tw, Bianchi:1988fr, Horava:1989vt}, as shown in Fig. \ref{FigMoebius}. In the limit of an infinitely long tube, only massless states propagate and, in the NS-NS sector, and the resulting diagram involves the product of the tensions D9 branes and O9 planes, times the on-shell propagator at zero momentum. Similarly, in the R-R sector only the non-dynamical 10-form potential probes both the D9 branes and the O9 planes and, therefore, the amplitude is proportional to the product of the two charges. The sign ambiguity $\epsilon$ reflects the possibility that the tensions and charges of D-branes and O-planes have the same or opposite sign. 

As usual, an $S$-modular transformation brings the direct (loop) channel representation \eqref{OrientAnnulus} into the transverse (tree-level) representation
  \begin{equation}
 	\tilde{\mathscr{A}} = 2^{-5}\,N^2 \,\int_0^\infty d\ell \, \frac{1}{\eta^8(i\ell)} \, (V_8 - S_8)(i\ell) \,.
 \end{equation}
In the case of the M\"obius amplitude, this map is realised by 
 the transformation 
 \begin{equation}
 	P: \quad \tfrac{1}{2}+\tfrac{i\tau_2}{2} \to \tfrac{1}{2} + \tfrac{i}{2\tau_2} \,,
 \end{equation}
with\footnote{Actually, on the hatted characters this transformation acts as $P=T^{1/2}ST^2 ST^{1/2}$.} $P = TST^2 S$ \cite{Bianchi:1990yu}, and $S,T$ being the standard generators of $\text{SL}(2;\mathbb{Z})$.
 The resulting transverse channel M\"obius amplitude, thus, reads
\begin{equation}
	\tilde{\mathscr{M}} = 2\epsilon\, N \,\int_0^\infty d\ell \, \frac{1}{\hat\eta^8(\frac{1}{2}+i\ell)} \, (\hat V_8 - \hat S_8)(\tfrac{1}{2}+i\ell) \,.
\end{equation}

\begin{figure}
	\begin{center}
		\includegraphics[width=0.8\textwidth]{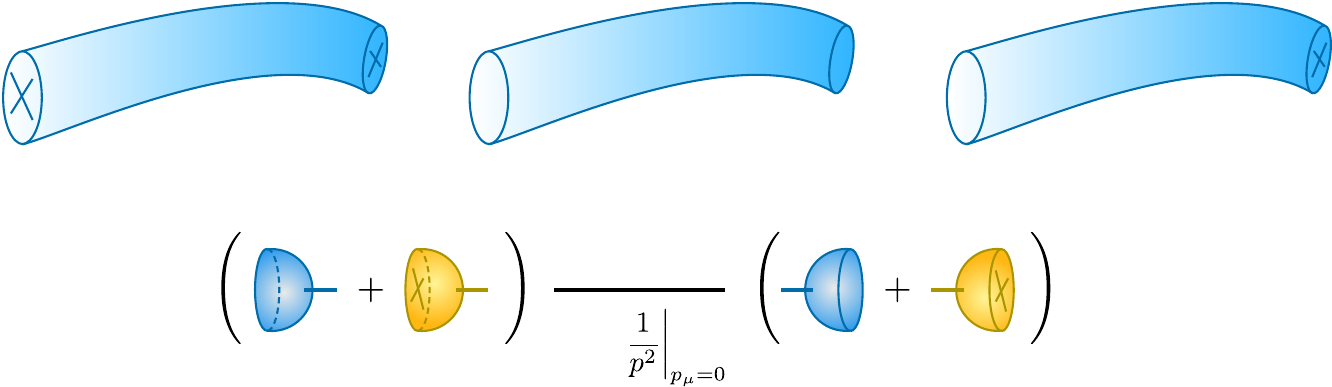}
	\end{center}
	\caption{In the $\ell \to \infty$ limit, the transverse-channel Klein bottle, annulus and M\"obius strip amplitudes probe the tension and the charge of O-planes and D-branes. The contributions $\tilde{\mathscr{K}}+\tilde{\mathscr{A}}+\tilde{\mathscr{M}}$ factorise into the square of the tadpoles times the zero-momentum propagator of the corresponding massless state.}
	\label{FigTadpole}
\end{figure}

In the $\ell \to \infty$ limit, aside from a multiplicative overall divergence ascribed to the massless propagator at zero momentum,  $\tilde{\mathscr{K}}+\tilde{\mathscr{A}}+\tilde{\mathscr{M}}$ yield the perfect square
\begin{equation}
	2^{5}+2^{-5}N^2 +2\epsilon N  = 2^{-5} \left( N+ 2^5\epsilon \right)^2 \,.
\end{equation}
The term inside the parenthesis on the r.h.s. is nothing but the overall tension of D-branes and O-planes in the NS-NS sector, while, it equals the overall 10-form charge in the R-R sector. The consistency of the $C_{10}$ equations of motion requires the \emph{tadpole cancellation condition}
\begin{equation}
	N+ 2^5\epsilon  = 0\,,
	\label{TadopeCancellation}
\end{equation}
which guarantees a neutral configuration in accordance with Gauss' law. The unique solution is $\epsilon=-1$ and $N=32$. In this supersymmetric setup, the cancellation of R-R tadpoles also guarantees the cancellation of NS-NS ones, since they are related by supersymmetry transformations. Assuming that the physical D-branes have positive tension and charge, this implies that the orientifold planes involved in this construction have negative tension and charge and are known in the literature as $\text{O}9_{-}$ planes. 
 
 Returning to the direct channel annulus and M\"obius strip amplitudes, the tadpole condition uniquely fixes the spectrum which, at the massless level, comprises an $\mathscr{N}=(1,0)$ vector multiplet in the adjoint representation of SO(32). Together with the (unoriented) closed string spectrum this yields the field content of type I superstrings, which is free of irreducible gravitational and gauge anomalies, as guaranteed by the cancellation of R-R tadpoles \cite{Aldazabal:1999nu, Scrucca:1999uz, Morales:1998ux, Bianchi:2000de}. Although the type I superstring shares the same massless spectrum as the SO(32) heterotic string, the elementary degrees of freedom are obviously completely different. Because the various couplings in the low energy effective actions emerge from different Riemann surfaces, the map between the two is non-perturbative since it inverts the string coupling constants, $g_{s,\text{het}} = 1/g_{s,\text{I}}$ \cite{Hull:1994ys, Townsend:1995kk, Witten:1995ex, Polchinski:1995df}.
 
 Actually, this is not the only vacuum that one may construct as an orientifold of the type IIB theory. In fact, although R-R tadpoles must always be cancelled for the theory to be unitary, NS-NS ones are not associated to any conservation law and, therefore, can be relaxed at the cost of modifying the background \cite{Fischler:1986ci, Fischler:1986tb}. This new vacuum has the same Klein bottle \eqref{OrientKlein} and annulus \eqref{OrientAnnulus} amplitudes, but the M\"obius amplitudes now read 
 \begin{equation}
 	\mathscr{M} =  N \,\int_0^\infty \frac{d\tau_2}{\tau_2^6} \, \frac{1}{\hat\eta^8(\frac{1}{2}+\frac{i\tau_2}{2})} \, (\hat V_8 + \hat S_8)(\tfrac{1}{2}+\tfrac{i\tau_2}{2}) \,,
 \end{equation} 
 and
 \begin{equation}
	\tilde{\mathscr{M}} = 2 N \,\int_0^\infty d\ell \, \frac{1}{\hat\eta^8(\frac{1}{2}+i\ell)} \, (\hat V_8 + \hat S_8)(\tfrac{1}{2}+i\ell) \,.
\end{equation}
From the latter, we can extract that the tensions of O-planes and D-branes have the same sign, while their R-R charges have opposite signs. Indeed, this vacuum involves what is known as $\text{O}9_{+}$ planes (with positive tension and R-R charge) and anti-D9 branes (with positive tension and negative R-R charge). The R-R tadpole cancellation conditions still select $N=32$, while the NS-NS tadpoles are not cancelled and induce the potential 
\begin{equation}
	\int d^{10}x\, \sqrt{-G}\,V(\Phi) \sim (N+32)\int d^{10}x\, \sqrt{-G}\,e^{-\Phi} \,,
	\label{DilatonPotential}
\end{equation}
 in the string frame. The dependence on the dilaton originates from the coupling to the disk and the cross-caps, which both have Euler characteristic $\chi=1$. This vacuum is known as the Sugimoto model \cite{Sugimoto:1999tx}. It is characterised by a supersymmetric closed string spectrum containing the massless $\mathscr{N}=(1,0)$ supergravity multiplet, while supersymmetry is explicitly broken in the open string sector, whose massless excitations now include a vector in the adjoint representation of $\text{USp}(32)$ and a left-handed fermion in the $\textbf{496}$ anti-symmetric representation. Also in this case the cancellation of the R-R tadpoles fix the gauge group and the representation of the fermions and ensures the cancellation of irreducible anomalies \cite{Aldazabal:1999nu, Scrucca:1999uz, Morales:1998ux, Bianchi:2000de}. Notice that, in this case, the anti-symmetric representation $\textbf{496}$ is reducible and decomposes as $\textbf{496} = \textbf{495}+\textbf{1}$. The singlet fermion plays the role of the Goldstino and, together with the dilaton potential \eqref{DilatonPotential}, implies that in the open string sector supersymmetry is still present, albeit non-linearly realised \cite{Dudas:2000nv, Pradisi:2001yv}. 
 
 This vacuum is free of tachyonic instabilities. However, the emergence of the dilaton potential, and thus a non-trivial interaction between $\text{O}_{+}$-planes and anti-branes, implies a back-reaction on the Minkowski vacuum breaking the SO(1,9) down to SO(1,8) \cite{Dudas:2000ff}. 
 
 The type IIB superstring is not the only closed string theory which is left-right symmetric. In fact, although heterotic strings are asymmetric by construction and type IIA involves different GSO projections for the left and right movers, the type 0 theories are $\Omega$-invariant and admit an orientifold projection. In the following, we shall focus on type 0B orientifolds, and refer the interested reader to the original work \cite{Bianchi:1990yu} and to \cite{Angelantonj:2002ct} for a discussion of the 0A case. 

The type 0B theory \eqref{10dSuperstrings} involves the two R-R sectors $|S_8|^2$ and $|C_8|^2$ and, therefore, two different (non-dynamical) 10-form potentials are present. As a result, the O9 planes and D9 branes of the 0B theory are charged with respect to both of them, in addition to carrying a non-trivial tension. This allows for a richer pattern of O-planes and D-branes with vanishing R-R charges. 

The standard orientifold projection of $\mathscr{T}_{\text{0B}}$ starts with the Klein bottle amplitude \cite{Bianchi:1990yu}
\begin{equation}
	\mathscr{K} = O_8+V_8-S_8-C_8 \,, 
\end{equation}
where, to lighten the notation, we henceforth do not explicitly display the integral nor the contribution of the world-sheet bosons, since they are unambiguous and may easily be reconstructed.
We will, however, take them into account when deriving the transverse channel amplitudes. The projected closed string spectrum in $(\mathscr{T}_{\text{0B}}+\mathscr{K})/2$ comprises the metric, the dilaton, two R-R 2-forms, as well as a tachyon, which clearly makes the system unstable. This orientifold is consistent and does not require the introduction of D-branes. In fact, the transverse channel Klein bottle reads
\begin{equation}
	\tilde{\mathscr{K}} = 2^6 \, V_8 \,,
	\label{tildeKlein0B1}
\end{equation}
and only NS-NS states propagate in the tube between the two cross-caps. As a result, the configuration of O-planes present in this construction has an overall tension but vanishing R-R charges. Hence, Gauss' law for both $C_{10}$'s is no longer violated, and the construction is self-consistent, although the dilaton tadpole in \eqref{tildeKlein0B1} induces a potential of the form \eqref{DilatonPotential}, as in the Sugimoto model. Open strings can be added at the cost of introducing additional tachyons. In fact, the configuration of the D-branes involved must be neutral with respect to the two $C_{10}$ forms present in the 0B spectrum, and this can only happen if brane-anti-brane pairs are present, which inevitably contain a tachyon stretching between a brane and an anti-brane. This open string sector was built in \cite{Bianchi:1990yu}, and we refer the reader to the original work for further details and to \cite{Angelantonj:2002ct} for a description in terms of D-branes.

Actually, for type 0B orientifolds the action of $\Omega$ on the closed string sector is not unique and, in fact, two more choices are possible \cite{Sagnotti:1995ga, Sagnotti:1996qj}. This is an instance of a more general framework whereby different choices of Klein bottle projections are allowed whenever simple currents are present in the CFT \cite{Fioravanti:1993hf, Pradisi:1995qy, Pradisi:1995pp, Pradisi:1996yd, Huiszoon:1999xq, Huiszoon:2000ge, Fuchs:2000cm}. In the first case, $\Omega$ symmetrises states both in the NS-NS and R-R sectors and is associated to the Klein bottle amplitude
\begin{equation}
	\mathscr{K}' = O_8+V_8+S_8+C_8 \,.
\end{equation}
The unoriented closed-string spectrum now comprises a tachyon, the graviton, the dilaton, two R-R scalars and a 4-form potential. Also in this case, the vacuum is consistent and does not require the introduction of open strings, since 
\begin{equation}
	\tilde{\mathscr{K}}' = 2^6 \, O_8 \,,
	\label{tildeKlein0B2}
\end{equation}
indicates once more that the configuration of O-planes has vanishing R-R charges, which guarantees compatibility with Gauss' law for the $C_{10}$ potentials. Nevertheless, as in the previous case, open strings can be added by introducing pairs of branes and anti-branes, which induce further tachyonic instabilities, as discussed in \cite{Sagnotti:1995ga, Sagnotti:1996qj}. 

In the second case, the Klein bottle instead reads\footnote{Clearly, there is an equivalent option, whereby the $S_8$ and $C_8$ characters are interchanged.}
\begin{equation}
	\mathscr{K}'' = -O_8+V_8+S_8-C_8 \,,
\end{equation}
and has the virtue of projecting away the tachyon, since it is now odd under $\Omega$. Therefore, the closed string sector is non-tachyonic and its massless spectrum includes the graviton and the dilaton from the NS-NS sector, as well as a scalar, a 2-form and a self-dual 4-form from the R-R sectors. This chiral spectrum is anomalous as reflected in the non-trivial R-R tadpole in
\begin{equation}
	\tilde{\mathscr{K}}'' = - 2^6 \, C_8 \,.
	\label{tildeKlein0B3}
\end{equation}
To cancel it, open strings must now be added and involve two types of D9 branes with charges $(+,+)$ and $(-,+)$ with respect to the two 10-forms originating from $|S_8|^2$ and $|C_8|^2$, respectively. This is reflected in the coefficients of the $S_8$ and $C_8$ characters in the transverse channel annulus amplitude 
\begin{equation}
	\tilde{\mathscr{A}}'' = 2^{-6}\, \left[ (N+\bar N)^2 \,(V_8 - C_8) - (N-\bar N)^2\, (O_8 - S_8) \right] \,.
\end{equation}
Together with
\begin{equation}
	\tilde{\mathscr{M}}'' = 2 (N+\bar N)\, \hat{C}_8 \,,
\end{equation}
$\tilde{\mathscr{K}}''$ and $\tilde{\mathscr{A}}''$  yield the tadpole conditions
\begin{equation}
	N+\bar{N} = 64 \,,\qquad N-\bar{N} = 0\,.
\end{equation}
The direct channel annulus and M\"obius strip amplitudes read
\begin{equation}
	\mathscr{A}'' = 2N\bar N\, V_8 - (N^2+\bar{N}^2)\,C_8 \,,
\end{equation}
and
\begin{equation}
	\mathscr{M}'' = (N+\bar{N})\,\hat{C}_8 \,.
\end{equation}
The open string spectrum is also non-tachyonic and contains gauge bosons in the adjoint representation of $\text{U}(32)$ and right-handed fermions in the anti-symmetric $\textbf{496} \oplus \overline{\textbf{496}}$ representations, precisely as needed to cancel the contribution of the self-dual 4-form to the gravitational anomaly. Being in the anti-symmetric representations of $\text{U}(32)$, the theory is also free from gauge anomalies, although R-R forms of different degree are at work to cancel the non-Abelian and Abelian parts, thus generalising the Green-Schwarz mechanisms which is normally at work in ten \cite{Green:1984sg} and six  \cite{Sagnotti:1992qw} dimensions. This is the celebrated type $0'\text{B}$ vacuum of Sagnotti \cite{Sagnotti:1995ga, Sagnotti:1996qj}. Together with the $\text{SO}(16)\times\text{SO}(16)$ heterotic string and the Sugimoto model, they are the only non-supersymmetric, non-tachyonic vacua in ten dimensions.

Orientifold vacua in lower dimensions exhibit a much richer structure involving O-planes and D-branes of different dimensionality and various patterns of (partial) supersymmetry breaking. However, they will not be considered here and we refer the reader to 
\cite{Angelantonj:2002ct} and \cite{Dudas:2000bn} for reviews.


\section*{Acknowledgments}
We are deeply indebted to our mentors Massimo Bianchi, Costas Kounnas and Augusto Sagnotti, as well as our friends and collaborators for the many insightful discussions that, over the years, have shaped our understanding and view of string theory.

\bibliographystyle{utphys}

\begin{thebibliography}{100}

\bibitem{Veneziano:1968yb}
G.~Veneziano, ``{Construction of a crossing-symmetric, Regge behaved amplitude
  for linearly rising trajectories},''
  \href{http://dx.doi.org/10.1007/BF02824451}{{\em Nuovo Cim. A} {\bfseries 57}
  (1968) 190--197}.

\bibitem{Nambu:1970xx}
Y.~Nambu, ``{Quark model and the factorization of the Veneziano amplitude},''
  in {\em {Symmetries and Quark Models, ed. R. Chand (Gordon and Breach)}}.
\newblock 1970.

\bibitem{Nielsen:1970xx}
H.~B. Nielsen, ``{An almost physical interpretation of the integrand of the
  $n$-point Veneziano model},'' in {\em {Submitted to the 15th International
  Conference on High Energy Physics (Kiev)}}.
\newblock 1970.

\bibitem{Susskind:1970xm}
L.~Susskind, ``{Dual symmetric theory of hadrons. 1.},''
  \href{http://dx.doi.org/10.1007/BF02726485}{{\em Nuovo Cim. A} {\bfseries 69}
  (1970) 457--496}.

\bibitem{Susskind:1970qz}
L.~Susskind, ``{Structure of hadrons implied by duality},''
  \href{http://dx.doi.org/10.1103/PhysRevD.1.1182}{{\em Phys. Rev. D}
  {\bfseries 1} (1970) 1182--1186}.

\bibitem{Yoneya:1973ca}
T.~Yoneya, ``{Quantum gravity and the zero slope limit of the generalized
  Virasoro model},'' \href{http://dx.doi.org/10.1007/BF02727806}{{\em Lett.
  Nuovo Cim.} {\bfseries 8} (1973) 951--955}.

\bibitem{Yoneya:1974jg}
T.~Yoneya, ``{Connection of Dual Models to Electrodynamics and
  Gravidynamics},'' \href{http://dx.doi.org/10.1143/PTP.51.1907}{{\em Prog.
  Theor. Phys.} {\bfseries 51} (1974) 1907--1920}.

\bibitem{Scherk:1974ca}
J.~Scherk and J.~H. Schwarz, ``{Dual Models for Nonhadrons},''
  \href{http://dx.doi.org/10.1016/0550-3213(74)90010-8}{{\em Nucl. Phys. B}
  {\bfseries 81} (1974) 118--144}.

\bibitem{Scherk:1975xx}
J.~Scherk and J.~H. Schwarz, ``{Dual Model Approach To A Renormalizable Theory
  Of Gravitation},'' in {\em {Honorable mention in the 1975 essay competition
  of the Gravity Research Foundation}}.
\newblock 1975.

\bibitem{Green:1987sp}
M.~B. Green, J.~H. Schwarz, and E.~Witten, {\em {Superstring Theory. Vol. 1:
  Introduction}}.
\newblock Cambridge Monographs on Mathematical Physics. 7, 1988.

\bibitem{Green:1987mn}
M.~B. Green, J.~H. Schwarz, and E.~Witten, {\em {Superstring Theory. Vol. 2:
  Loop Amplitudes, Anomalies And Phenomenology}}.
\newblock 7, 1988.

\bibitem{Lust:1989tj}
D.~L\"ust and S.~Theisen, \href{http://dx.doi.org/10.1007/BFb0113507}{{\em
  {Lectures on string theory}}}, vol.~346 of {\em Lecture Notes in Physics}.
\newblock Springer, 1989.

\bibitem{Polchinski:1998rq}
J.~Polchinski, \href{http://dx.doi.org/10.1017/CBO9780511816079}{{\em {String
  theory. Vol. 1: An introduction to the bosonic string}}}.
\newblock Cambridge Monographs on Mathematical Physics. Cambridge University
  Press, 12, 2007.

\bibitem{Polchinski:1998rr}
J.~Polchinski, \href{http://dx.doi.org/10.1017/CBO9780511618123}{{\em {String
  theory. Vol. 2: Superstring theory and beyond}}}.
\newblock Cambridge Monographs on Mathematical Physics. Cambridge University
  Press, 12, 2007.

\bibitem{Johnson:2003glb}
C.~V. Johnson, \href{http://dx.doi.org/10.1017/CBO9780511606540}{{\em
  {D-branes}}}.
\newblock Cambridge Monographs on Mathematical Physics. Cambridge University
  Press, 2005.

\bibitem{Zwiebach:2004tj}
B.~Zwiebach, {\em {A first course in string theory}}.
\newblock Cambridge University Press, 7, 2006.

\bibitem{Ortin:2015hya}
T.~Ortin, \href{http://dx.doi.org/10.1017/CBO9781139019750}{{\em {Gravity and
  Strings}}}.
\newblock Cambridge Monographs on Mathematical Physics. Cambridge University
  Press, 2nd ed.~ed., 7, 2015.

\bibitem{Becker:2006dvp}
K.~Becker, M.~Becker, and J.~H. Schwarz,
  \href{http://dx.doi.org/10.1017/CBO9780511816086}{{\em {String theory and
  M-theory: A modern introduction}}}.
\newblock Cambridge University Press, 12, 2006.

\bibitem{Kiritsis:2019npv}
E.~Kiritsis, {\em {String Theory in a Nutshell: Second Edition}}.
\newblock Princeton University Press, USA, 4, 2019.

\bibitem{Gubser:2010zz}
S.~S. Gubser, {\em {The Little Book of String Theory}}.
\newblock Princeton University Press, 2, 2010.

\bibitem{Dine:2007zp}
M.~Dine, \href{http://dx.doi.org/10.1017/9781009290883}{{\em {Supersymmetry and
  String Theory : Beyond the Standard Model}}}.
\newblock Cambridge University Press, 2022.

\bibitem{Ibanez:2012zz}
L.~E. Ib{\'a}{\~n}ez and A.~M. Uranga, {\em {String theory and particle
  physics: An introduction to string phenomenology}}.
\newblock Cambridge University Press, 2, 2012.

\bibitem{West:2012vka}
P.~West, {\em {Introduction to strings and branes}}.
\newblock Cambridge University Press, 7, 2012.

\bibitem{Blumenhagen:2013fgp}
R.~Blumenhagen, D.~L\"ust, and S.~Theisen,
  \href{http://dx.doi.org/10.1007/978-3-642-29497-6}{{\em {Basic concepts of
  string theory}}}.
\newblock Theoretical and Mathematical Physics. Springer, Heidelberg, Germany,
  2013.

\bibitem{Schomerus:2017ycz}
V.~Schomerus, \href{http://dx.doi.org/10.1017/9781316672631}{{\em {A primer on
  string theory}}}.
\newblock Cambridge University Press, Cambridge, 2017.

\bibitem{Szabo:2020ipd}
R.~Szabo, \href{http://dx.doi.org/10.1142/q0006}{{\em {An Introduction to
  String Theory and D-Brane Dynamics}}}.
\newblock Advanced Textbooks in Physics. WSP, 2, 2020.

\bibitem{Mohaupt:2022uqx}
T.~Mohaupt, \href{http://dx.doi.org/10.1017/9781108611619}{{\em {A Short
  Introduction to String Theory}}}.
\newblock Cambridge University Press, 3, 2022.

\bibitem{Schwarz:1982jn}
J.~H. Schwarz, ``{Superstring Theory},''
  \href{http://dx.doi.org/10.1016/0370-1573(82)90087-4}{{\em Phys. Rept.}
  {\bfseries 89} (1982) 223--322}.

\bibitem{Green:1986wn}
M.~B. Green, ``{Introduction To String And Superstring Theory. 1},'' in {\em
  {Theoretical Advanced Study Institute in Particle Physics - TASI 86}}, H.~E.
  Haber, ed., pp.~139--275.
\newblock 1986.

\bibitem{Peskin:1987rz}
M.~E. Peskin, ``{Introduction To String And Superstring Theory. 2.},'' in {\em
  {Theoretical Advanced Study Institute in Particle Physics - TASI 86}}.
\newblock 3, 1987.

\bibitem{DHoker:1988pdl}
E.~D'Hoker and D.~H. Phong, ``{The Geometry of String Perturbation Theory},''
  \href{http://dx.doi.org/10.1103/RevModPhys.60.917}{{\em Rev. Mod. Phys.}
  {\bfseries 60} (1988) 917}.

\bibitem{Polchinski:1994mb}
J.~Polchinski, ``{What is string theory?},'' in {\em {NATO Advanced Study
  Institute: Les Houches Summer School, Session 62: Fluctuating Geometries in
  Statistical Mechanics and Field Theory}}.
\newblock 11, 1994.
\newblock \href{http://arxiv.org/abs/hep-th/9411028}{{\ttfamily
  arXiv:hep-th/9411028}}.

\bibitem{Polchinski:1996fm}
J.~Polchinski, S.~Chaudhuri, and C.~V. Johnson, ``{Notes on D-branes},''
  \href{http://arxiv.org/abs/hep-th/9602052}{{\ttfamily arXiv:hep-th/9602052}}.

\bibitem{Polchinski:1996na}
J.~Polchinski, ``{Tasi lectures on D-branes},'' in {\em {Theoretical Advanced
  Study Institute in Elementary Particle Physics (TASI 96): Fields, Strings,
  and Duality}}, pp.~293--356.
\newblock 11, 1996.
\newblock \href{http://arxiv.org/abs/hep-th/9611050}{{\ttfamily
  arXiv:hep-th/9611050}}.

\bibitem{Bachas:1996sc}
C.~Bachas, \href{http://dx.doi.org/10.1142/9781848160927_0001}{``{(Half) a
  lecture on D-branes},''} in {\em {Workshop on Gauge Theories, Applied
  Supersymmetry and Quantum Gravity}}, pp.~3--22.
\newblock 7, 1996.
\newblock \href{http://arxiv.org/abs/hep-th/9701019}{{\ttfamily
  arXiv:hep-th/9701019}}.

\bibitem{Ooguri:1996ik}
H.~Ooguri and Z.~Yin, ``{TASI lectures on perturbative string theories},'' in
  {\em {Theoretical Advanced Study Institute in Elementary Particle Physics
  (TASI 96): Fields, Strings, and Duality}}, pp.~5--81.
\newblock 12, 1996.
\newblock \href{http://arxiv.org/abs/hep-th/9612254}{{\ttfamily
  arXiv:hep-th/9612254}}.

\bibitem{Vafa:1997pm}
C.~Vafa, ``{Lectures on strings and dualities},'' in {\em {ICTP Summer School
  in High-energy Physics and Cosmology}}, pp.~66--119.
\newblock 2, 1997.
\newblock \href{http://arxiv.org/abs/hep-th/9702201}{{\ttfamily
  arXiv:hep-th/9702201}}.

\bibitem{Dijkgraaf:1997ip}
R.~Dijkgraaf, ``{Les Houches lectures on fields, strings and duality},'' in
  {\em {NATO Advanced Study Institute: Les Houches Summer School on Theoretical
  Physics, Session 64: Quantum Symmetries}}, pp.~3--147.
\newblock 3, 1997.
\newblock \href{http://arxiv.org/abs/hep-th/9703136}{{\ttfamily
  arXiv:hep-th/9703136}}.

\bibitem{Kiritsis:1997gu}
E.~Kiritsis, ``{Introduction to nonperturbative string theory},''
  \href{http://dx.doi.org/10.1063/1.54695}{{\em AIP Conf. Proc.} {\bfseries
  419} no.~1, (1998) 265--308},
  \href{http://arxiv.org/abs/hep-th/9708130}{{\ttfamily arXiv:hep-th/9708130}}.

\bibitem{Sen:1998kr}
A.~Sen, ``{An Introduction to nonperturbative string theory},'' in {\em {A
  Newton Institute Euroconference on Duality and Supersymmetric Theories}},
  pp.~297--413.
\newblock 2, 1998.
\newblock \href{http://arxiv.org/abs/hep-th/9802051}{{\ttfamily
  arXiv:hep-th/9802051}}.

\bibitem{Bachas:1998rg}
C.~P. Bachas, ``{Lectures on D-branes},'' in {\em {A Newton Institute
  Euroconference on Duality and Supersymmetric Theories}}, pp.~414--473.
\newblock 6, 1998.
\newblock \href{http://arxiv.org/abs/hep-th/9806199}{{\ttfamily
  arXiv:hep-th/9806199}}.

\bibitem{Schwarz:2000ew}
J.~H. Schwarz, ``{Introduction to superstring theory},''
  \href{http://dx.doi.org/10.1007/978-94-010-0522-7_4}{{\em NATO Sci. Ser. C}
  {\bfseries 566} (2001) 143--187},
  \href{http://arxiv.org/abs/hep-ex/0008017}{{\ttfamily arXiv:hep-ex/0008017}}.

\bibitem{Forste:2001ah}
S.~Forste, ``{Strings, branes and extra dimensions},''
  \href{http://dx.doi.org/10.1002/1521-3978(200203)50:3/4<221::AID-PROP221>3.0.CO;2-P}{{\em
  Fortsch. Phys.} {\bfseries 50} (2002) 221--403},
  \href{http://arxiv.org/abs/hep-th/0110055}{{\ttfamily arXiv:hep-th/0110055}}.

\bibitem{Angelantonj:2002ct}
C.~Angelantonj and A.~Sagnotti, ``{Open strings},''
  \href{http://dx.doi.org/10.1016/S0370-1573(02)00273-9}{{\em Phys. Rept.}
  {\bfseries 371} (2002) 1--150},
  \href{http://arxiv.org/abs/hep-th/0204089}{{\ttfamily arXiv:hep-th/0204089}}.
  [Erratum: Phys.Rept. 376, 407 (2003)].

\bibitem{Wadia:2008yn}
S.~R. Wadia, ``{String Theory: A Framework for Quantum Gravity and Various
  Applications},'' \href{http://arxiv.org/abs/0809.1036}{{\ttfamily
  arXiv:0809.1036 [gr-qc]}}.

\bibitem{Brink:1976sc}
L.~Brink, P.~Di~Vecchia, and P.~S. Howe, ``{A Locally Supersymmetric and
  Reparametrization Invariant Action for the Spinning String},''
  \href{http://dx.doi.org/10.1016/0370-2693(76)90445-7}{{\em Phys. Lett. B}
  {\bfseries 65} (1976) 471--474}.

\bibitem{Deser:1976rb}
S.~Deser and B.~Zumino, ``{A Complete Action for the Spinning String},''
  \href{http://dx.doi.org/10.1016/0370-2693(76)90245-8}{{\em Phys. Lett. B}
  {\bfseries 65} (1976) 369--373}.

\bibitem{Goddard:1973qh}
P.~Goddard, J.~Goldstone, C.~Rebbi, and C.~B. Thorn, ``{Quantum dynamics of a
  massless relativistic string},''
  \href{http://dx.doi.org/10.1016/0550-3213(73)90223-X}{{\em Nucl. Phys. B}
  {\bfseries 56} (1973) 109--135}.

\bibitem{Shapiro:1972ph}
J.~A. Shapiro, ``{Loop graph in the dual tube model},''
  \href{http://dx.doi.org/10.1103/PhysRevD.5.1945}{{\em Phys. Rev. D}
  {\bfseries 5} (1972) 1945--1948}.

\bibitem{Friedan:1983xq}
D.~Friedan, Z.-a. Qiu, and S.~H. Shenker, ``{Conformal Invariance, Unitarity
  and Two-Dimensional Critical Exponents},''
  \href{http://dx.doi.org/10.1103/PhysRevLett.52.1575}{{\em Phys. Rev. Lett.}
  {\bfseries 52} (1984) 1575--1578}.

\bibitem{Friedan:1984rv}
D.~Friedan, Z.-a. Qiu, and S.~H. Shenker, ``{Superconformal Invariance in
  Two-Dimensions and the Tricritical Ising Model},''
  \href{http://dx.doi.org/10.1016/0370-2693(85)90819-6}{{\em Phys. Lett. B}
  {\bfseries 151} (1985) 37--43}.

\bibitem{Belavin:1984vu}
A.~A. Belavin, A.~M. Polyakov, and A.~B. Zamolodchikov, ``{Infinite Conformal
  Symmetry in Two-Dimensional Quantum Field Theory},''
  \href{http://dx.doi.org/10.1016/0550-3213(84)90052-X}{{\em Nucl. Phys. B}
  {\bfseries 241} (1984) 333--380}.

\bibitem{Friedan:1985ge}
D.~Friedan, E.~J. Martinec, and S.~H. Shenker, ``{Conformal Invariance,
  Supersymmetry and String Theory},''
  \href{http://dx.doi.org/10.1016/0550-3213(86)90356-1}{{\em Nucl. Phys. B}
  {\bfseries 271} (1986) 93--165}.

\bibitem{Ginsparg:1988ui}
P.~H. Ginsparg, ``{Applied Conformal Field Theory},'' in {\em {Les Houches
  Summer School in Theoretical Physics: Fields, Strings, Critical Phenomena}}.
\newblock 9, 1988.
\newblock \href{http://arxiv.org/abs/hep-th/9108028}{{\ttfamily
  arXiv:hep-th/9108028}}.

\bibitem{Itzykson:1989sx}
C.~Itzykson and J.~M. Drouffe,
  \href{http://dx.doi.org/10.1017/CBO9780511622779}{{\em {Statistical Field
  Theory. Vol. 1: From Brownian Motion To Renormalization And Lattice Gauge
  Theory}}}.
\newblock Cambridge Monographs on Mathematical Physics. CUP, 1989.

\bibitem{Itzykson:1989sy}
C.~Itzykson and J.~M. Drouffe,
  \href{http://dx.doi.org/10.1017/CBO9780511622786}{{\em {Statistical Field
  Theory. Vol. 2: Strong Coupling, Monte Carlo Methods, Conformal Field Theory,
  And Random Systems}}}.
\newblock Cambridge Monographs on Mathematical Physics. CUP, 1989.

\bibitem{DiFrancesco:1997nk}
P.~Di~Francesco, P.~Mathieu, and D.~Senechal,
  \href{http://dx.doi.org/10.1007/978-1-4612-2256-9}{{\em {Conformal Field
  Theory}}}.
\newblock Graduate Texts in Contemporary Physics. Springer-Verlag, New York,
  1997.

\bibitem{Gaberdiel:1999mc}
M.~R. Gaberdiel, ``{An Introduction to conformal field theory},''
  \href{http://dx.doi.org/10.1088/0034-4885/63/4/203}{{\em Rept. Prog. Phys.}
  {\bfseries 63} (2000) 607--667},
  \href{http://arxiv.org/abs/hep-th/9910156}{{\ttfamily arXiv:hep-th/9910156}}.

\bibitem{Blumenhagen:2009zz}
R.~Blumenhagen and E.~Plauschinn,
  \href{http://dx.doi.org/10.1007/978-3-642-00450-6}{{\em {Introduction to
  conformal field theory}: {with applications to String theory}}}, vol.~779 of
  {\em Lecture Notes in Physics}.
\newblock Springer, 2009.

\bibitem{Polyakov:1981rd}
A.~M. Polyakov, ``{Quantum Geometry of Bosonic Strings},''
  \href{http://dx.doi.org/10.1016/0370-2693(81)90743-7}{{\em Phys. Lett. B}
  {\bfseries 103} (1981) 207--210}.

\bibitem{Alvarez-Gaume:1980zra}
L.~Alvarez-Gaume and D.~Z. Freedman, ``{Kahler Geometry and the Renormalization
  of Supersymmetric Sigma Models},''
  \href{http://dx.doi.org/10.1103/PhysRevD.22.846}{{\em Phys. Rev. D}
  {\bfseries 22} (1980) 846}.

\bibitem{Alvarez-Gaume:1981exa}
L.~Alvarez-Gaume, D.~Z. Freedman, and S.~Mukhi, ``{The Background Field Method
  and the Ultraviolet Structure of the Supersymmetric Nonlinear Sigma Model},''
  \href{http://dx.doi.org/10.1016/0003-4916(81)90006-3}{{\em Annals Phys.}
  {\bfseries 134} (1981) 85}.

\bibitem{Alvarez-Gaume:1981exv}
L.~Alvarez-Gaume and D.~Z. Freedman, ``{Geometrical Structure and Ultraviolet
  Finiteness in the Supersymmetric Sigma Model},''
  \href{http://dx.doi.org/10.1007/BF01208280}{{\em Commun. Math. Phys.}
  {\bfseries 80} (1981) 443}.

\bibitem{Fradkin:1981dd}
E.~S. Fradkin and A.~A. Tseytlin, ``{Quantization of Two-Dimensional
  Supergravity and Critical Dimensions for String Models},''
  \href{http://dx.doi.org/10.1016/0370-2693(81)91081-9}{{\em Phys. Lett. B}
  {\bfseries 106} (1981) 63--68}.

\bibitem{Fradkin:1982ge}
E.~S. Fradkin and A.~A. Tseytlin, ``{On Quantized String Models},''
  \href{http://dx.doi.org/10.1016/0003-4916(82)90033-1}{{\em Annals Phys.}
  {\bfseries 143} (1982) 413}.

\bibitem{Lovelace:1983yv}
C.~Lovelace, ``{Strings in Curved Space},''
  \href{http://dx.doi.org/10.1016/0370-2693(84)90456-8}{{\em Phys. Lett. B}
  {\bfseries 135} (1984) 75--77}.

\bibitem{Callan:1985ia}
C.~G. Callan, Jr., E.~J. Martinec, M.~J. Perry, and D.~Friedan, ``{Strings in
  Background Fields},''
  \href{http://dx.doi.org/10.1016/0550-3213(85)90506-1}{{\em Nucl. Phys. B}
  {\bfseries 262} (1985) 593--609}.

\bibitem{Fradkin:1985qd}
E.~S. Fradkin and A.~A. Tseytlin, ``{Nonlinear Electrodynamics from Quantized
  Strings},'' \href{http://dx.doi.org/10.1016/0370-2693(85)90205-9}{{\em Phys.
  Lett. B} {\bfseries 163} (1985) 123--130}.

\bibitem{Virasoro:1969me}
M.~A. Virasoro, ``{Alternative constructions of crossing-symmetric amplitudes
  with Regge behavior},''
  \href{http://dx.doi.org/10.1103/PhysRev.177.2309}{{\em Phys. Rev.} {\bfseries
  177} (1969) 2309--2311}.

\bibitem{Shapiro:1969km}
J.~A. Shapiro, ``{Narrow-resonance model with Regge behavior for pi pi
  scattering},'' \href{http://dx.doi.org/10.1103/PhysRev.179.1345}{{\em Phys.
  Rev.} {\bfseries 179} (1969) 1345--1353}.

\bibitem{Polchinski:1985zf}
J.~Polchinski, ``{Evaluation of the One Loop String Path Integral},''
  \href{http://dx.doi.org/10.1007/BF01210791}{{\em Commun. Math. Phys.}
  {\bfseries 104} (1986) 37}.

\bibitem{Polyakov:1981re}
A.~M. Polyakov, ``{Quantum Geometry of Fermionic Strings},''
  \href{http://dx.doi.org/10.1016/0370-2693(81)90744-9}{{\em Phys. Lett. B}
  {\bfseries 103} (1981) 211--213}.

\bibitem{Gervais:1971ji}
J.-L. Gervais and B.~Sakita, ``{Field Theory Interpretation of Supergauges in
  Dual Models},'' \href{http://dx.doi.org/10.1016/0550-3213(71)90351-8}{{\em
  Nucl. Phys. B} {\bfseries 34} (1971) 632--639}.

\bibitem{Ramond:1971kx}
P.~Ramond, ``{An Interpretation of Dual Theories},''
  \href{http://dx.doi.org/10.1007/BF02731370}{{\em Nuovo Cim. A} {\bfseries 4}
  (1971) 544--548}.

\bibitem{Ramond:1971gb}
P.~Ramond, ``{Dual Theory for Free Fermions},''
  \href{http://dx.doi.org/10.1103/PhysRevD.3.2415}{{\em Phys. Rev. D}
  {\bfseries 3} (1971) 2415--2418}.

\bibitem{Neveu:1971rx}
A.~Neveu and J.~H. Schwarz, ``{Factorizable dual model of pions},''
  \href{http://dx.doi.org/10.1016/0550-3213(71)90448-2}{{\em Nucl. Phys. B}
  {\bfseries 31} (1971) 86--112}.

\bibitem{Neveu:1971iv}
A.~Neveu and J.~H. Schwarz, ``{Quark Model of Dual Pions},''
  \href{http://dx.doi.org/10.1103/PhysRevD.4.1109}{{\em Phys. Rev. D}
  {\bfseries 4} (1971) 1109--1111}.

\bibitem{Neveu:1971iw}
A.~Neveu, J.~H. Schwarz, and C.~B. Thorn, ``{Reformulation of the Dual Pion
  Model},'' \href{http://dx.doi.org/10.1016/0370-2693(71)90391-1}{{\em Phys.
  Lett. B} {\bfseries 35} (1971) 529--533}.

\bibitem{Scherk:1979zr}
J.~Scherk and J.~H. Schwarz, ``{How to Get Masses from Extra Dimensions},''
  \href{http://dx.doi.org/10.1016/0550-3213(79)90592-3}{{\em Nucl. Phys. B}
  {\bfseries 153} (1979) 61--88}.

\bibitem{Gliozzi:1976qd}
F.~Gliozzi, J.~Scherk, and D.~I. Olive, ``{Supersymmetry, Supergravity Theories
  and the Dual Spinor Model},''
  \href{http://dx.doi.org/10.1016/0550-3213(77)90206-1}{{\em Nucl. Phys. B}
  {\bfseries 122} (1977) 253--290}.

\bibitem{Bianchi:1990yu}
M.~Bianchi and A.~Sagnotti, ``{On the systematics of open string theories},''
  \href{http://dx.doi.org/10.1016/0370-2693(90)91894-H}{{\em Phys. Lett. B}
  {\bfseries 247} (1990) 517--524}.

\bibitem{Goddard:1986bp}
P.~Goddard and D.~I. Olive, ``{Kac-Moody and Virasoro Algebras in Relation to
  Quantum Physics},'' \href{http://dx.doi.org/10.1142/S0217751X86000149}{{\em
  Int. J. Mod. Phys. A} {\bfseries 1} (1986) 303}.

\bibitem{Dixon:1986iz}
L.~J. Dixon and J.~A. Harvey, ``{String Theories in Ten-Dimensions Without
  Space-Time Supersymmetry},''
  \href{http://dx.doi.org/10.1016/0550-3213(86)90619-X}{{\em Nucl. Phys. B}
  {\bfseries 274} (1986) 93--105}.

\bibitem{Seiberg:1986by}
N.~Seiberg and E.~Witten, ``{Spin Structures in String Theory},''
  \href{http://dx.doi.org/10.1016/0550-3213(86)90297-X}{{\em Nucl. Phys. B}
  {\bfseries 276} (1986) 272}.

\bibitem{Alvarez-Gaume:1983ihn}
L.~Alvarez-Gaume and E.~Witten, ``{Gravitational Anomalies},''
  \href{http://dx.doi.org/10.1016/0550-3213(84)90066-X}{{\em Nucl. Phys. B}
  {\bfseries 234} (1984) 269}.

\bibitem{Schellekens:1986yi}
A.~N. Schellekens and N.~P. Warner, ``{Anomalies and Modular Invariance in
  String Theory},'' \href{http://dx.doi.org/10.1016/0370-2693(86)90760-4}{{\em
  Phys. Lett. B} {\bfseries 177} (1986) 317--323}.

\bibitem{Schellekens:1986xh}
A.~N. Schellekens and N.~P. Warner, ``{Anomalies, Characters and Strings},''
  \href{http://dx.doi.org/10.1016/0550-3213(87)90108-8}{{\em Nucl. Phys. B}
  {\bfseries 287} (1987) 317}.

\bibitem{Lerche:1987sg}
W.~Lerche, B.~E.~W. Nilsson, and A.~N. Schellekens, ``{Heterotic String Loop
  Calculation of the Anomaly Cancelling Term},''
  \href{http://dx.doi.org/10.1016/0550-3213(87)90397-X}{{\em Nucl. Phys. B}
  {\bfseries 289} (1987) 609}.

\bibitem{Gross:1984dd}
D.~J. Gross, J.~A. Harvey, E.~J. Martinec, and R.~Rohm, ``{The Heterotic
  String},'' \href{http://dx.doi.org/10.1103/PhysRevLett.54.502}{{\em Phys.
  Rev. Lett.} {\bfseries 54} (1985) 502--505}.

\bibitem{Gross:1985fr}
D.~J. Gross, J.~A. Harvey, E.~J. Martinec, and R.~Rohm, ``{Heterotic String
  Theory. 1. The Free Heterotic String},''
  \href{http://dx.doi.org/10.1016/0550-3213(85)90394-3}{{\em Nucl. Phys. B}
  {\bfseries 256} (1985) 253}.

\bibitem{Gross:1985rr}
D.~J. Gross, J.~A. Harvey, E.~J. Martinec, and R.~Rohm, ``{Heterotic String
  Theory. 2. The Interacting Heterotic String},''
  \href{http://dx.doi.org/10.1016/0550-3213(86)90146-X}{{\em Nucl. Phys. B}
  {\bfseries 267} (1986) 75--124}.

\bibitem{Green:1984sg}
M.~B. Green and J.~H. Schwarz, ``{Anomaly Cancellation in Supersymmetric $D=10$
  Gauge Theory and Superstring Theory},''
  \href{http://dx.doi.org/10.1016/0370-2693(84)91565-X}{{\em Phys. Lett. B}
  {\bfseries 149} (1984) 117--122}.

\bibitem{Bilal:2008qx}
A.~Bilal, ``{Lectures on Anomalies},''
  \href{http://arxiv.org/abs/0802.0634}{{\ttfamily arXiv:0802.0634 [hep-th]}}.

\bibitem{Alvarez-Gaume:2022aak}
L.~Alvarez-Gaume and M.~A. Vazquez-Mozo, ``{Anomalies and the Green-Schwarz
  Mechanism},'' \href{http://arxiv.org/abs/2211.06467}{{\ttfamily
  arXiv:2211.06467 [hep-th]}}.

\bibitem{Coleman:1974bu}
S.~R. Coleman, ``{The Quantum Sine-Gordon Equation as the Massive Thirring
  Model},'' \href{http://dx.doi.org/10.1103/PhysRevD.11.2088}{{\em Phys. Rev.
  D} {\bfseries 11} (1975) 2088}.

\bibitem{Mandelstam:1975hb}
S.~Mandelstam, ``{Soliton Operators for the Quantized Sine-Gordon Equation},''
  \href{http://dx.doi.org/10.1103/PhysRevD.11.3026}{{\em Phys. Rev. D}
  {\bfseries 11} (1975) 3026}.

\bibitem{Alvarez-Gaume:1986ghj}
L.~Alvarez-Gaum{\'e}, P.~H. Ginsparg, G.~W. Moore, and C.~Vafa, ``{An
  $\text{O}(16) \times \text{O}(16)$ Heterotic String},''
  \href{http://dx.doi.org/10.1016/0370-2693(86)91524-8}{{\em Phys. Lett. B}
  {\bfseries 171} (1986) 155--162}.

\bibitem{Kawai:1986vd}
H.~Kawai, D.~C. Lewellen, and S.~H.~H. Tye, ``{Classification of Closed
  Fermionic String Models},''
  \href{http://dx.doi.org/10.1103/PhysRevD.34.3794}{{\em Phys. Rev. D}
  {\bfseries 34} (1986) 3794}.

\bibitem{Chaudhuri:1995fk}
S.~Chaudhuri, G.~Hockney, and J.~D. Lykken, ``{Maximally supersymmetric string
  theories in $D < 10$},''
  \href{http://dx.doi.org/10.1103/PhysRevLett.75.2264}{{\em Phys. Rev. Lett.}
  {\bfseries 75} (1995) 2264--2267},
  \href{http://arxiv.org/abs/hep-th/9505054}{{\ttfamily arXiv:hep-th/9505054}}.

\bibitem{Dai:1989ua}
J.~Dai, R.~G. Leigh, and J.~Polchinski, ``{New Connections Between String
  Theories},'' \href{http://dx.doi.org/10.1142/S0217732389002331}{{\em Mod.
  Phys. Lett. A} {\bfseries 4} (1989) 2073--2083}.

\bibitem{Leigh:1989jq}
R.~G. Leigh, ``{Dirac-Born-Infeld Action from Dirichlet Sigma Model},''
  \href{http://dx.doi.org/10.1142/S0217732389003099}{{\em Mod. Phys. Lett. A}
  {\bfseries 4} (1989) 2767}.

\bibitem{Horava:1989ga}
P.~Horava, ``{Background Duality of Open String Models},''
  \href{http://dx.doi.org/10.1016/0370-2693(89)90209-8}{{\em Phys. Lett. B}
  {\bfseries 231} (1989) 251--257}.

\bibitem{Polchinski:1987tu}
J.~Polchinski and Y.~Cai, ``{Consistency of Open Superstring Theories},''
  \href{http://dx.doi.org/10.1016/0550-3213(88)90382-3}{{\em Nucl. Phys. B}
  {\bfseries 296} (1988) 91--128}.

\bibitem{Sagnotti:1987tw}
A.~Sagnotti, ``{Open Strings and their Symmetry Groups},'' in {\em {NATO
  Advanced Summer Institute on Nonperturbative Quantum Field Theory (Cargese
  Summer Institute)}}.
\newblock 9, 1987.
\newblock \href{http://arxiv.org/abs/hep-th/0208020}{{\ttfamily
  arXiv:hep-th/0208020}}.

\bibitem{Bianchi:1988fr}
M.~Bianchi and A.~Sagnotti, ``{The Partition Function of the SO(8192) Bosonic
  String},'' \href{http://dx.doi.org/10.1016/0370-2693(88)91884-9}{{\em Phys.
  Lett. B} {\bfseries 211} (1988) 407--416}.

\bibitem{Horava:1989vt}
P.~Horava, ``{Strings on World Sheet Orbifolds},''
  \href{http://dx.doi.org/10.1016/0550-3213(89)90279-4}{{\em Nucl. Phys. B}
  {\bfseries 327} (1989) 461--484}.

\bibitem{Polchinski:1995mt}
J.~Polchinski, ``{Dirichlet Branes and Ramond-Ramond charges},''
  \href{http://dx.doi.org/10.1103/PhysRevLett.75.4724}{{\em Phys. Rev. Lett.}
  {\bfseries 75} (1995) 4724--4727},
  \href{http://arxiv.org/abs/hep-th/9510017}{{\ttfamily arXiv:hep-th/9510017}}.

\bibitem{Hull:1994ys}
C.~M. Hull and P.~K. Townsend, ``{Unity of superstring dualities},''
  \href{http://dx.doi.org/10.1016/0550-3213(94)00559-W}{{\em Nucl. Phys. B}
  {\bfseries 438} (1995) 109--137},
  \href{http://arxiv.org/abs/hep-th/9410167}{{\ttfamily arXiv:hep-th/9410167}}.

\bibitem{Townsend:1995kk}
P.~K. Townsend, ``{The eleven-dimensional supermembrane revisited},''
  \href{http://dx.doi.org/10.1016/0370-2693(95)00397-4}{{\em Phys. Lett. B}
  {\bfseries 350} (1995) 184--187},
  \href{http://arxiv.org/abs/hep-th/9501068}{{\ttfamily arXiv:hep-th/9501068}}.

\bibitem{Witten:1995ex}
E.~Witten, ``{String theory dynamics in various dimensions},''
  \href{http://dx.doi.org/10.1016/0550-3213(95)00158-O}{{\em Nucl. Phys. B}
  {\bfseries 443} (1995) 85--126},
  \href{http://arxiv.org/abs/hep-th/9503124}{{\ttfamily arXiv:hep-th/9503124}}.

\bibitem{Duff:1994an}
M.~J. Duff, R.~R. Khuri, and J.~X. Lu, ``{String solitons},''
  \href{http://dx.doi.org/10.1016/0370-1573(95)00002-X}{{\em Phys. Rept.}
  {\bfseries 259} (1995) 213--326},
  \href{http://arxiv.org/abs/hep-th/9412184}{{\ttfamily arXiv:hep-th/9412184}}.

\bibitem{Bianchi:1991eu}
M.~Bianchi, G.~Pradisi, and A.~Sagnotti, ``{Toroidal compactification and
  symmetry breaking in open string theories},''
  \href{http://dx.doi.org/10.1016/0550-3213(92)90129-Y}{{\em Nucl. Phys. B}
  {\bfseries 376} (1992) 365--386}.

\bibitem{Witten:1995im}
E.~Witten, ``{Bound states of strings and p-branes},''
  \href{http://dx.doi.org/10.1016/0550-3213(95)00610-9}{{\em Nucl. Phys. B}
  {\bfseries 460} (1996) 335--350},
  \href{http://arxiv.org/abs/hep-th/9510135}{{\ttfamily arXiv:hep-th/9510135}}.

\bibitem{Schwarz:1982md}
J.~H. Schwarz, ``{Gauge Groups for Type I Superstrings},'' in {\em {6th Johns
  Hopkins Workshop on Current Problems in Particle Theory}}.
\newblock 6, 1982.

\bibitem{Marcus:1982fr}
N.~Marcus and A.~Sagnotti, ``{Tree Level Constraints on Gauge Groups for Type I
  Superstrings},'' \href{http://dx.doi.org/10.1016/0370-2693(82)90253-2}{{\em
  Phys. Lett. B} {\bfseries 119} (1982) 97--99}.

\bibitem{Marcus:1986cm}
N.~Marcus and A.~Sagnotti, ``{Group Theory from Quarks at the Ends of
  Strings},'' \href{http://dx.doi.org/10.1016/0370-2693(87)90705-2}{{\em Phys.
  Lett. B} {\bfseries 188} (1987) 58--64}.

\bibitem{Pradisi:1988xd}
G.~Pradisi and A.~Sagnotti, ``{Open String Orbifolds},''
  \href{http://dx.doi.org/10.1016/0370-2693(89)91369-5}{{\em Phys. Lett. B}
  {\bfseries 216} (1989) 59--67}.

\bibitem{Bianchi:1990tb}
M.~Bianchi and A.~Sagnotti, ``{Twist symmetry and open string Wilson lines},''
  \href{http://dx.doi.org/10.1016/0550-3213(91)90271-X}{{\em Nucl. Phys. B}
  {\bfseries 361} (1991) 519--538}.

\bibitem{Dudas:2000bn}
E.~Dudas, ``{Theory and phenomenology of type I strings and M theory},''
  \href{http://dx.doi.org/10.1088/0264-9381/17/22/201}{{\em Class. Quant.
  Grav.} {\bfseries 17} (2000) R41--R116},
  \href{http://arxiv.org/abs/hep-ph/0006190}{{\ttfamily arXiv:hep-ph/0006190}}.

\bibitem{Douglas:1986eu}
M.~R. Douglas and B.~Grinstein, ``{Dilaton Tadpole for the Open Bosonic
  String},'' \href{http://dx.doi.org/10.1016/0370-2693(87)91416-X}{{\em Phys.
  Lett. B} {\bfseries 183} (1987) 52}. [Erratum: Phys.Lett.B 187, 442 (1987)].

\bibitem{Weinberg:1987ie}
S.~Weinberg, ``{Cancellation of One Loop Divergences in SO(8192) String
  Theory},'' \href{http://dx.doi.org/10.1016/0370-2693(87)91096-3}{{\em Phys.
  Lett. B} {\bfseries 187} (1987) 278--282}.

\bibitem{Paton:1969je}
J.~E. Paton and H.-M. Chan, ``{Generalized veneziano model with isospin},''
  \href{http://dx.doi.org/10.1016/0550-3213(69)90038-8}{{\em Nucl. Phys. B}
  {\bfseries 10} (1969) 516--520}.

\bibitem{Aldazabal:1999nu}
G.~Aldazabal, D.~Badagnani, L.~E. Ibanez, and A.~M. Uranga, ``{Tadpole versus
  anomaly cancellation in D = 4, D = 6 compact IIB orientifolds},''
  \href{http://dx.doi.org/10.1088/1126-6708/1999/06/031}{{\em JHEP} {\bfseries
  06} (1999) 031}, \href{http://arxiv.org/abs/hep-th/9904071}{{\ttfamily
  arXiv:hep-th/9904071}}.

\bibitem{Scrucca:1999uz}
C.~A. Scrucca and M.~Serone, ``{Anomalies and inflow on D-branes and O -
  planes},'' \href{http://dx.doi.org/10.1016/S0550-3213(99)00357-0}{{\em Nucl.
  Phys. B} {\bfseries 556} (1999) 197--221},
  \href{http://arxiv.org/abs/hep-th/9903145}{{\ttfamily arXiv:hep-th/9903145}}.

\bibitem{Morales:1998ux}
J.~F. Morales, C.~A. Scrucca, and M.~Serone, ``{Anomalous couplings for
  D-branes and O-planes},''
  \href{http://dx.doi.org/10.1016/S0550-3213(99)00217-5}{{\em Nucl. Phys. B}
  {\bfseries 552} (1999) 291--315},
  \href{http://arxiv.org/abs/hep-th/9812071}{{\ttfamily arXiv:hep-th/9812071}}.

\bibitem{Bianchi:2000de}
M.~Bianchi and J.~F. Morales, ``{Anomalies \& tadpoles},''
  \href{http://dx.doi.org/10.1088/1126-6708/2000/03/030}{{\em JHEP} {\bfseries
  03} (2000) 030}, \href{http://arxiv.org/abs/hep-th/0002149}{{\ttfamily
  arXiv:hep-th/0002149}}.

\bibitem{Polchinski:1995df}
J.~Polchinski and E.~Witten, ``{Evidence for heterotic - type I string
  duality},'' \href{http://dx.doi.org/10.1016/0550-3213(95)00614-1}{{\em Nucl.
  Phys. B} {\bfseries 460} (1996) 525--540},
  \href{http://arxiv.org/abs/hep-th/9510169}{{\ttfamily arXiv:hep-th/9510169}}.

\bibitem{Fischler:1986ci}
W.~Fischler and L.~Susskind, ``{Dilaton Tadpoles, String Condensates and Scale
  Invariance},'' \href{http://dx.doi.org/10.1016/0370-2693(86)91425-5}{{\em
  Phys. Lett. B} {\bfseries 171} (1986) 383--389}.

\bibitem{Fischler:1986tb}
W.~Fischler and L.~Susskind, ``{Dilaton Tadpoles, String Condensates and Scale
  Invariance. 2.},'' \href{http://dx.doi.org/10.1016/0370-2693(86)90514-9}{{\em
  Phys. Lett. B} {\bfseries 173} (1986) 262--264}.

\bibitem{Sugimoto:1999tx}
S.~Sugimoto, ``{Anomaly cancellations in type I D9 - anti-D9 system and the
  USp(32) string theory},'' \href{http://dx.doi.org/10.1143/PTP.102.685}{{\em
  Prog. Theor. Phys.} {\bfseries 102} (1999) 685--699},
  \href{http://arxiv.org/abs/hep-th/9905159}{{\ttfamily arXiv:hep-th/9905159}}.

\bibitem{Dudas:2000nv}
E.~Dudas and J.~Mourad, ``{Consistent gravitino couplings in nonsupersymmetric
  strings},'' \href{http://dx.doi.org/10.1016/S0370-2693(01)00777-8}{{\em Phys.
  Lett. B} {\bfseries 514} (2001) 173--182},
  \href{http://arxiv.org/abs/hep-th/0012071}{{\ttfamily arXiv:hep-th/0012071}}.

\bibitem{Pradisi:2001yv}
G.~Pradisi and F.~Riccioni, ``{Geometric couplings and brane supersymmetry
  breaking},'' \href{http://dx.doi.org/10.1016/S0550-3213(01)00441-2}{{\em
  Nucl. Phys. B} {\bfseries 615} (2001) 33--60},
  \href{http://arxiv.org/abs/hep-th/0107090}{{\ttfamily arXiv:hep-th/0107090}}.

\bibitem{Dudas:2000ff}
E.~Dudas and J.~Mourad, ``{Brane solutions in strings with broken supersymmetry
  and dilaton tadpoles},''
  \href{http://dx.doi.org/10.1016/S0370-2693(00)00734-6}{{\em Phys. Lett. B}
  {\bfseries 486} (2000) 172--178},
  \href{http://arxiv.org/abs/hep-th/0004165}{{\ttfamily arXiv:hep-th/0004165}}.

\bibitem{Sagnotti:1995ga}
A.~Sagnotti, ``{Some properties of open string theories},'' in {\em
  {International Workshop on Supersymmetry and Unification of Fundamental
  Interactions (SUSY 95)}}, pp.~473--484.
\newblock 9, 1995.
\newblock \href{http://arxiv.org/abs/hep-th/9509080}{{\ttfamily
  arXiv:hep-th/9509080}}.

\bibitem{Sagnotti:1996qj}
A.~Sagnotti, ``{Surprises in open string perturbation theory},''
  \href{http://dx.doi.org/10.1016/S0920-5632(97)00344-7}{{\em Nucl. Phys. B
  Proc. Suppl.} {\bfseries 56} (1997) 332--343},
  \href{http://arxiv.org/abs/hep-th/9702093}{{\ttfamily arXiv:hep-th/9702093}}.

\bibitem{Fioravanti:1993hf}
D.~Fioravanti, G.~Pradisi, and A.~Sagnotti, ``{Sewing constraints and
  nonorientable open strings},''
  \href{http://dx.doi.org/10.1016/0370-2693(94)90255-0}{{\em Phys. Lett. B}
  {\bfseries 321} (1994) 349--354},
  \href{http://arxiv.org/abs/hep-th/9311183}{{\ttfamily arXiv:hep-th/9311183}}.

\bibitem{Pradisi:1995qy}
G.~Pradisi, A.~Sagnotti, and Y.~S. Stanev, ``{Planar duality in SU(2) WZW
  models},'' \href{http://dx.doi.org/10.1016/0370-2693(95)00532-P}{{\em Phys.
  Lett. B} {\bfseries 354} (1995) 279--286},
  \href{http://arxiv.org/abs/hep-th/9503207}{{\ttfamily arXiv:hep-th/9503207}}.

\bibitem{Pradisi:1995pp}
G.~Pradisi, A.~Sagnotti, and Y.~S. Stanev, ``{The Open descendants of
  nondiagonal SU(2) WZW models},''
  \href{http://dx.doi.org/10.1016/0370-2693(95)00840-H}{{\em Phys. Lett. B}
  {\bfseries 356} (1995) 230--238},
  \href{http://arxiv.org/abs/hep-th/9506014}{{\ttfamily arXiv:hep-th/9506014}}.

\bibitem{Pradisi:1996yd}
G.~Pradisi, A.~Sagnotti, and Y.~S. Stanev, ``{Completeness conditions for
  boundary operators in 2-D conformal field theory},''
  \href{http://dx.doi.org/10.1016/0370-2693(96)00578-3}{{\em Phys. Lett. B}
  {\bfseries 381} (1996) 97--104},
  \href{http://arxiv.org/abs/hep-th/9603097}{{\ttfamily arXiv:hep-th/9603097}}.

\bibitem{Huiszoon:1999xq}
L.~R. Huiszoon, A.~N. Schellekens, and N.~Sousa, ``{Klein bottles and simple
  currents},'' \href{http://dx.doi.org/10.1016/S0370-2693(99)01241-1}{{\em
  Phys. Lett. B} {\bfseries 470} (1999) 95--102},
  \href{http://arxiv.org/abs/hep-th/9909114}{{\ttfamily arXiv:hep-th/9909114}}.

\bibitem{Huiszoon:2000ge}
L.~R. Huiszoon and A.~N. Schellekens, ``{Crosscaps, boundaries and T
  duality},'' \href{http://dx.doi.org/10.1016/S0550-3213(00)00320-5}{{\em Nucl.
  Phys. B} {\bfseries 584} (2000) 705--718},
  \href{http://arxiv.org/abs/hep-th/0004100}{{\ttfamily arXiv:hep-th/0004100}}.

\bibitem{Fuchs:2000cm}
J.~Fuchs, L.~R. Huiszoon, A.~N. Schellekens, C.~Schweigert, and J.~Walcher,
  ``{Boundaries, crosscaps and simple currents},''
  \href{http://dx.doi.org/10.1016/S0370-2693(00)01271-5}{{\em Phys. Lett. B}
  {\bfseries 495} (2000) 427--434},
  \href{http://arxiv.org/abs/hep-th/0007174}{{\ttfamily arXiv:hep-th/0007174}}.

\bibitem{Sagnotti:1992qw}
A.~Sagnotti, ``{A Note on the Green-Schwarz mechanism in open string
  theories},'' \href{http://dx.doi.org/10.1016/0370-2693(92)90682-T}{{\em Phys.
  Lett. B} {\bfseries 294} (1992) 196--203},
  \href{http://arxiv.org/abs/hep-th/9210127}{{\ttfamily arXiv:hep-th/9210127}}.

\end{thebibliography}
\providecommand{\href}[2]{#2}\begingroup\raggedright\endgroup

\end{document}